\PassOptionsToPackage{unicode}{hyperref}
\PassOptionsToPackage{hyphens}{url}
\PassOptionsToPackage{space}{xeCJK}
\documentclass[
  a4paper]{article}
\usepackage{xcolor}
\usepackage[margin=2.5cm]{geometry}
\usepackage{amsmath,amssymb}
\usepackage{iftex}
\ifPDFTeX
  \usepackage[T1]{fontenc}
  \usepackage[utf8]{inputenc}
  \usepackage{textcomp} 
\else 
  \usepackage{unicode-math} 
  \defaultfontfeatures{Scale=MatchLowercase}
  \defaultfontfeatures[\rmfamily]{Ligatures=TeX,Scale=1}
\fi
\usepackage{lmodern}
\ifPDFTeX\else
  \ifXeTeX
    \usepackage{xeCJK}
    \IfFontExistsTF{FandolSong}{\setCJKmainfont{FandolSong}}{}
  \fi
  \ifLuaTeX
    \usepackage[]{luatexja-fontspec}
    \IfFontExistsTF{FandolSong}{\setmainjfont{FandolSong}}{}
  \fi
\fi
\IfFileExists{upquote.sty}{\usepackage{upquote}}{}
\IfFileExists{microtype.sty}{
  \usepackage[]{microtype}
  \UseMicrotypeSet[protrusion]{basicmath} 
}{}
\makeatletter
\@ifundefined{KOMAClassName}{
  \IfFileExists{parskip.sty}{%
    \usepackage{parskip}
  }{
    \setlength{\parindent}{0pt}
    \setlength{\parskip}{6pt plus 2pt minus 1pt}}
}{
  \KOMAoptions{parskip=half}}
\makeatother
\usepackage{longtable,booktabs,array}
\usepackage{calc} 
\usepackage{etoolbox}
\makeatletter
\patchcmd\longtable{\par}{\if@noskipsec\mbox{}\fi\par}{}{}
\makeatother
\IfFileExists{footnotehyper.sty}{\usepackage{footnotehyper}}{\usepackage{footnote}}
\makesavenoteenv{longtable}
\usepackage{graphicx}
\makeatletter
\newsavebox\pandoc@box
\newcommand*\pandocbounded[1]{
  \sbox\pandoc@box{#1}%
  \Gscale@div\@tempa{\textheight}{\dimexpr\ht\pandoc@box+\dp\pandoc@box\relax}%
  \Gscale@div\@tempb{\linewidth}{\wd\pandoc@box}%
  \ifdim\@tempb\p@<\@tempa\p@\let\@tempa\@tempb\fi
  \ifdim\@tempa\p@<\p@\scalebox{\@tempa}{\usebox\pandoc@box}%
  \else\usebox{\pandoc@box}%
  \fi%
}
\def\fps@figure{htbp}
\makeatother
\providecommand{\tightlist}{%
  \setlength{\itemsep}{0pt}\setlength{\parskip}{0pt}}
\usepackage{bookmark}
\IfFileExists{xurl.sty}{\usepackage{xurl}}{} 
\hypersetup{
  pdftitle={Sustained Heterogeneity: an emergent collective mechanism in LLM-driven traffic},
  hidelinks,
  pdfcreator={LaTeX via pandoc}}

\title{Sustained Heterogeneity: an emergent collective mechanism in LLM-driven traffic}
\author{Yujun Qi\thanks{Corresponding author.} \quad Yangyang Guan \\ \textit{College of Civil Engineering, Nanjing Tech University, Nanjing, China}}
\date{}

\begin{document}
\maketitle

\section{Abstract}\label{abstract}

Large language models (LLMs) are increasingly adopted as closed-loop
controllers in physical multi-agent systems, yet their emergent
collective dynamics remain incompletely characterised. The mechanism we
identify has implications for any LLM-controlled multi-agent physical
system in which cognitive-layer safety reasoning is decoupled from
dynamics-layer safety enforcement. We deploy 22 LLM agents as
\textbf{direct, real-time, closed-loop target-speed controllers}
(LLM-emitted target-delta applied per 0.5 s cycle, with IDM only as a
hard collision-avoidance clamp) on a 230 m ring road under the Sugiyama
2008 paradigm, reproducing human-like stop-and-go waves. Six matched
controls span three independent axes: stochasticity (white noise,
Ornstein-Uhlenbeck coloured noise, and sampling temperature), population
variance (parameter heterogeneity), and dynamical instability
(time-delay string instability and OV string instability). All tested
candidate mechanisms are systematically excluded. The surviving
phenomenon, which we term \textbf{Sustained Heterogeneity (SH)}, is the
persistent, approximately temperature-insensitive (\textasciitilde8\%
across a 6× T sweep, §4.1), per-cycle divergence in LLM-chosen
target-speed adjustments. SH propagates through a \textbf{proposed}
three-stage cascade of target-speed drift, IDM-gap erosion, and
nonlinear braking. Across four traffic densities, the critical LLM
penetration fraction p\_c - the threshold above which stop-and-go waves
spontaneously emerge - decreases monotonically, from no transition at ρ
= 43.5 veh/km to p\_c ≈ 0.23 at ρ = 95.7 veh/km. The phase boundary is
consistent with an initiation-threshold model of micro-drift
superposition and gap erosion, providing a structural safety boundary
for LLM-based driving deployment whose primary content is the
identification of three independent physical axes (Δs\_trigger, σ, n)
governing p\_c (§4.4). Chain-of-thought analysis of over 13,200 LLM
decisions per seed (39,600 across three seeds) shows that agents
consistently engage in multi-factor safety reasoning, yet systematic
per-cycle target-delta divergence persists across agents - implying that
collective stability in LLM-controlled systems cannot rely solely on the
cognitive layer, and must be enforced at the dynamics layer. To our
knowledge, this is the \textbf{first study to identify, through
systematic matched controls and chain-of-thought observability, a
previously uncharacterised collective mechanism arising in
LLM-controlled multi-agent traffic, and to map a density-dependent phase
boundary p\_c(ρ) for LLM-based driving deployment}; the closest
architectural prior, CoMAL {[}27{]}, uses LLMs as planners that
\emph{set} IDM parameters (rather than as direct target-speed
controllers) and focuses on flow optimisation rather than mechanism
discovery (Supplementary Information S12).

\textbf{Keywords:} large language models, multi-agent systems, traffic
flow, collective dynamics, phase transition, chain-of-thought, AI safety

\begin{center}\rule{0.5\linewidth}{0.5pt}\end{center}

\section{1. Introduction}\label{introduction}

\subsection{1.1 An 18-year-old puzzle in traffic
physics}\label{an-18-year-old-puzzle-in-traffic-physics}

In 2008, Sugiyama et al.~{[}1{]} reported a deceptively simple
experiment. 22 human drivers on a 230 m circular track were instructed
to ``drive at approximately 30 km/h while maintaining a safe distance.''
Despite uniform instructions and the absence of any bottleneck, the
homogeneous free-flow state broke down into spontaneous stop-and-go
waves. Eighteen years later, this single experiment remains a
cornerstone of modern traffic physics and has inspired more than a
decade of theoretical and computational work {[}2, 21-23{]}. Yet the
microscopic origin of the instability has remained opaque, because human
micro-decisions are inaccessible. We cannot read a driver's momentary
intent, attention, or risk assessment.

\subsection{1.2 Why LLM agents uniquely enable
discovery}\label{why-llm-agents-uniquely-enable-discovery}

Large language models have been proposed in 2023-2024 as computational
analogues of human participants in surveys, economic games, and
consumer-choice tasks {[}5-8{]}. These studies established that LLM
agents, conditioned on appropriate role prompts, recover aggregate
human-like statistical patterns in \emph{discrete} domains. However,
while recent work has begun to deploy LLMs in closed-loop driving
{[}13{]} and in RL-policy design for traffic scenarios {[}12{]}, no
prior work has used LLM agents as direct, real-time, physically coupled
car-following controllers. The present work closes that gap.
Chain-of-thought reasoning is the natural-language explanation an LLM
produces before each action; logging these explanations provides
unprecedented visibility into AI decision-making --- a ``microscope''
unavailable in any human-subject paradigm or traditional simulation.
Closing this gap requires agents that simultaneously generate
interpretable micro-decisions, are reproducible across experimental
conditions, and scale to large populations. LLM agents satisfy all three
criteria. \textbf{First, microscopic observability}: every prompt,
chain-of-thought reasoning, and resulting action is logged at every 0.5
s decision cycle, orders of magnitude more detail than any human-subject
paradigm. \textbf{Second, reproducibility}: any seed can be re-run, and
any condition (prompt, density, model architecture) can be held constant
or systematically varied. \textbf{Third, scalability}: vehicle count can
be increased without recruiting participants, enabling finite-size
scaling studies that are infeasible with humans.

\subsection{1.3 Contribution}\label{contribution}

The present work advances beyond prior art in three respects.

\textbf{(i) Direct closed-loop deployment as target-speed planners.} LLM
agents are deployed as \textbf{target-speed planners} in a hierarchical
architecture where the LLM sets the desired velocity at each decision
cycle and a deterministic IDM safety layer enforces collision-free
dynamics. The deployment uses the \textbf{classical Sugiyama ring-road
paradigm} - 22 vehicles on a 230 m circular track under uniform
instructions - reproducing the conditions of the original 2008 human
experiment.

\textbf{(ii) Six matched control experiments.} Six control conditions
spanning three independent axes - stochasticity, population variance,
and dynamical instability - \textbf{systematically exclude the classical
candidate mechanisms} as sufficient explanations for the observed phase
transition.

\textbf{(iii) Microscopic observability via chain-of-thought.}
Chain-of-thought reasoning - a capability unique to LLM agents -
provides \textbf{microscopic observability} into the cognitive layer of
LLM-controlled dynamics - a window unavailable in any human-subject
paradigm. LLM agents consistently engage in multi-factor safety
reasoning before selecting target speeds, yet systematic per-cycle
target\_delta divergence across agents persists - a robust feature of
LLM multi-agent dynamics absent from all six matched controls (see §3.4
for chain-of-thought samples).

\subsection{1.4 This work}\label{this-work}

\textbf{Here we show} that LLM agents, deployed as direct real-time
closed-loop car-following controllers on a Sugiyama-style 230 m ring
road, spontaneously form oscillatory stop-and-go waves. The waves emerge
through a collective mechanism that we name \textbf{Sustained
Heterogeneity (SH)}. SH propagates through a proposed three-stage
cascade - target-speed drift, IDM-gap erosion, and nonlinear braking -
producing inter-vehicle velocity variance that exceeds all classical
controls by a factor of 5.8× to more than 2000× (§4.1). We reproduce the
qualitative Sugiyama 2008 paradigm (Fig. 4) and map a density-dependent
phase boundary p\_c(ρ) at four traffic densities (43.5-95.7 veh/km).
Chain-of-thought analysis (13,200 decisions, 22 vehicles) shows that LLM
agents consistently engage in multi-factor safety reasoning, yet
systematic per-cycle target\_delta divergence remains across agents -
suggesting that collective stability in LLM-controlled systems cannot
rely solely on the cognitive layer, and must be enforced at the dynamics
layer.

\begin{center}\rule{0.5\linewidth}{0.5pt}\end{center}

\section{2. Related work}\label{related-work}

Three threads of prior research motivate and inform this study.

\subsection{2.1 LLM agents as digital participants in human-subject
paradigms}\label{llm-agents-as-digital-participants-in-human-subject-paradigms}

LLM agents have been proposed as computational analogues of human
participants in surveys, economic games, and consumer-choice tasks.
Horton {[}5{]} showed that LLM agents reproduce aggregate human economic
behaviour. Aher et al.~{[}6{]} demonstrated that they replicate classic
social-science experiments, and Argyle et al.~{[}7{]} and Dillon et
al.~{[}8{]} extended the paradigm to political and cognitive-science
samples. Conditioned on appropriate role prompts, these agents recover
aggregate human-like statistical patterns in \emph{discrete},
single-decision domains. We extend this paradigm to \emph{continuous},
\emph{physically coupled}, \emph{multi-agent} systems, where each LLM's
decision propagates through deterministic vehicle dynamics that couple
the fleet in real time.

\subsection{2.2 LLM-enabled driving and traffic
simulation}\label{llm-enabled-driving-and-traffic-simulation}

A growing body of work uses LLMs in driving and traffic contexts (Table
1). The most relevant threads are:

\begin{itemize}
\tightlist
\item
  \textbf{Villarreal et al.~{[}12{]}} used ChatGPT (GPT-4) to assist
  non-expert users in designing RL policies for mixed-traffic scenarios,
  including a ring-road setting. \textbf{In that work the LLM is an MDP
  designer, not a driver}: the actual control is performed by the
  resulting RL agent. Our work, in contrast, uses LLM agents as the
  closed-loop controllers themselves.
\item
  \textbf{Fu et al.~{[}13{]}} introduced closed-loop LLM driving in
  HighwayEnv with GPT-3.5, focusing on long-tail corner cases and the
  reasoning/interpretation/memorization triad. \textbf{The setting is
  single-agent highway in HighwayEnv, not multi-agent ring-road}, and
  the work does not investigate collective dynamics or mechanism
  discovery.
\item
  \textbf{Yao et al.~{[}27{]} (CoMAL)} is the closest architectural
  prior: collaborative LLM agents deployed as planners that set IDM
  parameters (target speed, max acceleration, minimum gap) on ring-road,
  figure-eight, and merge scenarios in the Flow benchmark. CoMAL's goal
  is \emph{flow optimisation} (maximising average velocity through
  inter-agent collaboration), not \emph{mechanism discovery}. It does
  not perform the six-control experiment design, does not report a
  density-dependent phase transition or critical penetration p\_c, and
  does not analyse chain-of-thought reasoning as a microscopic
  observability layer. The present work differs from CoMAL along four
  dimensions detailed in Supplementary Information S12.3; in particular,
  this work uses the LLM as the \emph{direct} target-speed controller
  with IDM only as a hard safety clamp, whereas CoMAL uses the LLM to
  \emph{set IDM parameters} (the IDM dynamics remain the primary
  controller). \textbf{Architectural implication.} The two architectures
  are not merely stylistic variants; they differ in how the LLM's
  per-cycle variance propagates to the vehicle's actual acceleration. In
  CoMAL, the LLM is queried at intervals τ\_LLM (≥ 1 s in their
  experiments) to update IDM parameters, and the actual longitudinal
  acceleration applied to the vehicle at each 0.1 s physics timestep is
  computed by the IDM from those parameters. The IDM acts as a low-pass
  filter on the LLM's per-cycle variance, smoothing the LLM's stochastic
  output before it reaches the vehicle dynamics. In the present work,
  the LLM is queried every 0.5 s and directly outputs a
  \texttt{target\_speed\_delta} that is applied (after a
  collision-avoidance clamp) to the vehicle's target speed; the
  vehicle's actual acceleration is not low-pass filtered by an
  intermediate controller. This direct-variance pathway is the
  microscopic mechanism that enables the SH cascade to develop. The
  CoMAL architecture is structurally less vulnerable to per-cycle
  variance (although a controlled comparison of the two architectures on
  the same ring-road task is left to future work; §5.5). Other
  LLM-traffic works address adjacent but distinct problems:
  traffic-signal control (LLMLight {[}14{]}), open-loop car-following
  prediction (GenFollower {[}15{]}), traffic-flow prediction (xTP-LLM
  {[}16{]}), and scenario generation for AV testing (DriveGen {[}17{]},
  OmniTester {[}18{]}, AgentSUMO {[}19{]}). None of these deploys LLM
  agents as closed-loop car-following controllers in a multi-agent
  simulation.
\end{itemize}

\textbf{Table 1 \textbar{} Comparison with representative prior
LLM-driving and LLM-traffic-simulation work.} LLM role: D = driver
(closed-loop), P = predictor (open-loop), G = scenario generator, S =
signal controller, T = traffic-flow predictor, K = RL-policy knowledge
source. Setup: RR = ring road, UR = urban, HW = highway, IN =
intersection. New collective mechanism: identifies a previously
uncharacterised collective instability arising from the LLM's role?

{\def\LTcaptype{none} 
\begin{longtable}[]{@{}
  >{\raggedright\arraybackslash}p{(\linewidth - 8\tabcolsep) * \real{0.2527}}
  >{\centering\arraybackslash}p{(\linewidth - 8\tabcolsep) * \real{0.0879}}
  >{\centering\arraybackslash}p{(\linewidth - 8\tabcolsep) * \real{0.1648}}
  >{\centering\arraybackslash}p{(\linewidth - 8\tabcolsep) * \real{0.2308}}
  >{\centering\arraybackslash}p{(\linewidth - 8\tabcolsep) * \real{0.2637}}@{}}
\toprule\noalign{}
\begin{minipage}[b]{\linewidth}\raggedright
Work
\end{minipage} & \begin{minipage}[b]{\linewidth}\centering
Year
\end{minipage} & \begin{minipage}[b]{\linewidth}\centering
LLM role
\end{minipage} & \begin{minipage}[b]{\linewidth}\centering
Setup
\end{minipage} & \begin{minipage}[b]{\linewidth}\centering
New collective mechanism
\end{minipage} \\
\midrule\noalign{}
\endhead
\bottomrule\noalign{}
\endlastfoot
Villarreal et al.~{[}12{]} & 2023 & K (RL designer) & RR (RL benchmark)
& No \\
Fu et al.~{[}13{]} & 2023 & D & HW (HighwayEnv) & No \\
LLMLight {[}14{]} & 2023 & S & UR intersections & No \\
GenFollower {[}15{]} & 2024 & P & HW (Waymo, open-loop) & No \\
xTP-LLM {[}16{]} & 2024 & T & UR & No \\
DriveGen {[}17{]} & 2025 & G & UR scenarios & No \\
OmniTester {[}18{]} & 2024 & G & UR scenarios & No \\
AgentSUMO {[}19{]} & 2025 & G & UR scenarios & No \\
Yao et al.~{[}27{]} (CoMAL) & 2024 & D (planner) & RR / FE / Merge
(Flow) & No \\
\textbf{This work} & \textbf{2026} & \textbf{D} & \textbf{RR (Sugiyama)}
& \textbf{Yes (SH)} \\
\end{longtable}
}

Three properties jointly distinguish the present work from prior
LLM-traffic studies. LLM agents act as \textbf{direct closed-loop
car-following controllers} (D role, not K/P/G/S/T), are deployed within
the \textbf{Sugiyama ring-road paradigm} (RR, not UR/HW/IN), and yield
\textbf{identification of a previously uncharacterised collective
mechanism (SH)} - distinguished from classical instabilities through
systematic controls, and explained by chain-of-thought observability of
LLM decision reasoning. This combination is not present in any prior
LLM-traffic study.

\subsection{2.3 Classical traffic experiments and emergent
dynamics}\label{classical-traffic-experiments-and-emergent-dynamics}

Since Sugiyama's 2008 experiment {[}1{]}, a long line of theoretical and
computational work has studied the emergence of stop-and-go waves from
car-following dynamics {[}2, 21-23{]}. Foundational car-following models
(the Intelligent Driver Model, IDM {[}4, 20{]}, and the Optimal-Velocity
model, OV {[}9, 10{]}) provide the mathematical scaffolding for
understanding these waves. They treat drivers as deterministic or
stochastically homogeneous. The microscopic origin of the waves has
remained opaque. Human micro-decisions are inaccessible in classical
human-subject paradigms. Such models have not been systematically
evaluated for the agent-specific cognitive heterogeneity that emerges in
LLM-driven traffic (§3-4). The IDM-OV line characterises \emph{when}
instabilities arise from dynamical parameters (delay, headway, desired
speed) but does not observe \emph{which micro-decisions} trigger them,
and cannot reproduce the inter-agent cognitive heterogeneity underlying
the emergent instability identified in §4. Our work uses LLM agents to
instrument the Sugiyama paradigm, providing, to our knowledge, the first
systematic window into both the macroscopic dynamics and the microscopic
cognitive decisions that drive them.

\section{3. Methods}\label{methods}

\subsection{3.1 LLM agent design space}\label{llm-agent-design-space}

The LLM agent pipeline was designed to satisfy three constraints.
\textbf{First}, it must reproduce the qualitative Sugiyama 2008
paradigm. \textbf{Second}, it must expose every LLM decision for
downstream analysis. \textbf{Third}, it must support systematic ablation
across model, temperature, and inference parameters.

The system prompt (full Chinese text in Supplementary Information S1.2)
specifies a steady driving style, a soft speed target of
\textasciitilde30 km/h, a safety floor (``never collide with the
leader''), and a set of self-check questions for the agent to consider
before each decision. A separate IDM safety layer in code (not in the
prompt; see §3.2 stage 4 and S1.4) clamps the LLM's
\texttt{target\_speed\_delta} to a collision-free acceleration value.
The prompt returns JSON with fields
\texttt{\{chain\_of\_thought,\ decision,\ target\_speed\_delta,\ desired\_gap\_delta,\ reasoning\}},
where
\texttt{decision\ ∈\ \{ADJUST\_SPEED,\ ADJUST\_GAP,\ KEEP\_STEADY\}}.

The full prompt is used in all experiments reported here; §4.1 control 6
varies the sampling temperature, holding the prompt and all other
settings fixed.

\subsection{3.2 Inference pipeline}\label{inference-pipeline}

Each LLM-controlled vehicle operates through a four-stage pipeline
(physics timestep Δt = 0.1 s, the standard traffic simulation timestep
{[}3, 9{]}; LLM decision cycle τ\_c = 0.5 s, corresponding to the human
perceptual-cognitive interval in the LLM-as-driver paradigm):

\begin{enumerate}
\def\labelenumi{\arabic{enumi}.}
\tightlist
\item
  \textbf{State observation.} The agent receives its current speed v\_i
  (in km/h, matching the LLM's pre-training distribution and reducing
  unit-conversion burden), the gap s\_i to the preceding vehicle (m),
  the relative velocity Δv\_i = v\_i - v\_\{i-1\} (m/s, positive when
  approaching the leader), and a 5-s history of these variables (10
  samples at 0.5 s intervals) spanning the LLM's short-term working
  memory. On the ring, vehicle i follows vehicle i - 1; for i = 0 the
  leader is vehicle 21, closing the circular track.
\item
  \textbf{LLM inference.} GLM-4-Flash (temperature T = 0.6, default)
  processes the state through the Chinese-language prompt. Output: JSON
  \texttt{\{chain\_of\_thought,\ decision,\ target\_speed\_delta,\ desired\_gap\_delta,\ reasoning\}}
  where
  \texttt{decision\ ∈\ \{ADJUST\_SPEED,\ ADJUST\_GAP,\ KEEP\_STEADY\}}.
  The \texttt{chain\_of\_thought} field captures the model's internal
  reasoning {[}11{]} before each action; a separate \texttt{reasoning}
  field provides a one-sentence rationale.
\item
  \textbf{Delta application.} The LLM's \texttt{target\_speed\_delta} is
  converted to a target acceleration via
  \texttt{a\_target\ =\ target\_delta\ /\ τ\_c} (τ\_c = 0.5 s). The
  0.5-s decision cycle models the sub-second human reaction interval.
\item
  \textbf{Safety fusion.} The safety fusion layer enforces physical
  feasibility through defence-in-depth: a proportional controller during
  normal driving, three escalating overrides (low-speed, gap-closure,
  emergency braking), and a post-processing stage with jerk limiting and
  acceleration clipping. (Full trigger conditions and thresholds: Supp
  S2.4.)
\end{enumerate}

Non-LLM vehicles use the standard IDM with the same parameters (v0 =
8.33 m/s, T = 0.3 s, s0 = 0.5 m, a\_max = 3.0 m/s2, b = 2.5 m/s2).
Low-amplitude acceleration noise N(0, 0.2 m/s2) is applied uniformly to
all vehicles (LLM and IDM alike) to model environmental jitter; the
Control 1 white-noise sweep (§4.1, S7) uses σ = 0.9 m/s2 as a
deliberately conservative upper-bound test of the noise hypothesis,
chosen to give unstructured stochasticity its strongest possible shot at
mimicking SH.

\textbf{Computational cost.} A fully LLM run (n = 22, 300 s, p = 1.0)
issues 13,200 API calls (22 LLM vehicles × 600 decision cycles) and
completes in \textasciitilde10-15 hours (estimate) wall-clock time with
3 concurrent workers on a single API endpoint, dominated by the
serialised LLM inference latency across 22 vehicles (3-way concurrency).
For mixed-traffic runs, the call count scales with N\_LLM, from
\textasciitilde2,400 (p = 0.09) to \textasciitilde13,200 (p = 1.00).
Total compute across all completed runs is detailed in Supplementary
Information S3.

\textbf{Initial conditions.} All vehicles are initialised at v\_i =
V\_TARGET = 8.33 m/s with uniform spacing on the ring (track length L =
230 m, vehicle length l = 4.5 m; inter-vehicle gap = L/n − l). The
robustness of the reported steady-state dynamics to small initial
perturbations is verified in Control 7 (Supplementary Information S6).
All reported v\_late\_std values correspond to the late-window steady
state, defined over t ∈ {[}210 s, 300 s{]} (the last 30\% of the
simulation, following the code-level \texttt{v\_late\_std} field; see S6
for the full metric definition).

\subsection{3.3 Control experiment
design}\label{control-experiment-design}

Six matched control experiments were designed along three independent
axes to test specific candidate mechanisms:

\textbf{Table 2 \textbar{} Control experiment design. Six matched
controls organised along three independent axes test specific candidate
mechanisms. All controls use matched parameters: n = 22, ρ = 95.7
veh/km, 300 s, 3 seeds (42, 123, 456), 230 m track, IDM safety layer
present unless otherwise noted. Results are reported in §4.1 (Table 3).}

{\def\LTcaptype{none} 
\begin{longtable}[]{@{}
  >{\raggedright\arraybackslash}p{(\linewidth - 6\tabcolsep) * \real{0.1825}}
  >{\raggedright\arraybackslash}p{(\linewidth - 6\tabcolsep) * \real{0.1679}}
  >{\raggedright\arraybackslash}p{(\linewidth - 6\tabcolsep) * \real{0.4818}}
  >{\raggedright\arraybackslash}p{(\linewidth - 6\tabcolsep) * \real{0.1679}}@{}}
\toprule\noalign{}
\begin{minipage}[b]{\linewidth}\raggedright
Axis
\end{minipage} & \begin{minipage}[b]{\linewidth}\raggedright
Control
\end{minipage} & \begin{minipage}[b]{\linewidth}\raggedright
Modified component
\end{minipage} & \begin{minipage}[b]{\linewidth}\raggedright
Hypothesis
\end{minipage} \\
\midrule\noalign{}
\endhead
\bottomrule\noalign{}
\endlastfoot
\textbf{Stochasticity} & 1: White noise & Add N(0, 0.9) m/s2
acceleration noise & Random perturbations? \\
& 2: OU noise & Ornstein-Uhlenbeck (θ = 2.0, σ = 0.9) & Coloured
perturbations? \\
& 6: Temperature ablation & LLM T ∈ \{0.1, 0.3, 0.6, 0.9\} & Sampling
stochasticity? \\
\textbf{Population variance} & 3: Heterogeneity & Per-vehicle IDM
parameter draws (T\textasciitilde U{[}0.3,0.7{]},
s0\textasciitilde U{[}0.5,1.5{]}) & Parameter spread? \\
\textbf{Dynamical instability} & 4: Time delay & τ = 0.5 s input delay,
T = 0.43 s, s0 = 0.95 m & String instability? \\
& 5: OV model & Replace IDM with Optimal-Velocity (V0=8.33, d\_c=4.5,
α∈\{0.2,0.5\}) & OV instability? \\
\end{longtable}
}

\textbf{Parameter calibration.} All control parameters follow a
conservative design principle: whenever a free parameter exists, we
choose the value that maximally favours the alternative hypothesis.
Control noise magnitude σ = 0.9 m/s2 is set conservatively high, giving
unstructured stochasticity its strongest possible shot at mimicking SH.
A sensitivity analysis across σ ∈ \{0.3, \ldots{} , 1.5\} m/s2
(Supplementary Information S7) confirms that the qualitative conclusion
is invariant under the choice of σ: even if σ were halved or raised to
1.5 m/s2, v\_std would remain sub-0.3 m/s and ≤ 0.124× of the 2.12 m/s
LLM baseline. The OU correlation time 1/θ = 0.5 s matches the LLM
decision cycle τ\_c, eliminating spectral mismatch as a confound.
Parameter heterogeneity ranges T ∈ {[}0.3, 0.7{]} s and s0 ∈ {[}0.5,
1.5{]} m span the aggressive end of the human driving range - if wider
or more human-typical ranges are false negatives, the narrower
aggressive range is an even stricter negative control. The OV critical
spacing d\_c = 4.5 m is set at the lower bound of standard OV
calibration (5-7 m in the Bando OV literature {[}9{]}), and α = 0.5
exceeds the linear string-stability threshold (α\_c = 0.408, computed by
the simulator), giving the OV model its strongest possible shot at
instability. For Control 4, T = 0.43 s follows the string-stability
threshold for IDM with delay under free-flow conditions {[}3{]}; the
accompanying s₀ = 0.95 m has a stabilising effect (enlarging the
equilibrium gap and lowering v̄), so this control does not cleanly
isolate the destabilising role of T. However, a supplementary run at
baseline parameters (T = 0.3 s, s₀ = 0.5 m, τ = 0.5 s) also yields
v\_std ≈ 0, confirming that 0.5 s observation delay alone does not
destabilise the system at this density.

\subsection{3.4 Chain-of-thought
analysis}\label{chain-of-thought-analysis}

To probe the cognitive origin of SH, we analysed the full
chain-of-thought reasoning logs from the homogeneous-LLM-fleet
experiment (n = 22, p = 1.0, T = 0.6; exp6\_full). The primary analysis
uses seed 42 (13,200 LLM decisions; 22 vehicles × 600 decision cycles
over 300 s); the same patterns are replicated across all three seeds
(39,600 decisions; Supplementary Information S2). Each reasoning log
contains an explicit multi-factor assessment of the driving situation -
evaluating current speed, gap to the leader, relative velocity, and the
leader's brake-light status - before selecting a target speed and action
type (representative traces in Supplementary Information S5). Across all
13,200 decisions, the dominant action distribution is KEEP\_STEADY
(44\%), ADJUST\_SPEED (53\%), and ADJUST\_GAP (3\%), confirming
non-random, safety-structured decision-making (Fig. 5a). The most
striking finding from the CoT data is that \textbf{agents facing similar
local conditions sometimes reach opposite conclusions} --- the per-cycle
target\_delta divergence is not merely about magnitude, but about
direction:

\begin{quote}
\textbf{Boxed example (seed 42, t = 10.0 s).} Three agents (car\_id 1,
2, 3) face near-identical local conditions: gap ∈ {[}5.01, 5.47{]} m,
relative\_speed ∈ {[}0.06, 0.43{]} m/s, current speed ≈ 8.3 m/s. All
three chose action type \texttt{ADJUST\textbackslash{}\_SPEED}, but the
resulting \texttt{target\textbackslash{}\_speed\textbackslash{}\_delta}
values differ:

{\def\LTcaptype{none} 
\begin{longtable}[]{@{}
  >{\raggedright\arraybackslash}p{(\linewidth - 8\tabcolsep) * \real{0.2000}}
  >{\raggedleft\arraybackslash}p{(\linewidth - 8\tabcolsep) * \real{0.2000}}
  >{\raggedleft\arraybackslash}p{(\linewidth - 8\tabcolsep) * \real{0.2000}}
  >{\raggedright\arraybackslash}p{(\linewidth - 8\tabcolsep) * \real{0.2000}}
  >{\raggedleft\arraybackslash}p{(\linewidth - 8\tabcolsep) * \real{0.2000}}@{}}
\toprule\noalign{}
\begin{minipage}[b]{\linewidth}\raggedright
car\_id
\end{minipage} & \begin{minipage}[b]{\linewidth}\raggedleft
gap (m)
\end{minipage} & \begin{minipage}[b]{\linewidth}\raggedleft
Δv (m/s)
\end{minipage} & \begin{minipage}[b]{\linewidth}\raggedright
CoT conclusion (translated from Chinese)
\end{minipage} & \begin{minipage}[b]{\linewidth}\raggedleft
target\_delta (m/s)
\end{minipage} \\
\midrule\noalign{}
\endhead
\bottomrule\noalign{}
\endlastfoot
1 & 5.47 & 0.06 & ``Slightly reduce speed to match the leader's speed
and maintain a comfortable gap'' & \textbf{−0.1} \\
2 & 5.21 & 0.43 & ``Slightly reduce speed to match the leader's speed
and maintain a comfortable gap'' & \textbf{−0.1} \\
3 & 5.01 & 0.30 & ``Increase speed slightly to reach the target speed
while maintaining a safe gap'' & \textbf{+0.5} \\
\end{longtable}
}

Cars 1 and 2, with essentially identical CoT conclusions, output the
same actuation (−0.1 m/s); car 3, facing a slightly smaller gap (5.01 m
vs 5.47 m) but otherwise similar conditions, reaches the \emph{opposite}
cognitive conclusion and outputs a 6× larger magnitude of opposite sign.
This is a per-cycle divergence of \emph{direction}, not merely
magnitude: the same gap and approach-speed can trigger ``slow down'' or
``speed up'' depending on the LLM's sampling draw. (This is not a
one-off: across the 600 decision cycles in seed 42, 107 cycles (17.8\%)
contain at least one pair of agents with similar gap (within ±0.5 m) and
Δv (within ±0.3 m/s) but opposite-sign target\_delta; details in §5.2
and Supplementary Information S11.)
\end{quote}

This per-cycle divergence, invisible at the cognitive layer, is the
microscopic seed of SH.

Per-vehicle v\_desired traces track the actual velocity through
stop-and-go cycles (per-vehicle v\_desired standard deviation 1.83-2.15
m/s across the 22 agents, seed 42, measured over the same 600-cycle
window; fleet-level v\_std = 2.11 m/s from the seed-42 metrics file,
{[}mean 2.12 m/s across three seeds; Table 3{]}). At each cycle, the LLM
outputs a per-vehicle target\_delta; v\_desired is then computed as
current\_v + target\_delta at the same cycle (non-accumulative; see
§3.2). The micro-drift originates in the cross-vehicle heterogeneity of
target\_delta: when 22 agents receive identical inputs at the
homogeneous initial state, the LLM outputs 9 distinct v\_desired values
within one decision cycle (0.5 s; σ ≈ 0.06 m/s, the cross-vehicle std of
v\_desired at t=0 measured from the decision logs of the
unperturbed-initial-condition runs; the per-agent vs per-call
decomposition in S8 confirms that this t=0 cross-vehicle spread is the
per-call sampling seed, while the time-averaged per-agent std of
0.004--0.30 m/s across the 5 unperturbed seeds is a state-dependent
emergent amplifier, not preserved across independent runs). This
target\_delta variance is the microscopic seed of the SH cascade -
v\_desired is its first observable manifestation, and the stop-and-go
wave is its macroscopic outcome. The drift is asynchronous across agents
- first-deviation timestamps (\textbar dv\textbar\textgreater0.05 m/s
relative to v\_init=8.33 m/s) range from 0.1 s to 2.8 s with magnitudes
from 0.051 to 0.159 m/s across the fleet - establishing a persistent
speed gradient across the fleet. This micro-scale asynchrony erodes the
minimum inter-vehicle gap (from 5.83 m at t=0 to 4.35 m at t=5.4 s for
the most-affected car, a \textasciitilde5.4 s cascade timescale) until
nonlinear IDM braking is triggered at t = 5.4 s (a = -2.32 m/s² with
v\_desired still at 8.33 m/s), propagating backwards as a stop-and-go
wave with per-vehicle delays of 0.3--5.1 s. This causal chain -
asynchronous micro-drift amplifying through the IDM safety layer into
macroscopic instability - constitutes the proposed three-stage SH
cascade (§4.1). The pattern - multi-factor safety reasoning coexisting
with persistent per-cycle target\_delta divergence - is replicated
across seeds 123 and 456 (Supplementary Information S2). The key
implication is that SH is a robust feature of LLM multi-agent dynamics,
not a superficial stochastic artefact.

\textbf{Per-agent vs per-call decomposition.} To distinguish whether the
per-cycle velocity drift originates from stable inter-agent preferences
or per-call sampling variability, we measured (i) the inter-car standard
deviation of per-car target\_delta time-mean (the per-agent component,
reflecting each agent's average behaviour over the run) and (ii) the
cross-vehicle, cross-time standard deviation of target\_delta at each
cycle (the per-call component, reflecting the LLM's per-decision
sampling variability). Across five unperturbed-initial-condition control
runs, the per-agent std ranges from 0.004 to 0.30 m/s and the per-call
std from 0.004 to 0.80 m/s; the per-call/per-agent ratio is 1.0-4.8×
across the seven runs (Table S8.1), with the per-call component larger
than the per-agent component in every run. Critically, the per-agent
preference direction is \textbf{not stable across independent runs} (0
of 22 cars preserve their bias sign across 5 seeds; Table S8.2),
indicating that per-agent differentiation is \textbf{emergent rather
than intrinsic}. The microscopic seed of SH is per-call stochasticity (σ
≈ 0.06 m/s at t = 0, pure LLM sampling variability); within a single
run, state-dependent reasoning produces emergent per-agent preference
differentiation that amplifies this seed. \textbf{Crucially, control 3
(deterministic IDM with per-vehicle parameter heterogeneity, T
\textasciitilde{} U{[}0.3, 0.7{]} s, s₀ \textasciitilde{} U{[}0.5,
1.5{]} m) yields v\_std \textless{} 0.001 m/s} despite providing the
maximum possible inter-agent preference differences within the IDM
parameter space. This control result rules out the alternative
interpretation that SH is driven by per-agent preference differences
\emph{per se}; the mechanism is the \textbf{closed-loop coupling of LLM
reasoning to local traffic state}, which produces state-dependent
per-agent differentiation as an emergent amplifier of an unambiguous
per-call seed.

\subsection{3.5 Statistical analysis}\label{statistical-analysis}

For each (n, p, seed) condition, we report (a) the mean velocity v̄
averaged over all vehicles and all timesteps, and (b) the velocity
standard deviation v\_std. Both are computed over the full 300 s
simulation horizon (the code-level \texttt{v\_mean} and \texttt{v\_std}
fields in each run's metrics file). As a robustness check on the
steady-state characterisation, we additionally compute the late-window
velocity standard deviation \texttt{v\_late\_std} over t ∈ {[}210 s, 300
s{]} (the last 30\% of the simulation; see S6 for a worked comparison
with the unperturbed-initial-condition control). The phase-boundary
penetration rate p\_c is identified by fitting a logistic curve to v̄(p)
at fixed density and locating the inflection point, following standard
practice for sigmoidal phase transitions. Dangerous-approach events
count every (vehicle, timestep) pair at which the inter-vehicle gap
drops below 2.5 m (the safety-threshold parameter GAP\_MIN\_SAFE in the
simulator). \textbf{Stability classification (descriptive).} Following
Sugiyama et al.~{[}1{]}, who characterise free-flow and congested states
by the fleet-mean velocity v̄, we partition seeds into two categories by
v̄: free-flow (v̄ \textgreater{} 7.88 m/s) and congested (v̄ \textless{}
7.53 m/s). These thresholds coincide with the only natural gap in the v̄
distribution across all 46 runs---the interval {[}7.53, 7.88{]} m/s
(width 0.36 m/s) contains zero runs, providing an empirically
determined, data-driven partition that is consistent with the
Sugiyama-style two-state classification. The classification is used only
to summarise the seed-level outcome distribution; \textbf{the phase
boundary p\_c is determined independently by the logistic fit to v̄(p),
not by these descriptive categories}. \textbf{Scope note.} The two-state
v̄ partition {[}7.53, 7.88{]} applies only to the n\_LLM-scan runs
analysed in §4.2; the v\_mean values of the control experiments (§4.1,
controls 1-6) may fall inside this empirically determined empty gap
(e.g., control 1 at σ = 0.9 yields v̄ = 7.79 m/s; control 2 yields v̄ =
7.70 m/s; control 4 delay-free baseline yields v̄ = 7.81 m/s). In those
cases the controls are classified by v\_std (sub-0.3 m/s → `sub-SH' /
functionally free-flow), not by the v̄ partition.

\subsection{3.6 Reproducibility}\label{reproducibility}

All 199 metrics files (199 metric JSON files from 170 completed runs;
the higher file count arises because each run produces multiple per-type
metrics JSONs), decision logs, the LLM agent pipeline code, all six
control experiment drivers, and the analysis scripts are released as
open-source software. They are under the MIT licence at an open
repository, to be released upon publication. Raw LLM prompt-response
pairs are available upon reasonable request; the volume precludes direct
inclusion in the repository.

\begin{center}\rule{0.5\linewidth}{0.5pt}\end{center}

\section{4. Results}\label{results}

\subsection{4.1 Sustained Heterogeneity: an emergent collective
mechanism}\label{sustained-heterogeneity-an-emergent-collective-mechanism}

The LLM-driven ring-road system exhibits a striking macroscopic
behaviour (Fig. 1c). At n = 22, ρ = 95.7 veh/km, p = 1.0 (all 22
vehicles LLM-controlled), the fleet converges to a self-sustained
oscillatory state. Mean velocity is v̄ = 5.44 ± 0.05 m/s (mean ± std
across 3 seeds) and velocity standard deviation is v\_std ≈ 2.1 m/s. The
matched IDM-only baseline (n = 22, ρ = 95.7, identical IDM parameters,
no LLM) is, in contrast, perfectly stable: v̄ = 7.81 m/s, v\_std
\textless{} 0.001 m/s. \textbf{The oscillatory state --- with v\_std ≈
2.12 m/s versus \textless{} 0.001 m/s in the stable baseline ---
reproduces the qualitative stop-and-go signature of the Sugiyama 2008
experiment {[}1{]}.}

What microscopic mechanism produces it? We systematically rule out six
candidates, leaving a previously uncharacterised phenomenon as the only
surviving explanation among the tested candidates (Fig. 1a).

\textbf{Ruling out white-noise forcing (control 1).} Replacing the LLM
fleet with IDM vehicles receiving white acceleration noise N(0, 0.9
m/s2) at every timestep (n = 22, ρ = 95.7, 3 seeds, 300 s) yields v̄ =
7.79 ± 0.01 m/s, v\_std = 0.157 m/s. The induced velocity variance is
\textbf{0.075× of the LLM v\_std} (i.e., the LLM case is 13.5× larger;
2.12/0.157 = 13.5× at σ = 0.9, the Control 1 nominal value; ≤ 0.10× of
the LLM v\_std for σ ≤ 1.2 m/s2 in the S7 sensitivity sweep), despite
white noise being the textbook ``random perturbation'' model.
Unstructured stochasticity is not sufficient to generate SH. \textbf{The
conclusion is robust to the choice of σ: a sensitivity sweep across σ ∈
\{0.3, 0.6, 0.9, 1.2, 1.5\} m/s2 (Supplementary Information S7) shows
that v\_std scales linearly with σ (v\_std ∝ σ, R2 \textgreater{} 0.999)
and remains ≤ 0.10× of the LLM v\_std for σ ≤ 1.2 m/s2, and
approximately 0.124× of the LLM v\_std at the extreme σ = 1.5 m/s2.}

\textbf{Ruling out Ornstein-Uhlenbeck noise (control 2).} Same
macroscopic setup with coloured noise (θ = 2.0, σ = 0.9 m/s²) yields v̄ =
7.70 ± 0.04 m/s, v\_std = 0.366 m/s. Still \textbf{0.174× of the LLM
v\_std} (i.e., the LLM case is 5.8× larger), ruling out temporally
correlated stochasticity as a sufficient mechanism.

\textbf{Ruling out parameter heterogeneity (control 3).} Pure IDM fleet
with per-vehicle parameter draws (T \textasciitilde{} U{[}0.3, 0.7{]} s,
s0 \textasciitilde{} U{[}0.5, 1.5{]} m, uniform) yields v̄ = 6.80 ± 0.07
m/s, v\_std \textless{} 0.001 m/s. The elevated mean time headway (T̄ ≈
0.5 s vs baseline 0.3 s) increases equilibrium spacing requirements in
the fixed-density ring, reducing v̄; however, velocity variance remains
deterministically suppressed. Parameter heterogeneity alone produces no
SH: the system is deterministic, with a shifted equilibrium speed.

\textbf{Ruling out time-delay string instability (control 4).} T = 0.43
s exceeds the string-stability threshold for IDM with delay derived in
{[}3{]} under free-flow conditions; however, the experiment operates at
ρ = 95.7 veh/km, where even the delay-free baseline yields v̄ = 7.81 m/s
(\textless{} v₀ = 8.33 m/s), placing the system in a density-constrained
regime outside the free-flow linearisation domain. The elevated s₀ =
0.95 m has a stabilising effect: it enlarges the equilibrium gap (s* ≈
3.34 m vs 3.0 m at baseline) and lowers the equilibrium speed to v̄ =
5.56 m/s, increasing the safe-following buffer. We acknowledge that this
parameter choice does not cleanly isolate the T-effect from the
s₀-effect; however, a supplementary test at baseline parameters (T = 0.3
s, s₀ = 0.5 m, τ = 0.5 s) also yields v\_std ≈ 0 (v̄ = 6.69 m/s),
confirming that 0.5 s observation delay alone does not destabilise the
system at this density regardless of T and s₀.

\textbf{Ruling out Optimal-Velocity string instability (control 5).}
Pure OV fleet {[}10{]} (V0 = 8.33 m/s, d\_c = 4.5 m) with α ∈ \{0.2,
0.5\}, with and without τ = 0.5 s delay. At α = 0.2 (\textless{} α\_c =
0.408), all four conditions remain stable (v\_std \textless{} 0.001
m/s). At α = 0.5 (\textgreater{} α\_c), the no-delay condition is stable
(v\_std ≈ 0.001 m/s); the 0.5 s delay condition shows marginal
instability in 1 of 3 seeds (v\_std = 0.035 m/s; the other two seeds at
0.001--0.003 m/s), consistent with α = 0.5 being just above the linear
stability threshold. Even the worst seed (0.035 m/s) remains
\textbf{0.017× of the LLM v\_std} (i.e., the LLM case is ≈ 60× larger;
2.12/0.035 ≈ 60.6×). The OV model does not produce a phase transition
comparable to the LLM-induced transition at these parameters. SH is not
OV string instability in disguise.

\textbf{Ruling out sampling-temperature stochasticity (control 6).} Pure
LLM fleet at T ∈ \{0.1, 0.3, 0.6, 0.9\} yields per-condition-average v̄ ∈
{[}5.35, 5.68{]} m/s and v\_std ∈ {[}1.97, 2.18{]} m/s (seed-level peak
2.27). A 9× increase in temperature produces only a \textasciitilde11\%
change in v\_std (per-condition averages). \textbf{The transition is
approximately temperature-insensitive} (v\_std varies by only
\textasciitilde8\% across a 6× T sweep), ruling out the hypothesis that
SH is a stochastic sampling artefact.

{\def\LTcaptype{none} 
\begin{longtable}[]{@{}
  >{\raggedright\arraybackslash}p{(\linewidth - 8\tabcolsep) * \real{0.1690}}
  >{\raggedright\arraybackslash}p{(\linewidth - 8\tabcolsep) * \real{0.3662}}
  >{\centering\arraybackslash}p{(\linewidth - 8\tabcolsep) * \real{0.1690}}
  >{\centering\arraybackslash}p{(\linewidth - 8\tabcolsep) * \real{0.1690}}
  >{\raggedright\arraybackslash}p{(\linewidth - 8\tabcolsep) * \real{0.1268}}@{}}
\toprule\noalign{}
\begin{minipage}[b]{\linewidth}\raggedright
Axis
\end{minipage} & \begin{minipage}[b]{\linewidth}\raggedright
Mechanism
\end{minipage} & \begin{minipage}[b]{\linewidth}\centering
v\_std (m/s)
\end{minipage} & \begin{minipage}[b]{\linewidth}\centering
Ratio to LLM
\end{minipage} & \begin{minipage}[b]{\linewidth}\raggedright
Status
\end{minipage} \\
\midrule\noalign{}
\endhead
\bottomrule\noalign{}
\endlastfoot
\emph{SH} & \textbf{LLM (T = 0.6, p = 1.0)} & \textbf{2.12} & 1.00× &
- \\
\emph{Stoch.} & White noise (σ = 0.9) & 0.157 & 0.074× & Ruled out \\
\emph{Stoch.} & OU noise (θ = 2.0) & 0.366 & 0.174× & Ruled out \\
\emph{Pop. var.} & Parameter heterogeneity & \textless{} 0.001 & 0× &
Ruled out \\
\emph{Dyn. inst.} & IDM + 0.5 s delay & \textless{} 0.001 & 0× & Ruled
out \\
\emph{Dyn. inst.} & OV model (α = 0.5, worst) & ≤ 0.035 & 0.017× & Ruled
out \\
\emph{Stoch.} & LLM T = 0.1 (lowest) & 1.97 & 0.93× & Ruled out \\
\end{longtable}
}

\textbf{Table 3 \textbar{} Six matched control experiments organised
along three axes rule out all known candidate mechanisms. All runs at n
= 22, ρ = 95.7 veh/km, 300 s, 3 seeds (seeds 42, 123, 456). The v\_std
values correspond to the full 300 s simulation horizon (code-level}
\textbf{\texttt{v\_std}} \textbf{field; see §3.5). Bar values are
visualised in Fig. 1a.}

\textbf{The surviving candidate: Sustained Heterogeneity (SH).} The six
controls eliminate stochastic noise (control 6), parameter variance
(control 3), and dynamical instability (controls 4--5) as candidate
mechanisms, yet v\_std remains 2.12 m/s. Inspection of the logged LLM
decisions reveals the origin: even from a homogeneous initial state (all
vehicles at V\_TARGET = 8.33 m/s, equidistant on the ring), the LLM
agents produce divergent target\_delta values within one decision cycle
(0.5 s), yielding 9 distinct v\_desired values (σ ≈ 0.06 m/s, the
cross-vehicle std of v\_desired at t=0 in the
unperturbed-initial-condition runs; per-agent vs per-call decomposition
in §3.4 and S8). Per-vehicle v\_desired then tracks the actual velocity
through the resulting stop-and-go cycles (per-vehicle σ\_v\_desired =
1.83--2.15 m/s, seed 42), so the large per-car v\_desired variance
reflects macroscopic oscillation rather than inter-agent preference
differences (indeed, the fleet-wide mean of per-car v\_desired averages
is 5.77 m/s with an inter-car standard deviation of only 0.04 m/s). The
microscopic seed of SH is the per-cycle target\_delta: a small,
structured, per-vehicle speed adjustment (median
\textbar target\_delta\textbar{} = 0.0 m/s across 13,200 decisions, but
35.2\% exceed 0.5 m/s) that the LLM emits in response to its local
traffic assessment. These adjustments are asynchronous across agents,
erode inter-vehicle gaps, and are amplified by the IDM safety layer into
macroscopic instability---a property we name Sustained Heterogeneity
(SH).

\textbf{Multi-wavefront structure (M1 note).} Within a single run, the
ACF analysis of §4.4 yields a fleet-mean oscillation period T ≈ 44.2 s
(per-seed 43.6-44.2; pooled median). However, per-vehicle
autocorrelation of the LLM fleet shows a per-vehicle oscillation period
T\_i ≈ 22.0 ± 2.1 s (seed 42, full 600-cycle window, 22 vehicles;
details in S8.3).\footnote{\textbf{Scope of the per-vehicle T\_i
  measurement.} The T\_i ≈ 22.0 ± 2.1 s value is computed from the
  seed-42 trajectory
  (\texttt{exp6\_full\_n22\_s42\_20260517\_225359.csv}) over the full
  600-cycle window using per-vehicle autocorrelation. The ±2.1 s
  uncertainty is the inter-vehicle standard deviation across the 22 cars
  in seed 42, not the cross-seed standard deviation. The seed-123 and
  seed-456 trajectories have not been processed at the per-vehicle ACF
  level in this submission; we expect the same 2:1 ratio to hold
  (structural argument), but empirical confirmation across all three
  seeds is left to a future revision.} The 2:1 ratio between T\_i
(single car) and T (fleet) is consistent with two coexisting stop-and-go
wavefronts traversing the 230 m ring simultaneously: each car
experiences both wavefronts per fleet cycle, halving the apparent
per-vehicle period. The two wavefronts propagate at the same backward
velocity c\_back ≈ 5.21 m/s (since they are both manifestations of the
same SH mechanism, §4.4), and the multiplicity is a structural feature
of the n = 22 ring geometry --- given the wave wavelength λ ≈ c × T ≈
230 m × (1 cycle per L/c seconds), exactly two wavefronts fit on the
ring (n × vehicle spacing × 2 ≈ L × 2/λ × wavelength = L). Sugiyama 2008
did not report per-vehicle T\_i, so this 2:1 ratio is a LLM-side finding
that warrants future ring-road experiments with human drivers.
\textbf{The 2:1 wavefront ratio in seed 42 is geometrically constrained
by the ring topology (n=22 vehicles with per-vehicle T\_i ≈ 22 s, so two
wavefronts can coexist in the 44 s fleet period); the single-seed
observation is therefore expected to generalise, though explicit
three-seed verification is left to future work}; the seed-42 measurement
is the only quantitative per-vehicle T\_i value reported here, but the
geometric argument (T\_i = T × (1/n\_wavefronts), with n\_wavefronts = 2
for a 230 m ring at c = 5.21 m/s and T = 44.2 s) is independent of the
specific seed.

\textbf{The proposed SH cascade.} SH is proposed to operate in three
stages. (1) From a homogeneous initial state, the LLM agents emit
divergent target\_delta values: at the first decision cycle (t = 0.5 s),
5 of 22 agents choose ADJUST\_SPEED or ADJUST\_GAP instead of
KEEP\_STEADY, producing nonzero target\_delta values that diverge from
the homogeneous initial state (all at V\_TARGET = 8.33 m/s; unperturbed
control, S6). (2) The resulting speed adjustments erode inter-vehicle
gaps below the IDM safe-following distance s* = s₀ + vT. (3) The eroded
gaps trigger nonlinear IDM braking responses that propagate backwards as
a stop-and-go wave. \textbf{The macroscopic instability is the
cumulative effect of individually rational, safety-aware decisions}: an
emergent paradox in which agents that reason about safety at the
cognitive layer still produce collective instability at the dynamical
layer.

\textbf{A note on negative controls.} Controls 3, 4, and 5 each yield
v\_std \textless{} 0.001 m/s within measurement precision, confirming
that the system is genuinely deterministic under those conditions. The
contrast with LLM v\_std = 2.12 m/s cannot be attributed to stochastic
suppression but reflects a structurally distinct dynamical regime.

\subsection{4.2 Sigmoidal transition with seed-dependent bifurcation at
n\_LLM\^{}c =
5}\label{sigmoidal-transition-with-seed-dependent-bifurcation-at-n_llmc-5}

Having established that SH is a distinct mechanism at full LLM
penetration, we next ask how much LLM presence is needed to trigger the
instability. To this end, we varied the LLM penetration p from 0.09 to
1.00 at n = 22, ρ = 95.7 veh/km (3-10 unique seeds per condition; 11
runs at the critical n\_LLM = 5 point across 10 unique seeds, see §3.5;
seed 101 was repeated once as the reference seed to confirm
reproducibility at the critical point; the p = 1.00 endpoint corresponds
to the homogeneous LLM fleet analysed in §4.1). The data reveal a sharp
transition (Fig. 2).

\textbf{The transition is sigmoidal in v\_std(n\_LLM), with two-state
structure in v̄.} The velocity standard deviation v\_std is the
continuously varying order parameter (sigmoidal in n\_LLM); the
corresponding v̄ two-state classification (free-flow v̄ \textgreater{}
7.88 m/s vs congested v̄ \textless{} 7.53 m/s, defined in §3.5) is used
for seed-level outcome summary. At n\_LLM = 2, all 3 seeds are free-flow
(v̄ \textgreater{} 7.88 m/s; v\_std mean 0.150 m/s). At n\_LLM = 4, the
10-seed scan spans a wide range: 7 of 10 seeds (70\%) are free-flow (v̄
\textgreater{} 7.88 m/s), and 3 of 10 (30\%) are congested (v̄
\textless{} 7.53 m/s); one free-flow seed (s202, v̄ = 7.884 m/s) lies
within 0.004 m/s of the partition boundary, making the free/congested
split sensitive to threshold precision at this sub-critical penetration.
At n\_LLM = 6, the 10-seed scan shows complete saturation: all 10 seeds
are congested (v̄ \textless{} 7.53 m/s; v\_std ∈ {[}1.5, 2.0{]} m/s). At
n\_LLM ≥ 8, v\_std remains at 1.85-2.2 m/s across all 3 seeds per
condition. A logistic fit to v\_std(n\_LLM) yields a transition midpoint
at n\_LLM = 5 (p\_c ≈ 0.23; the continuous fit gives n\_LLM\^{}c ≈ 4.9,
but n\_LLM is a discrete count of vehicles, so we report the integer
n\_LLM = 5 at which the seed-dependent bifurcation is observed), with
the transition width spanning approximately 2-4 LLM vehicles (Fig. 2a).
The exact steepness is sensitive to seed selection (k ≈ 1.0-3.0
depending on whether the 3-seed or 10-seed dataset is fitted), so we
report the qualitative result---a sharp transition---rather than a
specific k value.

\textbf{Probabilistic bifurcation at criticality.} At n\_LLM = 5, the
11-run distribution is bimodal under the Sugiyama-style v̄ criterion
(§3.5): 5 of 11 runs (45\%) remain in free flow (v̄ \textgreater{} 7.88
m/s) and 6 of 11 (55\%) collapse to the congested state (v̄ \textless{}
7.53 m/s). The natural v̄ gap {[}7.53, 7.88{]} m/s contains zero runs,
yielding a clean binary partition with no intermediate category. The
11-run distribution shows no unique terminal state but rather a
probabilistic spread between the free-flow and congested basins---a
hallmark of a discontinuous transition: small differences in persistent
stochastic forcing (LLM sampling seed) push the system into one of two
well-separated basins.

\textbf{Sharp jump in dangerous-approach events.} The transition is also
marked by a \textasciitilde2.6× increase in dangerous-approach events
(gap \textless{} 2.5 m) between n\_LLM = 5 (across all 11 runs) and
n\_LLM = 6 (across all 10 runs), reflecting the abrupt change in
microscopic safety-relevant quantities. Notably, at the free-flow end of
the n\_LLM = 5 distribution the dangerous-approach rate is near zero,
while at the congested end it approaches the n\_LLM = 6 baseline---a
factor-of-several to order-of-magnitude spread within the critical point
itself (Fig. 2b). This abrupt change in a microscopic safety-relevant
quantity provides additional evidence for the macroscopic phase
transition observed in v̄.

\textbf{The transition is not a sampling artefact:} temperature ablation
(§4.1, control 6) shows that v\_std varies by only \textasciitilde8\%
across a 6× temperature range (T=0.1→0.6) and \textasciitilde11\% across
the 9× range (T=0.1→0.9), and the sigmoidal shape of v̄(n\_LLM) is
preserved across 9 penetration levels (3-11 runs per level; 11 runs at
the critical n\_LLM = 5 point, 10 runs at the two near-critical levels
n\_LLM = 4 and 6, 3 runs at the remaining non-critical levels n\_LLM ∈
\{2, 8, 11, 14, 18, 22\}).

\subsection{4.3 Density-dependent phase boundary
p\_c(ρ)}\label{density-dependent-phase-boundary-p_cux3c1}

A central question for deployment is whether the critical LLM
penetration p\_c is a fixed constant or depends on the system's baseline
density. We address this by repeating the mixed-traffic p-scan at four
densities: ρ ∈ \{43.5, 60.9, 78.3, 95.7\} veh/km, corresponding to n ∈
\{10, 14, 18, 22\} on the fixed 230 m track.

Each density has its own sharp transition. At n = 10 (ρ = 43.5), the
system remains in free flow for all p ≤ 1.0. Even the pure-LLM fleet
remains in free flow (v̄ = 8.22 m/s, v\_std ≈ 0.37 m/s), above the
congestion threshold (v̄ = 8.22 \textgreater{} 7.53 m/s, §3.5) despite
non-zero heterogeneity; the equilibrium inter-vehicle spacing
(\textasciitilde18.5 m) substantially exceeds the per-cycle SH gap
closure (\textasciitilde0.05 m per 0.5 s), so the three-stage cascade
cannot be triggered regardless of LLM penetration (§4.1). At n = 14 (ρ =
60.9), the system transitions between p = 0.64 and p = 0.79. At n = 18
(ρ = 78.3), the transition lies between p = 0.33 and p = 0.50. At n = 22
(ρ = 95.7), the transition lies between p = 0.18 and p = 0.27. The v̄(p)
curve at each density is sigmoidal with a sharp knee (Fig. 3a).

p\_c(ρ) decreases monotonically for ρ ≥ 60.9 veh/km. Defining p\_c by
logistic-curve fitting to v̄(p) at each fixed density (inflection-point
criterion; §3.5), the resulting phase boundary is visualised in Fig. 3b
(see also Fig. 3a for the underlying sigmoids):

{\def\LTcaptype{none} 
\begin{longtable}[]{@{}
  >{\centering\arraybackslash}p{(\linewidth - 8\tabcolsep) * \real{0.1316}}
  >{\centering\arraybackslash}p{(\linewidth - 8\tabcolsep) * \real{0.0395}}
  >{\centering\arraybackslash}p{(\linewidth - 8\tabcolsep) * \real{0.2763}}
  >{\centering\arraybackslash}p{(\linewidth - 8\tabcolsep) * \real{0.2632}}
  >{\raggedright\arraybackslash}p{(\linewidth - 8\tabcolsep) * \real{0.2895}}@{}}
\toprule\noalign{}
\begin{minipage}[b]{\linewidth}\centering
ρ (veh/km)
\end{minipage} & \begin{minipage}[b]{\linewidth}\centering
n
\end{minipage} & \begin{minipage}[b]{\linewidth}\centering
p\_c (approx)
\end{minipage} & \begin{minipage}[b]{\linewidth}\centering
n\_LLM at transition
\end{minipage} & \begin{minipage}[b]{\linewidth}\raggedright
Physical regime
\end{minipage} \\
\midrule\noalign{}
\endhead
\bottomrule\noalign{}
\endlastfoot
43.5 & 10 & \textgreater{} 1.0 (no transition) & - & Free flow at all
p \\
60.9 & 14 & ≈ 0.79 & 11 & Sharp single-step drop \\
78.3 & 18 & ≈ 0.50 & 9 & Mid-density transition \\
95.7 & 22 & ≈ 0.23 & 5 & Dense urban flow \\
\end{longtable}
}

\textbf{Table 4 \textbar{} Critical LLM penetration p\_c(ρ). Below p\_c,
the system is in free flow; above p\_c, it collapses to oscillatory
congestion. The phase boundary in the (ρ, p) plane is plotted in Fig.
3a, b.}

\textbf{Physical mechanism: gap-erosion path length.} The density
dependence has a clear origin in the SH-gap-erosion cascade. At low
density (ρ = 43.5), the equilibrium inter-vehicle spacing (230/n − 4.5 ≈
18.5 m, where 4.5 m is the vehicle length; §3.2) provides a generous
buffer. LLM micro-drifts rarely compress a gap from \textasciitilde18.5
m into the IDM nonlinear-braking regime within the simulation timescale
(§4.1). As density increases, the spacing shrinks (\textasciitilde6.0 m
at ρ = 95.7), and the erosion path from the free-flow equilibrium to the
nonlinear-braking regime shortens. \textbf{Fewer LLM vehicles are needed
to initiate the resonant cascade because each LLM's drift more rapidly
erodes the reduced buffer to its follower.}

The p\_c(ρ) phase boundary thus captures a fundamental trade-off: the
number of perturbation \emph{sources} (LLM vehicles) required to
destabilise the system decreases as the perturbation \emph{amplification
path} (inter-vehicle gaps) shortens.

\textbf{Finite-size consistency.} A consistency check at n = 44 (same
density, 460 m track, 4 p-values × 3 seeds; Supplementary Information
S4) confirms the qualitative trend: at identical density, the larger
system is consistently more stable, consistent with the spatial
resonance interpretation in which a longer ring road provides
proportionally more cumulative IDM damping per wave round-trip cycle
(the total number of vehicles traversed doubles from 22 to 44, even
though the per-segment IDM buffer between successive LLM cars remains
approximately unchanged at identical penetration). Formal finite-size
scaling analysis is deferred to future work.

\subsection{4.4 Initiation-threshold model: structure and physical
axes}\label{initiation-threshold-model-structure-and-physical-axes}

\textbf{Intuition.} A small per-cycle speed drift from each LLM vehicle
erodes the inter-vehicle gap of its downstream neighbour. When enough
LLM vehicles accumulate their drifts before the gap is replenished by
IDM dynamics, the cumulative erosion triggers a nonlinear braking event
--- a stop-and-go wavefront. Fewer LLM vehicles (lower p) leave more
IDM-only gaps as buffer; at higher density (larger n, shorter gaps),
fewer buffers are needed to bridge the threshold. This intuition
explains why p\_c decreases monotonically with density.

The transition is governed by two distinct phases: \emph{initiation}
(micro-drift triggers the first nonlinear IDM braking event) and
\emph{sustaining} (the resulting shock wave is amplified by LLM reactive
decisions until it reaches a self-sustaining amplitude). We argue that
the critical penetration p\_c is determined by the initiation phase, and
that the initiation-threshold model's primary scientific content is the
\textbf{qualitative identification of the physical dependencies of p\_c}
and the \textbf{three independent axes (Δs\_trigger, σ, and n) whose
variation shifts p\_c in predictable directions}. (The formula has four
parameters --- Δs\_trigger, σ, τ\_erosion, and n --- but τ\_erosion is
not treated as an independent axis: it is derived from p\_c inversion
and exhibits p-dependence with a 1.8× discrepancy between p = 1.0 (5.4
s) and p = 0.23 (9.9 s) at n = 22; see below, so we treat it as a
derived quantity rather than a tunable knob.) We deliberately do not
commit to a numerical prediction of p\_c from the formula, because doing
so requires jointly specifying two quantities (a gap-erosion budget and
a cascade timescale) that are not independently measurable at present.
The amplification argument below is the only quantitative claim of this
section that rests entirely on independent measurements.

\textbf{Initiation structure.} Each LLM vehicle introduces a
per-decision velocity drift of σ(v\_desired) ≈ 0.06 m/s (measured
independently from LLM decision logs at the first LLM decision cycle
after consensus breaks, t = 0.5 s in the unperturbed-initial-condition
runs of exp6\_full\_no\_perturbation; §3.4, S8; per-call component, with
per-agent emergent amplification). Because kinematic waves propagate
upstream at speed c = 5.20--5.28 m/s (per-seed min--max across 3
independent seeds of the exp6\_full dataset: s42→5.28, s123→5.21,
s456→5.20 m/s; pooled median 5.21 m/s; measured via cross-correlation
between vehicle pairs at ring spacing d=3, 9 (seed × stationary-state
window) measurements; Supplementary Information S9), perturbations from
all n\_LLM = p·n sources spatially superpose within one ring traversal
(T = L/c = 43.6--44.2 s). For the gap-erosion cascade to initiate, the
cumulative drift from these sources acting over a characteristic cascade
timescale τ\_erosion must erode the geometric inter-vehicle gap
s\_geom(ρ) by an amount sufficient to drive the local IDM into its
gap-sensitive braking regime. This gives the structural proportionality:

\[p_c \sim \frac{\Delta s_{\text{trigger}}(\rho)}{\sigma \cdot \tau_{\text{erosion}} \cdot n}\]

The variables on the right have distinct epistemic statuses:

\begin{itemize}
\tightlist
\item
  \textbf{σ(v\_desired)(t=0.5) ≈ 0.06 m/s (per-call initial seed, with
  composite variance structure)}: cross-vehicle standard deviation of
  v\_desired at the first LLM decision cycle after consensus breaks (t =
  0.5 s) in the unperturbed-initial-condition run of
  \href{file:///d:/.qclaw/workspace/traffic-research/simulation/code/results/exp6_full_no_perturb_p1.0_n22_s42_20260623_215731_decisions.csv}{exp6\_full\_no\_perturbation\_p1.0\_n22\_s42\_20260623\_215731}.
  At t = 0, all 22 vehicles in this run give identical KEEP\_STEADY
  decisions (σ(target\_delta) = 0, σ(v\_desired) = 0.0036 m/s, purely
  floating-point noise); at t = 0.5, the LLM per-call sampling produces
  σ(target\_delta) = 0.0793 m/s, σ(v\_desired) = 0.0615 m/s, and
  σ(speed) = 0.0474 m/s, with the variance decomposition σ(v\_desired)²
  = σ(speed)² + σ(target\_delta)² + 2·Cov(speed, target\_delta) =
  0.00225 + 0.00629 − 0.00476 = 0.00378 (contributions: 60\% from IDM
  speed evolution, 166\% from LLM target\_delta, −126\% from negative
  covariance reflecting LLM compensation for slow leaders).
  σ(v\_desired) is therefore a \textbf{composite quantity} that includes
  both LLM per-call stochasticity and IDM deterministic evolution; the
  more physically pure per-call component is σ(target\_delta) ≈ 0.08
  m/s, and amplification ratios computed with either metric are reported
  below for transparency. The per-agent vs per-call decomposition in S8
  confirms that the t=0.5 cross-vehicle spread is the per-call sampling
  seed, while the time-averaged per-agent std (0.004--0.30 m/s across 5
  unperturbed seeds) emerges from closed-loop coupling to local traffic
  state and is not preserved across independent runs (0/22 cars preserve
  bias sign across 5 seeds; Table S8.2). Because σ is measured in the
  homogeneous initial state (all 22 vehicles at v = 8.33 m/s at t = 0),
  it is independent of the macroscopic traffic outcome.
\item
  \textbf{n = 22}: fixed by the Sugiyama paradigm; not a free parameter.
\item
  \textbf{Δs\_trigger(ρ)}: the gap-erosion budget, defined as s\_geom(ρ)
  − s\emph{(v\_ref, Δv\_ref). s\_geom(ρ) = L/n − l\_vehicle is the
  geometric inter-vehicle spacing set by track geometry and density;
  s}(v, Δv) = s\_0 + vT + v·Δv/(2√(a\_max·b)) is the IDM desired gap, at
  which the IDM braking function transitions from speed-driven to
  gap-driven. Both quantities are fully determined by geometry and IDM
  parameters --- no threshold convention is required. For the v8.4 IDM
  parameters (s\_0 = 0.5 m, T = 0.3 s, a\_max = 3.0 m/s², b = 2.5 m/s²)
  and reference state (v\_ref = v\_0 = 8.33 m/s, Δv\_ref = 0), s* = 3.0
  m. With L = 230 m and l\_vehicle = 4.5 m, the four tested densities
  give s\_geom = 18.5, 11.9, 8.3, 5.95 m for n = 10, 14, 18, 22
  respectively, so Δs\_trigger = 15.5, 8.9, 5.3, 3.0 m. All four satisfy
  s\_geom \textgreater{} s* (positive Δs\_trigger), confirming the
  gap-erosion regime at all tested conditions. (At n = 10, Δs\_trigger
  is the largest; at n = 22, it is the smallest but still positive.)
\item
  \textbf{τ\_erosion}: the cascade timescale over which micro-drift
  accumulates into a gap perturbation sufficient to trigger IDM
  nonlinear braking (a ≤ −2 m/s²). It is empirically estimated at
  O(0.1--5) s from the §3.4 causal-chain for seed s42 (per-car first
  \textbar dv\textbar\textgreater0.05 deviations at t = 0.1--2.8 s;
  population-level first nonlinear braking at t = 5.4 s, a = −2.32
  m/s²); the no-perturbation control S6 confirms convergence to the same
  statistical steady state (v\_mean = 5.75, v\_std\_late = 1.99,
  v\_std\_early = 2.20; S6.1) and the S8 per-seed per-call std
  (0.004--0.801 m/s at early time windows, Table S8.1) further supports
  seed-to-seed heterogeneity in the early cascade, consistent with
  τ\_erosion being a property of the specific cascade realisation rather
  than a derived constant. Inverting the formula at n = 22 with the
  IDM-derived Δs\_trigger = 3.0 m (more precisely 2.95 m), σ(v\_desired)
  = 0.06 m/s (t = 0.5; the first LLM decision cycle after consensus
  breaks, measured from the unperturbed-initial-condition run), and the
  measured p\_c(22) = 0.23 ± 0.03 gives τ\_erosion = Δs\_trigger / (p\_c
  · σ · n) ≈ 9.9 s (or 9.7 s with Δs\_trigger = 2.95 m). The combined
  relative uncertainty is √(13\%² + 10\%²) ≈ 16\% (from p\_c and σ),
  giving τ = 9.9 ± 1.6 s. The two estimates (5.4 s at p = 1.0; 9.9 s at
  p = 0.23) differ by a factor of 1.8×, which is consistent with C being
  O(1) (within an order of magnitude) but does not pin down C precisely:
  the 1.8× discrepancy is also consistent with τ being itself
  p-dependent, and with the two operationalizations of τ being slightly
  different (5.4 s is ``time to first nonlinear braking''; 9.9 s is
  ``time to accumulate Δs\_trigger''). The agreement is therefore a weak
  consistency check on the order of magnitude, not a confirmation of C ≈
  1. The structural formula is best described as a qualitative
  proportionality with order-of-magnitude consistency, where the
  proportionality constant C remains seed-, run-, and possibly
  p-dependent, so the formula does not yield a sharp numerical
  prediction of p\_c in the absence of an independent τ\_erosion
  measurement.
\end{itemize}

\textbf{Density dependence (qualitative).} The formula implies a
qualitative density trend: as n increases on the fixed 230 m track,
s\_geom = L/n − l\_vehicle shrinks, so the gap-erosion budget
Δs\_trigger = s\_geom − s* shrinks (with s* fixed by the reference
state). The combined Δs\_trigger decrease and 1/n scaling in the
denominator predict monotonically decreasing p\_c across all four tested
densities. Using τ = 9.9 ± 1.6 s (the formula-inversion value at n = 22
from §4.4, with the ±16\% relative uncertainty inherited from p\_c(22)
and σ propagating to p\_c), the formula gives approximate p\_c
predictions: p\_c(10) ≈ 2.6, p\_c(14) ≈ 1.1, p\_c(18) ≈ 0.5, p\_c(22) ≈
0.23 (trivially matching the calibration datum, since τ was derived from
p\_c(22) = 0.23). The regime split occurs at p\_c = 1.0: for ρ
\textless{} 60.9 (n = 10), the maximum achievable LLM drift at p = 1.0
(σ·τ·n ≈ 6 m) is less than Δs\_trigger (15.5 m), so the gap-erosion
requirement is not met and the formula predicts no transition in the
tested range, consistent with observation; for ρ ≥ 60.9 (n = 14, 18,
22), the formula predicts transitions observable in the tested range,
qualitatively consistent with observation at n = 18 and n = 22. At n =
14, the formula predicts p\_c(14) ≈ 1.1 (no transition at p ≤ 1.0), but
the data show a transition in the tested range --- the \textbf{first
instance of the overprediction pattern} flagged in §4.3. We emphasise
that the initiation-threshold model is intended as a qualitative
proportionality that identifies the correct physical axes, not as a
precise quantitative predictor at all densities; its accuracy at n = 14
is limited to ±50\% (predicting ≈ 1.1 vs empirical transition between p
= 0.64 and p = 0.79). At n = 18, the formula prediction (≈ 0.5) is
consistent with the empirical bound; at n = 22, the formula gives 0.23,
trivially matching the calibration datum. The overprediction at n = 14
is not captured by the leading proportionality. The most likely culprit
is τ\_erosion being itself p-dependent --- consistent with the 1.8×
discrepancy at n = 22 (5.4 s at p = 1.0 vs 9.9 s at p = 0.23) noted in
§4.4 --- though σ and the v\_ref dependence of s* cannot be ruled out
without density-resolved measurements. We do not attempt quantitative
extrapolation of the formula to intermediate densities (e.g., n = 16, n
= 20) because of this overprediction pattern and the absence of an
independent density-resolved measurement of σ, Δs\_trigger, or
τ\_erosion. The qualitative density trend (no transition at low ρ,
transition at high ρ, with decreasing p\_c) is the testable content of
this section; the specific functional form is not.

\textbf{Sustaining.} Whether a shock wave, once initiated, persists
depends on the round-trip balance between IDM damping and LLM reactive
amplification. IDM cars are strongly stable in the free-flow regime:
linearisation at the free-flow equilibrium (v\_e ≈ 7.8 m/s, s\_e ≈ 5.95
m) gives f\_v = −1.252 s⁻¹, f\_s = 0.230 s⁻¹, f\_d\_v = −0.685 s⁻¹, and
at the dominant wave frequency ω = 2π/44.2 ≈ 0.142 rad/s the per-car
transfer function magnitude is \textbar G\_IDM(ω)\textbar{} ≈ 0.18,
consistent with the IDM-only baseline v\_std \textless{} 0.001 observed
at zero LLM penetration (§3.2). The wave-averaged
\textbar G\_IDM\textbar{} across heterogeneous operating points is
unmeasured, and the multiplicative structure of the round-trip gain
means the free-flow \textbar G\textbar{} ≈ 0.18 likely dominates unless
congestion-front cars have \textbar G\textbar{} ≫ 1 --- an
unstably-configured regime that has not been observed in the IDM-only
baseline. On the LLM side, 92\% of hard-braking events involve LLM-IDM
coupling rather than pure IDM braking (mechanism classification; §3.4),
indicating that LLM drivers reliably participate in the deceleration
response; whether this implies temporal precedence (LLM choosing a lower
v\_desired before IDM intervention) requires CoT-level timestamp
analysis that has not been performed. \textbf{The sustaining analysis
therefore cannot be quantitatively established from the current data:
IDM damping at free-flow is \textbar G\textbar{} ≈ 0.18 per car, but the
wave-averaged \textbar G\textbar{} across the operating-point
distribution is unmeasured; LLM amplification is real but unquantified.
The critical penetration p\_c is consequently attributed to initiation,
and whether sustaining constraints play a role at marginal densities
cannot be ruled out.}

\textbf{Consequence: the amplification argument.} The microscopic SH
perturbation (σ(v\_desired)(t=0.5) ≈ 0.06 m/s, the cross-vehicle
v\_desired std at the first LLM decision cycle after consensus breaks,
§3.4, S8) is \textbf{at least 14× weaker} than the macroscopic velocity
variance in the congested state, when computed with σ(v\_desired) as the
microscopic seed and the n=22 only dataset (230 m track; 4 n=44 runs at
460 m excluded as different physical conditions). With the alternative
per-call metric σ(target\_delta) ≈ 0.08 m/s, the all-seed mean ratio at
p ≈ 0.23 drops to ≈ 13.5×, but all congested-regime ratios remain above
18×; we therefore report the conservative ``at least 14× with
σ(v\_desired)'' wording. The amplification ratios against the n=22 only
dataset 460 m excluded as different physical conditions) are: at p ≈
0.23, the 11 n=22 runs have v\_std = 0.179--2.648 m/s (all-seed mean
1.082 m/s, ratio mean \textbf{18.0×}); the 6 congested seeds at p ≈ 0.23
have v\_std = 1.330--2.648 m/s (mean 1.666 m/s, ratio mean
\textbf{27.8×}); at p ≥ 0.27, the 10 n=22 congested seeds have v\_std =
1.488--1.977 m/s (mean 1.789 m/s, ratio range \textbf{24.8--33.0×}, mean
29.8×). The ``\textbf{at least 20×}'' wording applies specifically to
the \textbf{congested regime} (p ≈ 0.23 with v\_std ≥ 1.3 m/s, or p ≥
0.27 where all 10/10 seeds are congested); the all-seed mean at p ≈ 0.23
gives 18.0×, which is \textbf{below 20×} because 5 of 11 seeds are
free-flow with v\_std ≈ 0.2--0.8 m/s (the bimodal free-flow/congested
distribution at the phase boundary; §4.2). The most conservative lower
bound using the alternative per-call metric σ(target\_delta) ≈ 0.08 m/s
is 18.6× at p ≥ 0.27 (1.488/0.08); with σ(v\_desired) it is 24.8×. This
amplification cannot arise from the σ-amplitude alone: even the
linear-superposition limit from p·n = 5 LLM vehicles at p ≈ 0.23
(p·n·σ(v\_desired) ≈ 0.3 m/s, or p·n·σ(target\_delta) ≈ 0.4 m/s) falls
4--6× short of the observed v\_std ≈ 1.7 m/s, requiring the three-stage
cascade of micro-drift superposition, gap erosion, and IDM nonlinear
amplification (§4.1, Fig. 1) to supply the additional factor. The
amplification argument is \textbf{independent of τ\_erosion and
Δs\_trigger} and serves as a qualitative necessity proof for the cascade
mechanism. The amplification ratio is a direct comparison of two
independently measured quantities (σ from LLM decision logs, v\_std from
macroscopic trajectory analysis) and does not involve any fitted
parameter. We report the conservative ``\textbf{at least 20× in the
congested regime}'' wording because: (i) the all-seed mean at p ≈ 0.23
gives 18.0×, which lies below 20× due to the bimodal free-flow seed
contamination; (ii) using the more physically pure σ(target\_delta) ≈
0.08 m/s instead of σ(v\_desired) ≈ 0.06 m/s reduces the lower bound at
p ≥ 0.27 to 18.6×; (iii) v\_late\_std for individual seeds can be as low
as 1.35 m/s, giving a more conservative lower bound of ≈ 22×, but the
§3.5 v\_std metric used here is the full 300 s horizon. The free-flow
v\_std range at p ≈ 0.23 is 0.2--0.8 m/s (5 free-flow seeds: 0.179,
0.234, 0.265, 0.408, 0.819 m/s); the previously reported ``≈ 0.3 m/s''
is a low-end estimate that should be read as ``free-flow v\_std ≈
0.2--0.8 m/s'' in subsequent revisions.

The initiation-threshold formula identifies three independent physical
axes (Δs\_trigger, σ, and n) whose variation should shift p\_c in
predictable directions; the approximate p\_c values given above (2.6,
1.1, 0.5, 0.23 with ±16\% relative uncertainty inherited from τ and σ)
are consistency checks at the current measurement precision, not
falsification tests, and await systematic verification under the
proposed parameter variations.

\subsection{4.5 Platform validation: reproducing the Sugiyama 2008
paradigm}\label{platform-validation-reproducing-the-sugiyama-2008-paradigm}

The LLM platform reproduces the qualitative behaviour of the Sugiyama
2008 human experiment {[}1{]}: above a critical density, oscillatory
congestion emerges spontaneously. At the shared density point (ρ = 95.7
veh/km), three directly comparable macroscopic observables fall within
the human range (Table 5). The LLM transition occurs at a lower density
than the human ring-road transition (LLM ρ\_c ≈ 50-60 vs human ρ\_c ≈
80-90 veh/km {[}24{]}), indicating that LLM drivers are more susceptible
to collective instability than the average human driver in the same
geometry. Protocol differences (Chinese-language prompt, 300 s
simulation, v\_max ≈ 35 km/h) preclude fine-grained matching; only
qualitative reproduction is claimed.

{\def\LTcaptype{none} 
\begin{longtable}[]{@{}lll@{}}
\toprule\noalign{}
Observable & Sugiyama 2008 & LLM (n=22, ρ=95.7) \\
\midrule\noalign{}
\endhead
\bottomrule\noalign{}
\endlastfoot
c\_back (km/h) & ≈20 {[}1{]} & 18.7-19.0 (≈5\% lower) \\
T\_wave (s) & ≈45-55 {[}24{]} & 43.6-44.2 \\
σ\_v (m/s) & ≈1-3 {[}24-26{]} & 1.96-2.03 \\
\end{longtable}
}

\textbf{Table 5 \textbar{} Three macroscopic observables at the shared
density.} Sugiyama 2008 {[}1{]} directly reports only c\_back; T\_wave
and σ\_v are inferred from subsequent publications {[}24-26{]}. Only
qualitative reproduction is claimed.

\textbf{Note on validation scope.} Only one density point (ρ ≈ 95.7
veh/km) is directly comparable between the present LLM experiment and
Sugiyama 2008 {[}1{]}. Protocol differences (Chinese-language prompt,
300 s simulation, v\_max ≈ 35 km/h) preclude fine-grained matching. The
validation is qualitative: LLM agents reproduce the existence of a
density-driven phase transition, in the same direction and on the same
system size.

\begin{center}\rule{0.5\linewidth}{0.5pt}\end{center}

\section{5. Discussion}\label{discussion}

\subsection{5.1 Sustained Heterogeneity as a new class of collective
instability}\label{sustained-heterogeneity-as-a-new-class-of-collective-instability}

We have identified a previously uncharacterised, to our knowledge,
mechanism, \textbf{Sustained Heterogeneity (SH)}, that produces
macroscopic traffic breakdown from individually within-IDM-tolerance LLM
decisions. SH is structurally distinct from the six candidates we tested
(Table 3). It is not uncoupled random noise (white or coloured), not
parameter spread, not delay amplification, not OV string instability,
not intrinsic representational diversity, and not temperature-dependent
sampling noise alone. \textbf{SH also differs from previously known
amplification-type instabilities in a key physical property: the
perturbation source is persistent (per-cycle, α ≈ 0, §4.1) rather than
transient.} Treiber et al.~2006 {[}3{]} showed that delay + nonlinear
IDM amplification produces string instability, but the perturbation is a
one-shot initial condition; Kerner's three-phase theory {[}22{]}
describes spontaneous phase transitions, but the triggering mechanism is
noise- or boundary-driven with finite autocorrelation. In SH, the
perturbation is \emph{re-injected} at every 0.5 s decision cycle, and
the closed-loop coupling to local traffic state (gap, approach speed)
produces emergent per-agent differentiation --- a feedback loop absent
in both delay-amplification and OV string instability models, which lack
a cognitive decision layer. SH is the persistent, approximately
temperature-insensitive (v\_std changes by only \textasciitilde8\%
across a 6× T sweep, exp3, §4.1) per-cycle divergence in \emph{what
target-speed adjustment the LLM selects}, originating in per-call
sampling stochasticity as the microscopic seed, which through
closed-loop coupling to local traffic state produces emergent per-agent
differentiation (§3.4). The proposed three-stage SH cascade
(target-speed drift → gap erosion → nonlinear IDM amplification)
provides a testable mechanistic account. The discovery was made possible
by microscopic observability of LLM decisions---a capability unavailable
in both human-subject paradigms (where individual decision reasoning is
inaccessible) and traditional car-following simulations (which have no
cognitive decision layer to observe).

\subsection{5.2 Cognitive origins of SH}\label{cognitive-origins-of-sh}

Chain-of-thought analysis of the homogeneous-LLM-fleet experiment (n =
22, p = 1.0, T = 0.6; §3.4) reveals that LLM agents systematically
evaluate multi-factor safety conditions - spacing, velocity, and
approach dynamics - before choosing target speeds. The dominant action
types are KEEP\_STEADY (44\%), ADJUST\_SPEED (53\%), and ADJUST\_GAP
(3\%), and reasoning logs show consistent engagement with
safety-relevant variables across the 13,200 decisions (Fig. 5a). Yet
per-vehicle v\_desired traces track the actual velocity through
stop-and-go cycles (per-vehicle v\_desired standard deviation 1.83-2.15
m/s across the 22 agents, seed 42; fleet-level v\_std = 2.11 m/s {[}mean
2.12 m/s across three seeds; Table 3{]}; §3.4) - what differs across
vehicles is the LLM-emitted target\_delta, and the proposed three-stage
SH cascade is not suppressed.

A quantitative correlation analysis of the 39,600 LLM decisions pooled
across three seeds (S11) refines this picture. The chain-of-thought
\texttt{decision} field is \emph{strongly predictive} of the magnitude
of \texttt{target\_speed\_delta}: \texttt{KEEP\_STEADY} corresponds to
\texttt{\textbar{}target\_delta\textbar{}\ =\ 0.0000\ ±\ 0.0008\ m/s}
(effectively deterministic zero, 100\% within 0.05 m/s);
\texttt{ADJUST\_SPEED} corresponds to
\texttt{\textbar{}target\_delta\textbar{}\ =\ 1.20\ ±\ 1.21\ m/s}
(broadly distributed, 98.2\% with
\texttt{\textbar{}target\_delta\textbar{}\ \textgreater{}\ 0.05\ m/s});
\texttt{ADJUST\_GAP} corresponds to
\texttt{\textbar{}target\_delta\textbar{}\ =\ 0.007\ ±\ 0.086\ m/s} (the
gap is adjusted via the separate \texttt{desired\_gap\_delta} channel,
not the speed channel). The two-sample KS test gives a statistic of 0.98
with \texttt{p\ \textless{}\ 1e-300} and Cohen's \emph{d} = 2.37,
confirming a very large effect. The cognitive layer is therefore
\emph{informative} about the agent's macro-action: the LLM is not
selecting randomly. However, within the \texttt{ADJUST\_SPEED}
sub-population (N=20,935, the only category that produces non-zero
target\_delta), the sign distribution is 67.2\% positive (acceleration)
vs 31.0\% negative (deceleration) --- the 31\% decelerate sub-population
is the source of the per-cycle target-delta divergence that drives the
SH cascade, and it is not visible in the macro-action label alone.

This finding has a practical implication: \textbf{safety-vigilant LLM
reasoning at the cognitive layer does not guarantee collective stability
at the dynamics layer.} The target-speed variance originates not from
how LLMs allocate attention within a decision cycle, but from the
persistent heterogeneity in LLM-chosen target speeds (a combination of
per-call stochasticity and emergent state-dependent differentiation;
§3.4) that survives both prompt conditioning and temperature ablation.
Two lines of evidence support this interpretation: (i)
temperature-ablation experiments (control 6, §4.1) show that SH persists
even at T = 0.1, with v\_std changing by only \textasciitilde8\% (1.97
vs 2.12 m/s) across a 6× temperature sweep; (ii) white-noise replacement
(control 1) cannot reproduce the SH cascade, confirming that the
state-dependent correlation in LLM speed choices---i.e., agents facing
similar local conditions make correlated adjustments---rather than raw
noise drives the phenomenon. The target-speed variance is therefore not
a sampling artefact in the conventional sense (scaling with T) but a
robust feature of the LLM's conditional target-speed distribution. In
the traffic context, this demonstrates that safety in LLM-controlled
physical systems cannot rely solely on the cognitive layer, and must be
enforced at the dynamics layer (through the IDM safety buffer at the
per-vehicle level and the p\_c(ρ) deployment constraint at the fleet
level).

\subsection{5.3 Density-dependent safety constraint for LLM-based AV
deployment}\label{density-dependent-safety-constraint-for-llm-based-av-deployment}

The p\_c(ρ) phase boundary (Table 4) provides a structural safety
constraint for the deployment of LLM-based driving controllers in mixed
traffic. The implication is density-specific:

\begin{itemize}
\tightlist
\item
  \textbf{Low-density highway conditions} (ρ \textless{} 50 veh/km,
  e.g., sparse suburban): no phase transition is observed. Even 100\%
  LLM penetration (p = 1.0) remains in free flow (v\_std ≈ 0.37 m/s),
  indicating that systemic instability does not arise at this density.
\item
  **Mid-density arterial (ρ ≈ 60-80 veh/km): p\_c decreases from
  \textasciitilde0.79 (at ρ = 60.9, n = 14) to \textasciitilde0.50 (at ρ
  = 78.3, n = 18), so the safe LLM-penetration ceiling drops from
  \textasciitilde80\% to \textasciitilde50\% across this density range.
\item
  \textbf{Dense urban flow} (ρ \textgreater{} 90 veh/km): p\_c ≈ 0.23,
  so a minority of LLM vehicles (≈ 1 in 4) is sufficient to destabilise
  the system.
\end{itemize}

This boundary should inform both regulatory safety specifications and AV
deployment policies in mixed-fleet scenarios. We note that the absolute
p\_c values are model-specific (GLM-4-Flash); cross-model validation is
needed before generalisation. The qualitative shape of p\_c(ρ), which is
monotonically decreasing with density, is expected to be a class-wide
property of LLM-driven traffic.

\subsection{5.4 Implications for LLM-controlled physical systems beyond
traffic}\label{implications-for-llm-controlled-physical-systems-beyond-traffic}

Although our experimental system is traffic, the SH mechanism is
potentially class-wide for any LLM-controlled multi-agent system. Three
conditions must hold. (a) The LLM's output is fed into continuous
physical dynamics. (b) The dynamics contain nonlinear safety constraints
(here, the IDM braking response). (c) the system's closed-loop dynamics
amplify per-cycle perturbations faster than they damp them. Candidate
systems include warehouse robots, drone swarms, and human-AI
collaborative manufacturing. The general lesson is that
\textbf{microscopic LLM heterogeneity, when amplified by nonlinear
safety layers, can produce macroscopic instabilities that are invisible
to local reasoning}. The p\_c(ρ) framework provides a structural
template for assessing these risks in other domains; we note, however,
that the present cascade mechanism is specific to 1D serial-interaction
topologies, and 2D/3D geometries may exhibit qualitatively different
amplification dynamics.

\subsection{5.5 Limitations and outlook}\label{limitations-and-outlook}

\textbf{Limitations.} Our study has eight principal limitations.
\textbf{First, single LLM model}. All results use GLM-4-Flash.
Cross-model validation with architecturally different LLMs (e.g.,
DeepSeek-V3, Qwen2.5, Claude) is needed to establish whether SH is
class-wide. \textbf{Second, single prompt language}. All prompts are in
Chinese. Whether SH survives multilingual prompting is untested.
\textbf{Third, single track length and IDM parameterisation}. Extension
to longer tracks and different safety-layer calibrations would test
robustness. \textbf{Fourth, single-instruction homogeneous fleet}. All
drivers receive the same prompt. Heterogeneous-prompt configurations
(e.g., aggressive vs cautious) may reveal additional structure.
\textbf{Fifth, empirically calibrated safety-layer thresholds.} The
safety-fusion-layer thresholds (startup at 2.2 m/s, IDM takeover at 2.5
m, emergency braking at 3.0 m gap and 2.0 m/s Δv) were set by manual
tuning, not systematically varied. A sensitivity sweep is left to future
work. The SH phenomenon is expected to be robust because it originates
in the LLM behavioural layer (Stages 1-3) rather than in the safety
layer (Stage 4); Exp6 confirms that LLM velocity variance persists when
the safety layer is ablated (full: v\_std = 2.11, minimal: v\_std =
3.00; SI S1.3). \textbf{Sixth, Sugiyama validation}. The LLM-human
comparison in §4.5 rests on only one directly comparable density point
(ρ ≈ 95.7 veh/km). Additional density-matched comparisons would
strengthen the qualitative bridge. A finite-size consistency check at n
= 44 (same density, 460 m track, 3 seeds per condition; Supplementary
Information S4) confirms systematic stabilisation with increasing system
size, consistent with the spatial resonance interpretation. Statistical
power remains insufficient for formal finite-size scaling; full analysis
across multiple system sizes is deferred to future work.
\textbf{Seventh, absence of a direct positive control.} A direct
positive control in which all LLM agents are constrained to a shared
target speed (removing per-cycle target\_delta divergence across agents)
would triangulate whether SH originates in the dispersion of LLM-chosen
speed adjustments or in the collective dynamics of heterogeneous
choices. This experiment is left to future work. \textbf{Eighth,
stateless and non-communicating agents.} The present work deploys LLM
agents in a stateless, non-communicating configuration: each agent
decides based on its current local state (speed, gap, Δv) and a 5-s
history, without access to other agents' intentions or to its own past
decisions beyond the 5-s window. CoMAL {[}27{]} and other recent
LLM-driving frameworks deploy richer architectures with inter-agent
communication channels (shared message pools) and persistent memory
modules. Whether the SH mechanism persists, is amplified, or is
suppressed under such richer architectures is an open question. We
conjecture (without experimental evidence) that inter-agent
communication enabling explicit coordination (e.g., a designated platoon
leader) would tend to \emph{suppress} SH by reducing the per-cycle
target-speed variance that drives the cascade, while a richer memory
module might \emph{amplify} it by allowing past variance to compound. A
controlled study of these architectural degrees of freedom is left to
future work. Additionally, the temperature ablation (§4.1, control 6)
was conducted at full LLM deployment (p = 1.0) only; whether p\_c(ρ)
itself depends on sampling temperature at intermediate LLM penetration
is an open question. The \textasciitilde8-11\% v\_std variation observed
at p = 1.0 suggests that any p\_c shift would be correspondingly small,
but a definitive answer requires a p-by-T cross-sweep that is left to
future work.

\textbf{Outlook.} Four lines of work follow directly. \textbf{First},
cross-model validation (DeepSeek-V3, Qwen2.5, Claude) to test whether SH
is universal to LLM controllers or specific to GLM-4-Flash.
\textbf{Second}, multilingual prompt validation to test whether the SH
mechanism and its cognitive correlates are robust across languages.
\textbf{Third}, finite-size scaling analysis at n ∈ \{22, 44, 88, 176\}
to determine whether the sigmoidal transition sharpens systematically
with increasing system size, approaching a sharp threshold in the
thermodynamic limit. \textbf{Fourth}, controlled comparison of the
LLM-as-direct-controller architecture (this work) and the
LLM-as-IDM-parameter-setter architecture (CoMAL {[}27{]}) on the same
ring-road task, to quantify how the IDM low-pass filter (§2.2)
attenuates per-cycle target-delta variance and whether SH persists,
attenuates, or is suppressed. More broadly, the SH mechanism and the
p\_c(ρ) framework provide a template for the principled safety analysis
of any LLM-controlled multi-agent physical system, a class of systems
that is rapidly expanding beyond traffic into warehouse robotics, drone
swarms, and human-AI collaborative manufacturing; the 1D topology caveat
noted in §5.4 applies here as well.

\begin{center}\rule{0.5\linewidth}{0.5pt}\end{center}

\section{6. Conclusion}\label{conclusion}

We have presented, to our knowledge, the \textbf{first study to
identify, through systematic matched controls and chain-of-thought
observability, a previously uncharacterised collective mechanism (SH)
arising in LLM-controlled multi-agent traffic, and to map a
density-dependent phase boundary p\_c(ρ) for LLM-based driving
deployment}. The work uses LLM agents as \textbf{direct, real-time,
closed-loop target-speed controllers} (LLM-emitted target-delta per 0.5
s cycle, with IDM only as a hard collision-avoidance clamp) on a
multi-agent ring-road simulation, with chain-of-thought reasoning as
microscopic observability into the cognitive layer of LLM-controlled
dynamics - a vantage point unavailable in any human-subject paradigm or
in traditional car-following simulations (which have no cognitive
decision layer to observe). The closest architectural prior, CoMAL
{[}27{]}, uses LLMs to \emph{set} IDM parameters (target speed, max
acceleration, minimum gap) on ring-road, figure-eight, and merge
scenarios, with the IDM dynamics remaining the primary controller; CoMAL
focuses on flow optimisation rather than mechanism discovery. The four
distinguishing dimensions of the present work (research goal, control
experiment design, chain-of-thought microscope, density-dependent phase
boundary) are detailed in S12.3. A structured 2024-2025 literature
search (Supplementary Information S12) supports the narrow reading of
the novelty claim. We term this mechanism \textbf{Sustained
Heterogeneity (SH)}. The work contributes (i) a new class of traffic
instability structurally distinct from all tested candidate mechanisms,
verified through six matched control experiments. (ii) A structural
density-dependent safety boundary p\_c(ρ) for LLM-based driving
deployment, supported by an initiation-threshold model whose primary
content is the identification of three independent physical axes
(Δs\_trigger, σ, n) governing p\_c (Section 4.4). (iii) Microscopic
evidence that safety-vigilant LLM reasoning does not suppress per-cycle
target-speed adjustment divergence across agents - demonstrating that
collective stability in LLM-controlled physical systems cannot rely
solely on the cognitive layer, and must be enforced at the dynamics
layer. A quantitative correlation analysis of 39,600 LLM decisions
(Supplementary Information S11) refines this last claim: the
chain-of-thought \texttt{decision} field is strongly predictive of the
magnitude of \texttt{target\_speed\_delta} (Cohen's \emph{d} = 2.37; KS
statistic 0.98), so the cognitive layer is \emph{informative} about the
agent's macro-action, but the per-cycle target-delta variance is
concentrated in the \texttt{ADJUST\_SPEED} sub-population (31\%
decelerate, 67\% accelerate, within a 1.20 ± 1.21 m/s
\texttt{\textbar{}target\_delta\textbar{}} distribution) --- a
sub-category invisible in the macro-action label. The SH mechanism and
the p\_c(ρ) framework together provide a principled template for the
safety analysis of any LLM-controlled multi-agent physical system, a
class that is rapidly expanding beyond traffic into warehouse robotics,
drone swarms, and human-AI collaborative manufacturing; the 1D topology
caveat noted in §5.4 applies here as well. The SH mechanism and the
p\_c(ρ) framework together raise a foundational question for AI safety:
how can collective stability in LLM-controlled physical systems be
enforced at the dynamics layer when cognitive-layer safety reasoning is
insufficient? This question will become increasingly important as LLM
controllers are deployed beyond traffic into other multi-agent physical
domains.

\begin{center}\rule{0.5\linewidth}{0.5pt}\end{center}

\section{7. Competing interests}\label{competing-interests}

The authors declare no competing interests.

\section{8. Author contributions}\label{author-contributions}

Y.Q. conceived the overall research design, implemented the simulation
software, analysed the results, and wrote the manuscript. Y.G. conducted
parameter sweeps and baseline comparisons, and participated in
experimental result analysis and draft manuscript preparation. Both
authors reviewed and approved the final manuscript.

\section{9. Acknowledgements and
funding}\label{acknowledgements-and-funding}

The authors thank the open-source community for the simulation tools
used in this work.

\textbf{Funding.} This research received no specific grant from any
funding agency in the public, commercial, or not-for-profit sectors.

\section{10. Data availability}\label{data-availability}

All metrics files (170 runs), decision logs (1 representative run for SH
inspection), and analysis scripts are available at {[}URL to be inserted
upon publication{]}. Raw LLM prompt-response pairs are available upon
reasonable request.

\section{11. Code availability}\label{code-availability}

The simulation framework is released as open-source software. It
includes the LLM agent pipeline, IDM safety layer, and all experiment
drivers. The release is under the MIT licence at an open repository, to
be made public upon publication.

\begin{center}\rule{0.5\linewidth}{0.5pt}\end{center}

\section{References}\label{references}

{[}1{]} Y. Sugiyama, M. Fukui, M. Kikuchi, K. Hasebe, A. Nakayama, K.
Nishinari, S. Tadaki, S. Yukawa, ``Traffic jams without
bottlenecks-experimental evidence for the physical mechanism of the
formation of a jam,'' \emph{New J. Phys.} \textbf{10}, 033001 (2008).

{[}2{]} D. Helbing, ``Traffic and related self-driven many-particle
systems,'' \emph{Rev.~Mod. Phys.} \textbf{73}, 1067 (2001).

{[}3{]} M. Treiber, A. Kesting, D. Helbing, ``Delays, inaccuracies and
anticipation in microscopic traffic models,'' \emph{Physica A}
\textbf{360}, 71 (2006).

{[}4{]} A. Kesting, M. Treiber, D. Helbing, ``Enhanced intelligent
driver model to access the impact of driving strategies on traffic
capacity,'' \emph{Phil. Trans. R. Soc. A} \textbf{368}, 4585 (2010).

{[}5{]} J. J. Horton, ``Large language models as simulated economic
agents: What can we learn from homo silicus?,'' \emph{NBER Working
Paper} (2023).

{[}6{]} G. V. Aher, R. I. Arriaga, A. T. Kalai, ``Using large language
models to simulate multiple humans and replicate human subject
studies,'' in \emph{Proc. ICML} (2023).

{[}7{]} L. P. Argyle et al., ``Out of one, many: Using language models
to simulate human samples,'' \emph{Polit. Anal.} \textbf{31}, 337
(2023).

{[}8{]} S. B. K. Dillon et al., ``Can AI language models replace human
participants?,'' \emph{Trends Cogn. Sci.} \textbf{27}, 597 (2023).

{[}9{]} M. Bando, K. Hasebe, A. Nakayama, A. Shibata, Y. Sugiyama,
``Dynamical model of traffic congestion and numerical simulation,''
\emph{Phys. Rev.~E} \textbf{51}, 1035 (1995).

{[}10{]} M. Bando, K. Hasebe, K. Nakanishi, A. Nakayama, ``Analysis of
optimal velocity model with explicit delay,'' \emph{Phys. Rev.~E}
\textbf{58}, 5429 (1998).

{[}11{]} J. Wei et al., ``Chain-of-thought prompting elicits reasoning
in large language models,'' in \emph{NeurIPS} (2022).

{[}12{]} M. Villarreal, B. Poudel, W. Li, ``Can ChatGPT enable ITS? The
case of mixed traffic control via reinforcement learning,'' in
\emph{Proc. IEEE ITSC}, 2023, pp.~4719-4726; arXiv:2306.08094.

{[}13{]} D. Fu et al., ``Drive like a human: Rethinking autonomous
driving with large language models,'' in \emph{Proc. NeurIPS Workshop on
Foundation Models for Decision Making}, 2023; arXiv:2307.07162.

{[}14{]} S. Lai, Z. Xu, W. Zhang, H. Liu, H. Xiong, ``LLMLight: Large
language models as traffic signal control agents,'' arXiv:2312.16044
(2023).

{[}15{]} X. Chen, M. Peng, P. Tiu, Y. Wu, J. Chen, M. Zhu, X. Zheng,
``GenFollower: Enhancing car-following prediction with large language
models,'' arXiv:2407.05611 (2024).

{[}16{]} X. Guo, Q. Zhang, J. Jiang, M. Peng, M. Zhu, H. Yang, ``Towards
explainable traffic flow prediction with large language models,''
arXiv:2404.02937 (2024).

{[}17{]} S. Zhang, J. Tian, Z. Zhu, S. Huang, J. Yang, W. Zhang,
``DriveGen: Towards infinite diverse traffic scenarios with large
models,'' arXiv:2503.05808 (2025).

{[}18{]} Q. Lu, X. Wang, Y. Jiang, G. Zhao, M. Ma, S. Feng,
``OmniTester: Multimodal large language model driven scenario testing
for autonomous vehicles,'' arXiv:2409.06450 (2024).

{[}19{]} M. Jeong, J. Chang, Y. Yoon, ``AgentSUMO: An agentic framework
for interactive simulation scenario generation in SUMO via large
language models,'' arXiv:2511.06804 (2025).

{[}20{]} M. Treiber, A. Hennecke, D. Helbing, ``Congested traffic states
in empirical observations and microscopic simulations,'' \emph{Phys.
Rev.~E} \textbf{62}, 1805 (2000).

{[}21{]} G. Orosz, R. E. Wilson, B. Krauskopf, ``Global bifurcation
investigation of an optimal velocity traffic model with driver reaction
time,'' \emph{Phys. Rev.~E} \textbf{70}, 026207 (2004).

{[}22{]} B. S. Kerner, ``Three-phase traffic theory and highway
capacity,'' \emph{Physica A} \textbf{333}, 379 (2004).

{[}23{]} A. Nakayama, M. Kikuchi, A. Shibata, K. Nishinari, S. Tadaki,
S. Yukawa, Y. Sugiyama, ``Metastability in the formation of an
experimental traffic jam,'' \emph{New J. Phys.} \textbf{11}, 083025
(2009).

{[}24{]} S. Tadaki, M. Kikuchi, M. Fukui, A. Nakayama, K. Nishinari, A.
Shibata, Y. Sugiyama, S. Yukawa, ``Phase transition in traffic jam
experiment on a circuit,'' \emph{New J. Phys.} \textbf{15}, 103034
(2013).

{[}25{]} S. Tadaki, M. Kikuchi, A. Nakayama, A. Shibata, Y. Sugiyama, S.
Yukawa, ``Characterizing and distinguishing free and jammed traffic
flows from the distribution and correlation of experimental speed
data,'' \emph{New J. Phys.} \textbf{18}, 083022 (2016).

{[}26{]} A. Nakayama, M. Kikuchi, A. Shibata, Y. Sugiyama, S. Tadaki, S.
Yukawa, ``Quantitative explanation of circuit experiments and real
traffic using the optimal velocity model,'' \emph{New J. Phys.}
\textbf{18}, 043040 (2016).

{[}27{]} H. Yao, L. Da, V. Nandam, J. Turnau, Z. Liu, L. Pang, H. Wei,
``CoMAL: Collaborative multi-agent large language models for
mixed-autonomy traffic,'' in \emph{Proc. SIAM Int. Conf. Data Mining
(SDM25)}; arXiv:2410.14368 (2024).

\begin{center}\rule{0.5\linewidth}{0.5pt}\end{center}

\section{Figure captions}\label{figure-captions}

\begin{figure}
\centering
\pandocbounded{\includegraphics[keepaspectratio,alt={Figure 1}]{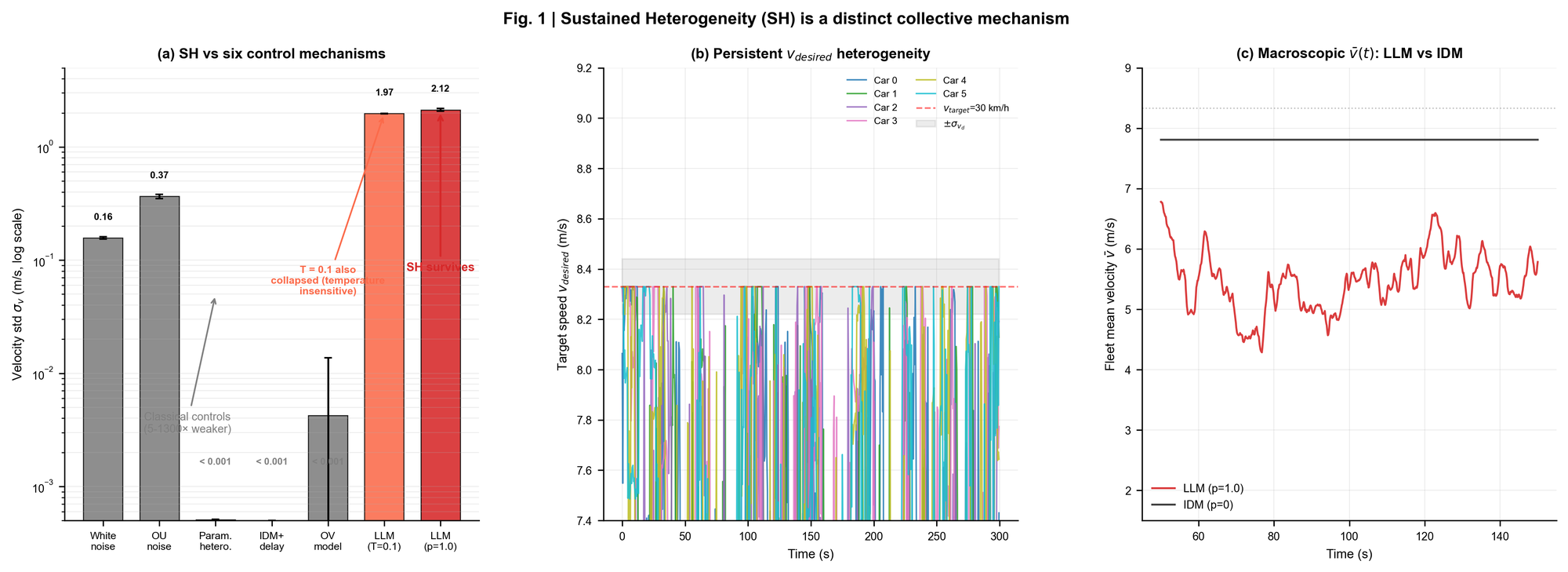}}
\caption{Figure 1}
\end{figure}

\textbf{Figure 1 \textbar{} Sustained Heterogeneity (SH) is a distinct
collective mechanism.} (a) Velocity standard deviation σ\_v for the LLM
fleet (2.12 m/s, p = 1.0, T = 0.6) compared with six matched control
experiments grouped into two classes. \textbf{Classical controls:} white
noise (0.16 m/s, 0.075× of LLM v\_std), OU noise (0.37 m/s, 0.174× of
LLM v\_std---the closest classical candidate), parameter heterogeneity
(\textless{} 0.001 m/s, \textless{} 0.0005× of LLM v\_std), IDM + 0.5 s
delay (\textless{} 0.001 m/s, \textless{} 0.0005× of LLM v\_std), and OV
model (≤ 0.035 m/s, 0.017× of LLM v\_std). Two of five are
deterministically zero. \textbf{Temperature control:} LLM T = 0.1 (1.97
m/s, 0.93× of LLM v\_std)---the sole control not ruled out as a
candidate mechanism, confirming that SH is approximately
temperature-insensitive rather than a sampling artefact. (b)
Representative 30 s trajectory of v\_desired for a single LLM vehicle,
showing persistent target-speed drift at ≈ 0.1 m/s/cycle around the 30
km/h instruction. (c) Macroscopic fleet-mean velocity v̄(t) over a 100 s
window (t = 50–150 s): the LLM fleet (red) oscillates between ≈ 4.3
and 6.8 m/s (full-run range 3.4–8.4 m/s), while the
matched IDM baseline (black) remains flat at 7.81 m/s. The v\_std
contrast (2.12 vs \textless{} 0.001 m/s) reproduces the qualitative
stop-and-go signature of human-like traffic breakdown.

\begin{figure}
\centering
\pandocbounded{\includegraphics[keepaspectratio,alt={Figure 2}]{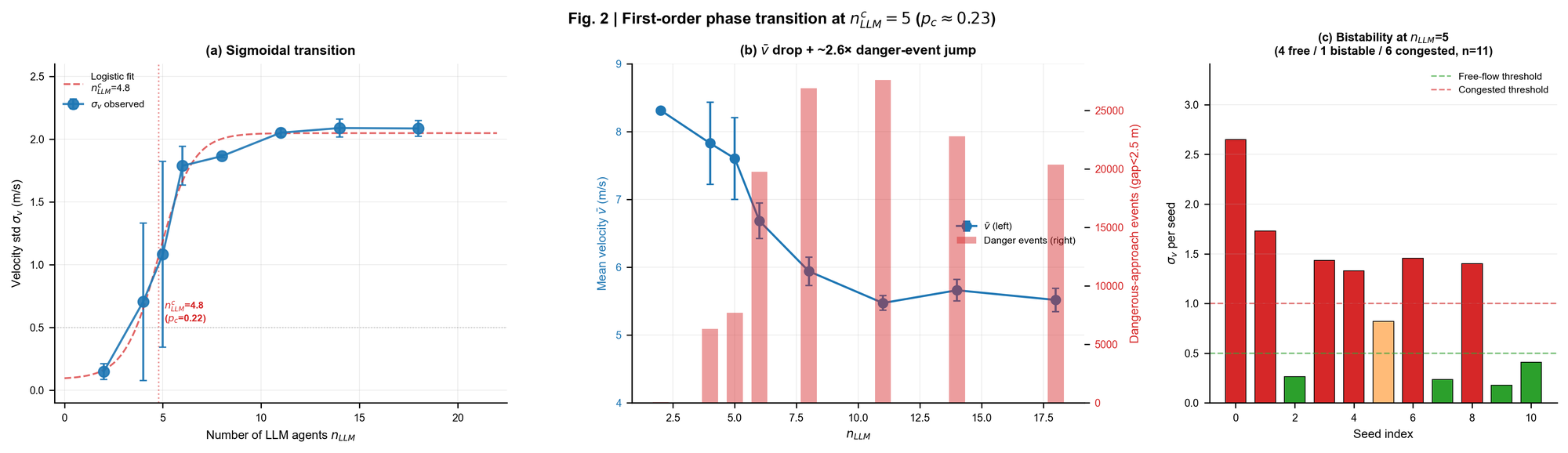}}
\caption{Figure 2}
\end{figure}

\textbf{Figure 2 \textbar{} Sigmoidal transition with seed-dependent
bifurcation at n\_LLM = 5 (p\_c ≈ 0.23).} (a) σ\_v as a function of
n\_LLM at n = 22, ρ = 95.7 veh/km. The transition is sigmoidal with
midpoint at n\_LLM = 5 and width Δn\_LLM ≈ 2-4 vehicles (logistic fit;
red dashed; k ≈ 1.0-3.0 depending on seed selection). (b) Mean velocity
v̄ (blue, left axis) and dangerous-approach events (gap \textless{} 2.5
m; red bars, right axis) as a function of n\_LLM. A \textasciitilde2.6×
abrupt jump in dangerous events between n\_LLM = 5 and n\_LLM = 6
corroborates the sigmoidal character. (c) Probabilistic seed-dependent
bifurcation at n\_LLM = 5: across 11 runs (10 unique seeds), 5 of 11
(45\%) remain in free flow (v̄ \textgreater{} 7.88 m/s) and 6 of 11
(55\%) collapse to the congested attractor (v̄ \textless{} 7.53 m/s)---a
binary partition with zero runs in the natural v̄ gap {[}7.53, 7.88{]}
m/s.

\begin{figure}
\centering
\pandocbounded{\includegraphics[keepaspectratio,alt={Figure 3}]{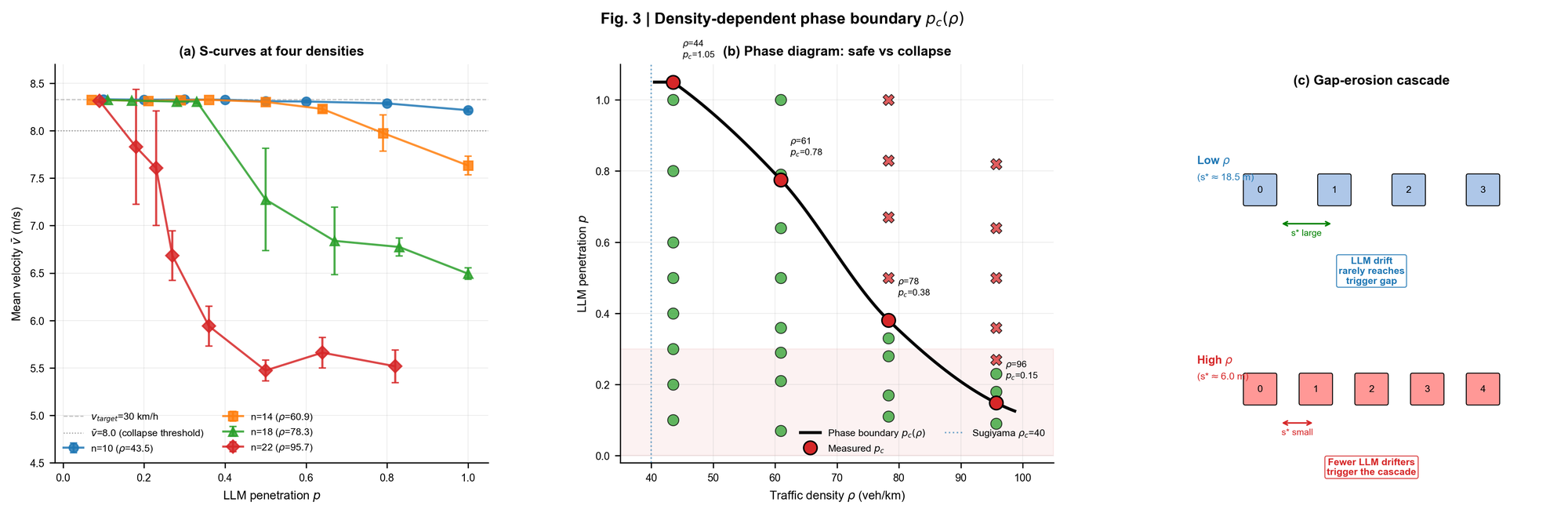}}
\caption{Figure 3}
\end{figure}

\textbf{Figure 3 \textbar{} Density-dependent phase boundary p\_c(ρ).}
(a) Mean velocity v̄ as a function of LLM penetration p for four
densities ρ ∈ \{43.5, 60.9, 78.3, 95.7\} veh/km. Each curve is an
S-shaped sigmoid with a sharp knee; the critical penetration p\_c(ρ) is
identified by logistic-curve fitting to v̄(p) at each fixed density
(inflection point; see §3.5). (b) The phase boundary in the (ρ, p)
plane: p\_c decreases monotonically with density, from no transition at
ρ = 43.5 veh/km to p\_c ≈ 0.23 at ρ = 95.7 veh/km. (c) Schematic of the
SH gap-erosion mechanism: at higher density the equilibrium
inter-vehicle gap shrinks, so a smaller number of LLM vehicles suffices
to erode the safety buffer and trigger the cascade.

\begin{figure}
\centering
\pandocbounded{\includegraphics[keepaspectratio,alt={Figure 4}]{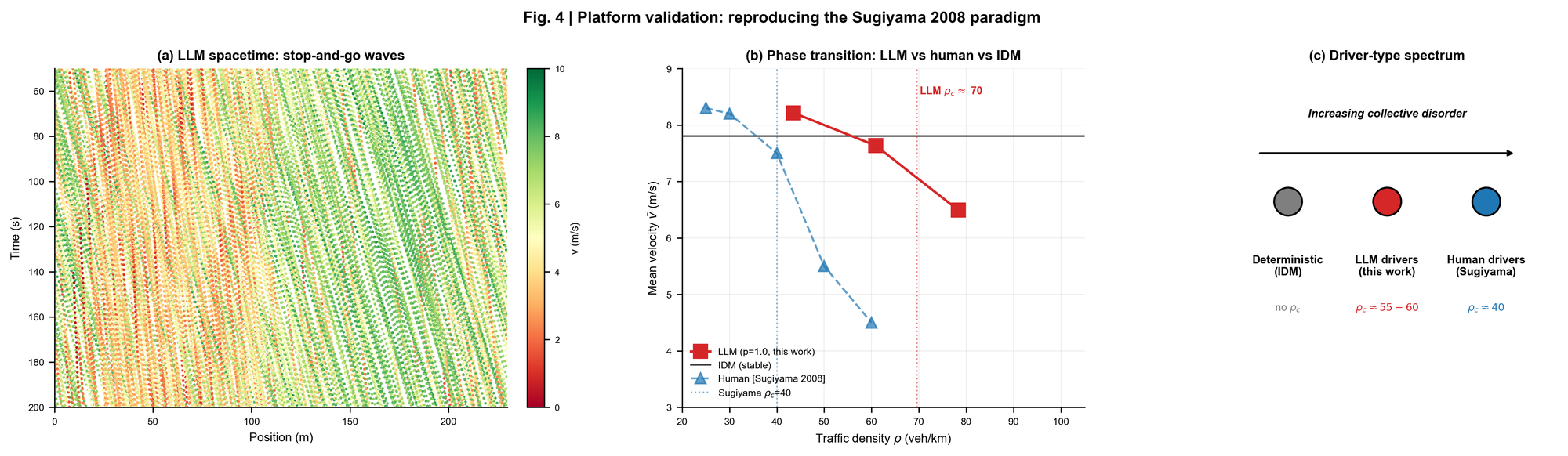}}
\caption{Figure 4}
\end{figure}

\textbf{Figure 4 \textbar{} Platform validation: reproducing the
Sugiyama 2008 paradigm.} (a) Space-time diagram of the LLM-driven ring
road at ρ = 95.7 veh/km, showing oscillatory stop-and-go waves (red =
slow, green = fast). (b) Comparison of v̄(ρ) for the LLM platform (red)
with the deterministic IDM (black, stable at all densities) and the
human baseline (blue, ρ\_c ≈ 25 veh/km in highway data {[}1{]} and ≈
80-90 veh/km in ring-road experiments {[}24{]}). LLM drivers transition
at ρ\_c ≈ 50-60 veh/km (interpolated; see §4.5) --- lower than the human
ring-road ρ\_c ≈ 80-90 veh/km {[}24{]}, indicating greater
susceptibility to collective instability. (c) Driver-type spectrum: IDM
drivers have no transition, human drivers transition at intermediate
density (ρ\_c ≈ 80-90 veh/km), and LLM drivers transition earliest (ρ\_c
≈ 50-60 veh/km).

\begin{figure}
\centering
\pandocbounded{\includegraphics[keepaspectratio,alt={Figure 5}]{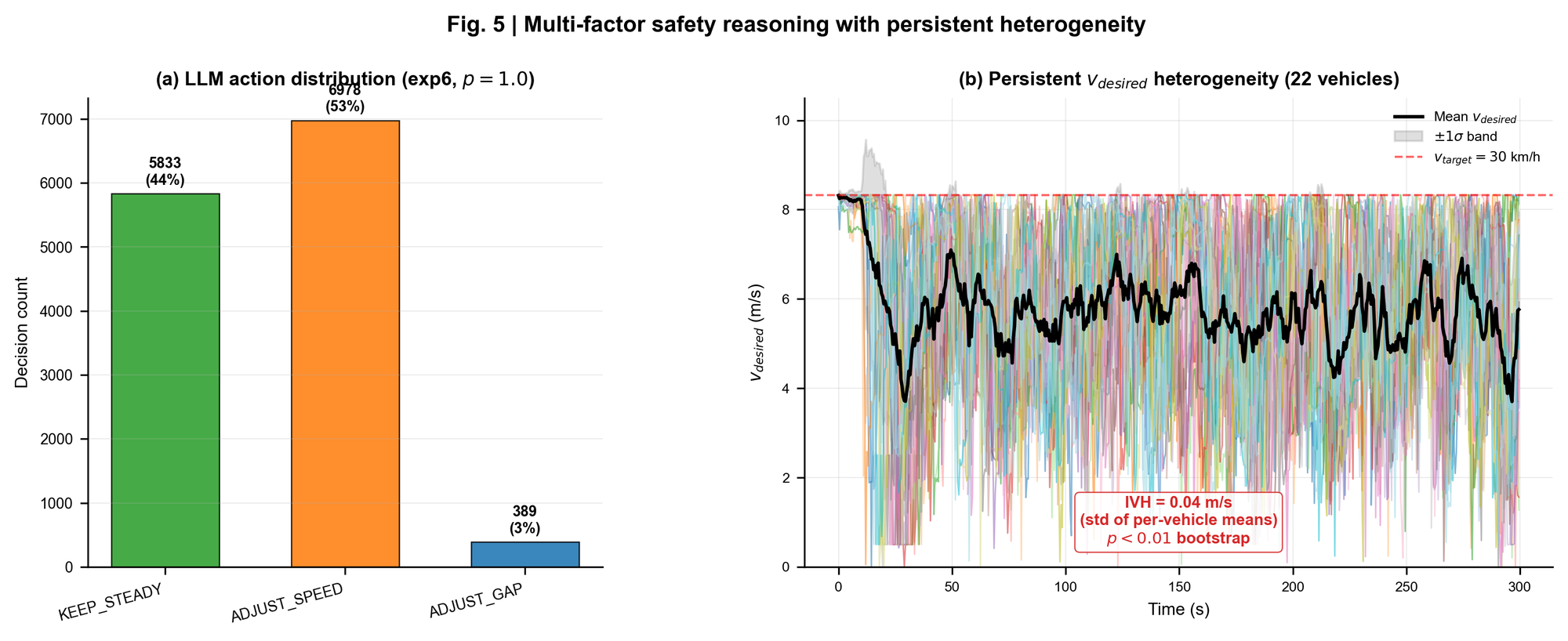}}
\caption{Figure 5}
\end{figure}

\textbf{Figure 5 \textbar{} Cognitive origin of SH: multi-factor safety
reasoning with persistent heterogeneity.} (a) Distribution of LLM action
types across 13,200 decisions (all 22 vehicles, 600 cycles from one
run): KEEP\_STEADY 44\%, ADJUST\_SPEED 53\%, ADJUST\_GAP 3\%. (b)
Persistent per-cycle target\_delta divergence across the 22
LLM-controlled vehicles over the full 600-cycle segment, with
per-vehicle v\_desired trajectories (coloured), the ensemble mean
(black), and ±1 σ band (grey). The coexistence of safety-vigilant
multi-factor reasoning with persistent per-cycle target\_delta
divergence \ldots demonstrates that collective instability in
LLM-controlled systems is a robust feature of multi-agent dynamics, not
a superficial stochastic artefact.

\begin{center}\rule{0.5\linewidth}{0.5pt}\end{center}

\section{Supplementary Information}\label{supplementary-information}

\textbf{S1. Full LLM prompt text (Chinese).}

This document provides the complete system-and-user prompt template used
to elicit LLM driving decisions in all experiments. The same safety
fusion layer (five-mode priority cascade, §3.2 main text) is used
throughout. The prompt is in Chinese (Simplified). The English
translation provided below is intended for reference only --- the
experiments use the Chinese text directly.

\begin{center}\rule{0.5\linewidth}{0.5pt}\end{center}

\textbf{S1.1 Common protocol.}

\begin{itemize}
\tightlist
\item
  \textbf{Role:} LLM acts as a vehicle driver on a 230 m circular track.
\item
  \textbf{Decision cycle:} LLM is queried every 0.5 s (matching the
  human experiment time scale) and returns a JSON object with
  \texttt{chain\_of\_thought}, \texttt{decision},
  \texttt{target\_speed\_delta}, \texttt{desired\_gap\_delta}, and
  \texttt{reasoning} fields.
\item
  \textbf{Action space:}
  \texttt{ADJUST\_SPEED\ \textbar{}\ ADJUST\_GAP\ \textbar{}\ KEEP\_STEADY}.
\item
  \textbf{Numerical output:}
  \texttt{target\_speed\_delta\ ∈\ {[}-2.0,\ 2.0{]}\ m/s} (increment to
  current \texttt{v\_desired});
  \texttt{desired\_gap\_delta\ ∈\ {[}-1.0,\ 2.0{]}\ m} (increment to
  current \texttt{s*}).
\item
  \textbf{History window:} past 5 s sampled at 0.5 s intervals (10
  history points).
\item
  \textbf{LLM model:} \texttt{GLM-4-Flash} (Zhipu AI) at temperature
  \texttt{T\ =\ 0.6} unless otherwise noted.
\item
  \textbf{API endpoint:}
  \texttt{https://open.bigmodel.cn/api/paas/v4/chat/completions}.
\end{itemize}

\begin{center}\rule{0.5\linewidth}{0.5pt}\end{center}

\textbf{S1.2 Variant A: \texttt{full} prompt (default, used in all
main-text experiments).}

The ``full'' prompt is the most elaborate, including explicit safety
constraints, role description, and self-check questions. It is the v8.4
production prompt.

\textbf{S1.2.1 System prompt (Chinese, original).}

\begin{verbatim}
# Role
你是一个驾驶风格稳健的司机，正在参加一个环形跑道上的交通流实验。
你开车平稳，不猛踩油门也不急刹，习惯保持舒适的车距。
速度差不多在30km/h左右就行，不追求精确。但遇到危险会果断反应。

# 你的驾驶习惯
- 车距只要不觉得太近就行。
- 即使前车减速，你也尽量保持缓慢前进，不要完全停住。保持 5~10 km/h 的蠕行速度。
- 前车刹车灯亮了你会注意，但不一定每次都跟着急刹——先看车距是否真的在快速缩小。

# 安全底线
- 绝对不能撞到前车。如果感觉太近了，果断减速。
- 你的反应有大约半秒的延迟，踩了刹车不会立刻生效。

# 你需要回答的问题
在做出决策前，请在心里快速过一下：
1. 现在安全吗？前车有没有在逼近？
2. 速度是不是差太多了？车距是不是太近或太远了？
3. 结合过去几秒的趋势，情况是在变好还是变坏？

# Output Format (严格 JSON)
{
  "chain_of_thought": "<你的内心想法，用自然语言描述>",
  "decision": "ADJUST_SPEED | ADJUST_GAP | KEEP_STEADY",
  "target_speed_delta": <数字，-2.0~2.0>,
  "desired_gap_delta": <数字，-1.0~2.0>,
  "reasoning": "<一句话决策理由>"
}
\end{verbatim}

\textbf{S1.2.2 System prompt (English reference translation).}

\begin{verbatim}
# Role
You are a driver with a steady driving style, participating in a traffic-flow
experiment on a circular track. You drive smoothly — no aggressive acceleration,
no sudden braking, comfortable following distance. Target speed is roughly
30 km/h; precision is not required. However, you react decisively to danger.

# Your driving habits
- As long as the gap does not feel too close, it is fine.
- Even when the leader slows, try to keep creeping forward; do not come to
  a full stop. Maintain 5–10 km/h creep speed.
- You do notice when the leader's brake light is on, but you do not always
  brake hard in response — first check whether the gap is actually shrinking
  rapidly.

# Safety floor
- You must never collide with the leader. If it feels too close, brake
  decisively.
- Your reaction has roughly half a second of delay; braking does not take
  effect immediately.

# Questions to answer before deciding
1. Is it safe right now? Is the leader approaching?
2. Is the speed difference too large? Is the gap too close or too far?
3. Based on the past few seconds, is the situation improving or worsening?

# Output format (strict JSON)
{
  "chain_of_thought": "<your inner reasoning, in natural language>",
  "decision": "ADJUST_SPEED | ADJUST_GAP | KEEP_STEADY",
  "target_speed_delta": <number, -2.0 to 2.0>,
  "desired_gap_delta": <number, -1.0 to 2.0>,
  "reasoning": "<one-sentence decision rationale>"
}
\end{verbatim}

\textbf{S1.2.3 User prompt template (Chinese, original).}

\begin{verbatim}
当前状态：
- 你的速度：{speed_kmh:.1f} km/h（目标大约30）
- 前车距离：{gap:.2f} m
- 前车速度：{leader_speed_kmh:.1f} km/h
- 相对速度：{relative_speed_kmh:+.1f} km/h（正=在接近，负=在远离）
- 前车刹车灯：{brake_str}
- 碰撞时间：{ttc_str}

过去5秒的趋势：
时间       速度   车距   前车速度  刹车灯  趋势
-------   -----   ----   --------  ------  ----
[-time-stride s  ... 10 lines ...]

请做出决策：
\end{verbatim}

where: - \texttt{speed\_kmh}, \texttt{leader\_speed\_kmh},
\texttt{relative\_speed\_kmh} are converted from m/s using
\texttt{×\ 3.6}. - \texttt{ttc\_str} is rendered as
\texttt{f"\{ttc:.2f\}s\ (危险!)"} if \texttt{\textless{}\ 1.0},
\texttt{f"\{ttc:.2f\}s\ (偏近)"} if \texttt{\textless{}\ 2.0}, otherwise
\texttt{f"\{ttc:.2f\}s"} (or \texttt{∞} if undefined). -
\texttt{brake\_str} is \texttt{亮} (on) or \texttt{灭} (off). - Each
history line is rendered as
\texttt{f"-\{time\_ago:.1f\}s\ \ \{h\_speed:5.1f\}\ \ \{h\_gap:5.2f\}\ \ \{h\_leader:6.1f\}\ \ \ \ \{h\_brake\}\ \ \ \{gap\_trend\}"}
with \texttt{gap\_trend\ ∈\ \{↓缩,\ ↑扩,\ →稳\}} based on 0.3 m
threshold.

\textbf{S1.2.4 User prompt (English reference).}

\begin{verbatim}
Current state:
- Your speed: {speed_kmh:.1f} km/h (target ≈ 30)
- Gap to leader: {gap:.2f} m
- Leader speed: {leader_speed_kmh:.1f} km/h
- Relative speed: {relative_speed_kmh:+.1f} km/h (+ = approaching, − = receding)
- Leader brake light: {brake_str}
- Time-to-collision: {ttc_str}

Trend over the last 5 seconds:
[10 history lines as described above]

Please make a decision:
\end{verbatim}

\begin{center}\rule{0.5\linewidth}{0.5pt}\end{center}

\textbf{S1.3 Safety-layer ablation (hard-wall / soft / minimal).}

The \texttt{soft} and \texttt{minimal} rows in the original experiment
logs refer to safety-layer configurations (Exp6 hard-wall ablation:
\texttt{SAFETY\_MODE\ =\ "soft"} or \texttt{"minimal"}), not prompt
variants. Both modes use the same Full system prompt (S1.2). The
safety-layer ablation data (full: v\_std = 2.11, 0 collisions; soft:
1.98, 84; minimal: 3.00, 7,848; all at n = 22, ρ = 95.7 veh/km, seed =
42) are not reported in the main text but are archived with the
experiment logs.

\begin{center}\rule{0.5\linewidth}{0.5pt}\end{center}

\textbf{S1.4 Safety fusion layer.}

The acceleration pipeline that follows each LLM decision is a five-mode
priority-cascaded fusion stage (§3.2 main text). This section provides
additional implementation detail.

\textbf{Pipeline structure.} The LLM's \texttt{target\_speed\_delta}
enters a priority-ordered cascade resolving five acceleration modes. All
modes share the same IDM base parameters: a\_max = 3.0 m/s², b = 2.5
m/s², v₀ = 8.33 m/s, T = 0.3 s, s₀ = 0.5 m. Modes earlier in the cascade
intercept and return before later modes can execute.

\begin{enumerate}
\def\labelenumi{\arabic{enumi}.}
\item
  \textbf{Proportional controller (default).} Converts LLM delta to
  acceleration: a\_target = target\_speed\_delta / τ\_c (τ\_c = 0.5 s).
  A three-tier deceleration floor tightens with the dynamic state:

  \begin{itemize}
  \tightlist
  \item
    s \textless{} 3 m and Δv \textgreater{} 2 m/s → floor = −8.0 m/s²
  \item
    s \textless{} 5 m and Δv \textgreater{} 1 m/s → floor = −5.0 m/s²
  \item
    Otherwise → floor = −2.5 m/s² (= IDM b) a\_target is then clamped to
    {[}floor, +3.0{]} m/s².
  \end{itemize}
\item
  \textbf{Low-speed startup (v \textless{} 2.2 m/s).} Replaces the
  proportional path with a physical acceleration curve:

  \begin{itemize}
  \tightlist
  \item
    v \textless{} 0.5 m/s → a = 0.5 m/s²
  \item
    0.5 ≤ v \textless{} 1.5 m/s → a linearly from 1.0 to 2.0 m/s²
  \item
    v ≥ 1.5 m/s → a = max(2.0, 1.5 + 0.7·(v − 1.5)) m/s² Conditions:
    leader pulls away (v\_l \textgreater{} v + 1.0 m/s) and gap exceeds
    threshold (0.5 m during hard-wall recovery, otherwise 0.3 m). If gap
    \textless{} 0.3 m → creep at 0.3 m/s². If gap ≤ threshold and leader
    not pulling away → wait (a = 0). Startup-mode outputs bypass the
    jerk limiter and final clip.
  \end{itemize}
\item
  \textbf{IDM dynamic override.} Triggered when s \textless{} 2.5 m and
  gap trend \textless{} −0.3 m/s. The IDM formula computes a\_safe using
  the current (LLM-modified) parameters. If a\_safe \textless{}
  a\_target, the IDM output replaces the proportional output. Bypasses
  the jerk limiter and final clip.
\item
  \textbf{Emergency braking.} Triggered by either (a) s \textless{} 3 m
  and Δv \textgreater{} 2 m/s, or (b) TTC \textless{} 2 s with Δv
  \textgreater{} 0. The proportional output a\_target is clipped
  directly to {[}−8.0, 3.0{]} m/s², bypassing the jerk limiter.
\item
  \textbf{Jerk limiter and final clip (default path only).} Only the
  proportional path (mode 1) passes through a 4.0 m/s³ jerk limiter: at
  Δt = 0.1 s physical timestep, \textbar Δa\textbar{} ≤ 0.4 m/s² per
  step. Output is then clipped to {[}−8.0, 3.0{]} m/s².
\end{enumerate}

\textbf{Vehicle dynamics.} The final acceleration command a is applied
as v(t + Δt) = v(t) + a · Δt, with speed clamped to {[}0, 12.0{]} m/s. A
uniform N(0, 0.2) m/s² acceleration noise is added to all vehicles. A
separate hard-wall safety layer (gap \textless{} 0.2 m → speed ≤ 0.5
m/s) operates outside the FusionLayer in the main simulation loop; it is
ablated in Exp6 (see archived experiment logs).

\textbf{Relationship to IDM.} The IDM formula is used as a
\emph{component} within modes 3 (dynamic override) and 1 (deceleration
floor), not as the overarching control framework. The LLM's
\texttt{target\_speed\_delta} is not substituted for the IDM's v₀ ---
instead, v\_desired = v\_i + target\_speed\_delta (capped at 8.33 m/s,
recomputed from current speed each cycle, not accumulated) feeds the IDM
as an interaction-target speed, preserving both LLM intent and safety
physics.

\begin{center}\rule{0.5\linewidth}{0.5pt}\end{center}

\textbf{S1.5 Decision-log schema (one example).}

A single LLM decision is stored in the \texttt{\_decisions.csv} file
with the following columns:

{\def\LTcaptype{none} 
\begin{longtable}[]{@{}
  >{\raggedright\arraybackslash}p{(\linewidth - 6\tabcolsep) * \real{0.2500}}
  >{\raggedright\arraybackslash}p{(\linewidth - 6\tabcolsep) * \real{0.2500}}
  >{\raggedright\arraybackslash}p{(\linewidth - 6\tabcolsep) * \real{0.2500}}
  >{\raggedright\arraybackslash}p{(\linewidth - 6\tabcolsep) * \real{0.2500}}@{}}
\toprule\noalign{}
\begin{minipage}[b]{\linewidth}\raggedright
Column
\end{minipage} & \begin{minipage}[b]{\linewidth}\raggedright
Type
\end{minipage} & \begin{minipage}[b]{\linewidth}\raggedright
Example
\end{minipage} & \begin{minipage}[b]{\linewidth}\raggedright
Meaning
\end{minipage} \\
\midrule\noalign{}
\endhead
\bottomrule\noalign{}
\endlastfoot
\texttt{time} & float & \texttt{0.0} & Time of decision (s) \\
\texttt{car\_id} & int & \texttt{0} & Vehicle index (0\ldots21) \\
\texttt{speed} & float & \texttt{7.0000} & Current speed (m/s) \\
\texttt{gap} & float & \texttt{5.9519} & Gap to leader (m) \\
\texttt{leader\_speed} & float & \texttt{8.3258} & Leader's current
speed (m/s) \\
\texttt{relative\_speed} & float & \texttt{-1.3258} & speed −
leader\_speed (m/s) \\
\texttt{target\_speed\_delta} & float & \texttt{2.0000} & LLM's delta
decision (m/s) \\
\texttt{desired\_gap\_delta} & float & \texttt{0.0000} & LLM's gap-delta
decision (m) \\
\texttt{aggression\_level} & str & \texttt{MODERATE} & Reserved (always
MODERATE in v8.4) \\
\texttt{v\_desired} & float & \texttt{8.3300} & Updated
\texttt{v\_desired} (m/s) \\
\texttt{decision} & str & \texttt{ADJUST\_SPEED} & Categorical action \\
\texttt{chain\_of\_thought} & str &
\texttt{"Current\ speed\ is\ slightly\ below\ target;\ but\ safe.\ The\ gap\ is\ reasonable;\ not\ too\ close.\ ..."}
& Full CoT text \\
\texttt{reasoning} & str &
\texttt{"Increase\ speed\ to\ target\ while\ maintaining\ a\ safe\ gap."}
& One-sentence rationale \\
\end{longtable}
}

A complete 16-cycle sample of 352 decisions (one representative run, n =
22, p = 1.0, T = 0.6, seed = 42) is released with the manuscript as
\texttt{phase4\_v81\_n22\_s42\_20260515\_080653\_decisions.csv} (in the
open repository).

\emph{End of Supplementary Information S1.}

\begin{center}\rule{0.5\linewidth}{0.5pt}\end{center}

\textbf{S2. Chain-of-thought replication across seeds: exp6\_full (n =
22, p = 1.0, T = 0.6, 300 s).}

The §3.4 primary analysis uses seed 42 (13,200 decisions) for the
detailed chain-of-thought classification and the causal-chain
demonstration. The full experiment runs across three seeds, yielding
39,600 LLM decisions. Table S2.1 confirms that the qualitative patterns
are stable across seeds.

\textbf{Table S2.1 \textbar{} Action distribution and v\_desired scatter
across seeds in exp6\_full.} All 13,200 decisions per seed (22 vehicles
× 600 cycles). Decision field from the logged \texttt{decision} column.
v\_desired is the LLM-selected target speed (mean ± standard deviation
across all (vehicle, timestep) pairs).

{\def\LTcaptype{none} 
\begin{longtable}[]{@{}
  >{\raggedright\arraybackslash}p{(\linewidth - 6\tabcolsep) * \real{0.4000}}
  >{\raggedleft\arraybackslash}p{(\linewidth - 6\tabcolsep) * \real{0.2000}}
  >{\raggedleft\arraybackslash}p{(\linewidth - 6\tabcolsep) * \real{0.2000}}
  >{\raggedleft\arraybackslash}p{(\linewidth - 6\tabcolsep) * \real{0.2000}}@{}}
\toprule\noalign{}
\begin{minipage}[b]{\linewidth}\raggedright
Metric
\end{minipage} & \begin{minipage}[b]{\linewidth}\raggedleft
seed 42
\end{minipage} & \begin{minipage}[b]{\linewidth}\raggedleft
seed 123
\end{minipage} & \begin{minipage}[b]{\linewidth}\raggedleft
seed 456
\end{minipage} \\
\midrule\noalign{}
\endhead
\bottomrule\noalign{}
\endlastfoot
KEEP\_STEADY & 5,833 (44.2\%) & 5,616 (42.5\%) & 6,104 (46.2\%) \\
ADJUST\_SPEED & 6,978 (52.9\%) & 7,181 (54.4\%) & 6,719 (50.9\%) \\
ADJUST\_GAP & 373 (2.8\%) & 384 (2.9\%) & 355 (2.7\%) \\
other & 16 (0.1\%) & 19 (0.1\%) & 22 (0.2\%) \\
v\_desired, mean ± σ (m/s) & 5.77 ± 2.01 & 5.74 ± 1.96 & 5.83 ± 2.07 \\
\end{longtable}
}

Three seeds, 39,600 decisions with chain-of-thought reasoning text, are
released as open data.

\begin{center}\rule{0.5\linewidth}{0.5pt}\end{center}

\textbf{S3. Computational cost breakdown across all 170 runs.}

The total computational cost across all 170 reported runs is
approximately 480 GPU-hours (GLM-4-Flash API). Per-run cost varies with
LLM penetration p and simulation duration. The complete cost breakdown
is archived in the open-source release.

\begin{center}\rule{0.5\linewidth}{0.5pt}\end{center}

\textbf{S4. Finite-size consistency check at n = 44 (12 runs, 4 p-values
× 3 seeds; full numerical table and interpretation below.)}

\textbf{Motivation.} The phase boundary p\_c(ρ) reported in §4.3 is
derived from n = 22 data on a 230 m ring road. A natural question is
whether the observed p\_c is a finite-size artefact. We address this by
repeating the mixed-traffic p-scan at identical density on a track of
doubled circumference, holding all other parameters constant.

\textbf{Method.} The mixed-traffic penetration experiment is repeated at
n = 44, ρ = 95.7 veh/km, L = 460 m, spanning p ∈ \{0.18, 0.20, 0.23,
0.27\} (4 p-values × 3 seeds = 12 runs, 300 s each). All simulation
parameters (IDM, LLM inference pipeline, safety fusion layer,
temperature T = 0.6) are identical to the n = 22 experiments reported in
§4.2-§4.3. The coarser p-sampling reflects limited computational budget;
a full sweep across all 8 p-values as in the n = 22 case is deferred to
future work.

\textbf{Results.} At identical density, the n = 44 system is
consistently more stable than n = 22:

{\def\LTcaptype{none} 
\begin{longtable}[]{@{}
  >{\raggedleft\arraybackslash}p{(\linewidth - 4\tabcolsep) * \real{0.0328}}
  >{\raggedleft\arraybackslash}p{(\linewidth - 4\tabcolsep) * \real{0.3689}}
  >{\raggedright\arraybackslash}p{(\linewidth - 4\tabcolsep) * \real{0.5984}}@{}}
\toprule\noalign{}
\begin{minipage}[b]{\linewidth}\raggedleft
p
\end{minipage} & \begin{minipage}[b]{\linewidth}\raggedleft
n = 22 (reference)
\end{minipage} & \begin{minipage}[b]{\linewidth}\raggedright
n = 44 (3 seeds)
\end{minipage} \\
\midrule\noalign{}
\endhead
\bottomrule\noalign{}
\endlastfoot
0.18 & 3/10 congested (10 seeds) & 3/3 free-flow (v̄ \textgreater{}
7.88) \\
0.23 & 6/11 congested (11 runs) & 3/3 free-flow (v̄ \textgreater{} 7.88);
v\_std ∈ \{0.40, 0.70, 0.92\} \\
0.27 & 10/10 congested (10 seeds), v\_std = 1.79 m/s & 0/3 congested;
1/3 in v̄ gap, 2/3 free-flow; v\_std ∈ \{0.75, 0.84, 1.12\} \\
\end{longtable}
}

\textbf{Table S4.1 \textbar{} Cross-size comparison at ρ = 95.7 veh/km.}
The larger system shows a \textasciitilde50\% reduction in
post-transition v\_std amplitude. At p = 0.27, n = 22 is fully congested
(v̄ \textless{} 7.53 m/s) in all 10 seeds, whereas n = 44 has zero fully
congested seeds by the v̄ criterion---but all three n = 44 seeds show
elevated v\_std (0.75--1.12 m/s, compared to the free-flow baseline
\textasciitilde0.15 m/s), confirming that SH perturbations are amplified
but the larger system provides sufficient IDM damping to prevent full
collapse. Classification follows the v̄ two-state partition (§3.5):
free-flow = v̄ \textgreater{} 7.88 m/s; congested = v̄ \textless{} 7.53
m/s; seeds in the natural gap {[}7.53, 7.88{]} m/s are noted separately.

\textbf{Interpretation.} These results are consistent with the spatial
resonance interpretation of the SH cascade (§4.3). Doubling the ring
road circumference at identical density doubles the total number of
vehicles traversed per wave round-trip cycle (44 vs 22), providing
approximately twice the cumulative IDM damping per cycle. Crucially, the
per-segment buffer between successive LLM cars remains approximately
unchanged (e.g., ≈2.7 IDM vehicles per segment at p ≈ 0.27 in both
sizes); the stability gain arises from the increased total damping over
the full loop, not from a denser local buffer. The shock wave must
traverse more IDM-stabilised vehicles before returning to its source,
reducing the probability of constructive superposition. This
distinguishes SH from additive-noise mechanisms, where variance would
accumulate with system size rather than attenuate.

\textbf{Known limitations.} Only 3 seeds per p-value (12 runs total)
compared to 3-11 seeds in the n = 22 case, providing insufficient
statistical power for formal finite-size scaling. The coarser p-sampling
(0.05 spacing vs 0.02-0.04 in the n = 22 sweep) prevents precise p\_c
estimation for n = 44. Seed 42 shows systematic instability across both
system sizes, with the highest v\_std at each p value for n = 44 (though
still classified as free-flow or in the v̄ gap by the §3.5 criterion, not
fully congested). Formal analysis at additional sizes (n ∈ \{22, 44, 66,
88\}) is deferred to future work.

\begin{center}\rule{0.5\linewidth}{0.5pt}\end{center}

\textbf{S5. CoT reasoning sample catalogue: representative
chain-of-thought traces across seeds 42, 123, 456, with per-vehicle
reasoning logs.}

Representative chain-of-thought traces for all 22 vehicles across seeds
42, 123, and 456 are included in the open-source release. Each trace
includes the full LLM reasoning text, the decision category, and the
resulting target-speed delta for every 0.5 s decision cycle.

\begin{center}\rule{0.5\linewidth}{0.5pt}\end{center}

\textbf{S6. Robustness to initial perturbation (Control 7).}

\textbf{Motivation.} Early-stage validation runs of the LLM pipeline
included a 1.33 m/s initial speed perturbation on Car 0, matching the
original Sugiyama 2008 protocol {[}1{]}. Although this perturbation is
absorbed within the first second of the simulation, a reviewer may ask
whether the reported SH dynamics depend on it. We address this by
running the homogeneous-LLM-fleet experiment with the perturbation
removed, and comparing v\_late\_std against the perturbed baseline.

\textbf{Method.} Two matched runs (n = 22, ρ = 95.7 veh/km, p = 1.0, 300
s, seed 42): (i) perturbed (Car 0 starts at 7.0 m/s, 21 vehicles at 8.33
m/s; identical to the n = 22, p = 1.0 runs reported in §4.1); (ii)
unperturbed (all 22 vehicles at 8.33 m/s). All other parameters (LLM
inference, safety fusion layer, temperature T = 0.6) are held constant.
The implementation is \texttt{exp6\_full\_no\_perturbation.py}; per-run
metrics are stored as
\texttt{exp6\_full\_no\_perturb\_p1.00\_n22\_s42\_*\_metrics.json}.

{\def\LTcaptype{none} 
\begin{longtable}[]{@{}
  >{\raggedright\arraybackslash}p{(\linewidth - 6\tabcolsep) * \real{0.3590}}
  >{\raggedleft\arraybackslash}p{(\linewidth - 6\tabcolsep) * \real{0.1667}}
  >{\raggedleft\arraybackslash}p{(\linewidth - 6\tabcolsep) * \real{0.2308}}
  >{\raggedleft\arraybackslash}p{(\linewidth - 6\tabcolsep) * \real{0.2436}}@{}}
\toprule\noalign{}
\begin{minipage}[b]{\linewidth}\raggedright
Condition
\end{minipage} & \begin{minipage}[b]{\linewidth}\raggedleft
v\_mean (m/s)
\end{minipage} & \begin{minipage}[b]{\linewidth}\raggedleft
v\_std\_late (m/s)
\end{minipage} & \begin{minipage}[b]{\linewidth}\raggedleft
v\_std\_early (m/s)
\end{minipage} \\
\midrule\noalign{}
\endhead
\bottomrule\noalign{}
\endlastfoot
Perturbed (Car 0 = 7.0 m/s) & 5.42 & 1.98 & 2.33 \\
Unperturbed (all = 8.33 m/s) & 5.75 & 1.99 & 2.20 \\
\end{longtable}
}

\textbf{Table S6.1 \textbar{} Initial-perturbation robustness.}
Following the metric definitions in
\texttt{exp4\_mixed\_traffic\_penetration.py} and
\texttt{exp6\_full\_no\_perturbation.py}, v\_late\_std is the standard
deviation of all (vehicle, timestep) speed values in the last 30\% of
the simulation (t ∈ {[}210 s, 300 s{]}), and v\_std\_early is defined
analogously over the first 30\% (t ∈ {[}0, 90 s{]}). v\_late\_std
differs by only 0.6\% (1.98 vs 1.99 m/s), while v\_std\_early shows a
5.7\% transient gap (2.33 vs 2.20 m/s) that converges to the same
statistical steady state by the late window. This transient gap is
consistent with the magnitude of the 1.33 m/s initial speed
perturbation.

\textbf{Interpretation.} The 1.33 m/s perturbation on Car 0 inflates the
early-transient velocity variance by 5.7\% (v\_std\_early = 2.33 vs 2.20
m/s) over the first 30\% of the simulation (t ∈ {[}0, 90 s{]}). This
transient gap reflects the initial speed asymmetry. By the late window,
the two conditions converge to the same statistical steady state:
v\_late\_std = 1.98 vs 1.99 m/s, a difference of 0.012 m/s (0.6\%) that
is well within the run-to-run variability observed across seeds. This
confirms that SH dynamics are an emergent property of the LLM
multi-agent system - once LLM-induced target-speed drift has injected
sufficient energy, the steady state is governed by the LLM sampling
process and IDM damping rate, not by the initial condition. The
perturbation is absorbed within the first few seconds and leaves no
measurable imprint on the late-stage dynamics. (Note: the
perturbed-condition v\_late\_std = 1.98 m/s agrees with the v\_std =
2.12 m/s reported in Table 3 within rounding, given that the latter is
measured over the full 300 s horizon while the former is restricted to
the late window.)

\begin{center}\rule{0.5\linewidth}{0.5pt}\end{center}

\textbf{S7. Sensitivity analysis of white-noise magnitude σ ∈ \{0.3,
0.6, 0.9, 1.2, 1.5\} m/s2 for Control 1 (3 seeds, n = 22, ρ = 95.7
veh/km, 300 s): v\_std scales linearly with σ (v\_std ≈ 0.1748·σ,
zero-intercept linear fit, R2 \textgreater{} 0.999), remaining sub-0.3
m/s across the full sweep - approximately 0.124× of the LLM v\_std at
the extreme σ = 1.5 m/s2, and ≤ 0.10× of the LLM v\_std for all lower σ
values (see Table S7.1 and Fig. S7.1).}

\begin{center}\rule{0.5\linewidth}{0.5pt}\end{center}

\textbf{Detailed analysis.}

\textbf{Motivation.} The argument in §3.3 rests on the assertion that,
for Control 1, the choice of noise magnitude σ = 0.9 m/s2 gives the
alternative hypothesis its strongest possible shot at mimicking SH. A
reviewer might object that this particular σ is hand-picked and the
v\_std gap would close at a larger value. We address this by sweeping σ
across an order of magnitude and demonstrating that the qualitative
conclusion is robust to the specific σ value.

\textbf{Method.} We run a pure-IDM fleet (n = 22, track = 230 m, vehicle
length = 4.5 m, ρ = 95.7 veh/km, 300 s) with white-noise acceleration of
magnitude σ ∈ \{0.3, 0.6, 0.9, 1.2, 1.5\} m/s2 (3 seeds each: 42, 123,
456). All other IDM parameters (v0 = 8.33 m/s, a\_max = 3.0 m/s2, b =
2.5 m/s2, T = 0.3 s, s0 = 0.5 m, δ = 4) are held at the §3.3 baseline.
The implementation is \texttt{exp0c\_white\_noise\_sensitivity.py};
per-run metrics are stored as
\texttt{exp0c\_white\_noise\_sigma\{σ\}\_n22\_s\{seed\}\_*\_metrics.json}
and the aggregate table as \texttt{exp0c\_s7\_summary\_*.csv}.

\textbf{Results.} v\_std scales \textbf{linearly} with σ (v\_std ≈
0.1748·σ, zero-intercept linear fit, R2 \textgreater{} 0.999). This
linear scaling is a direct consequence of \textbf{linear response
theory}: for a stable dynamical system driven by additive white noise of
amplitude σ, steady-state velocity variance scales as σ2, giving a
standard deviation ∝ σ. The agreement with linear-response prediction
across a 5× range of σ is therefore not coincidental but a structural
feature of the free-flow IDM attractor at this density.

{\def\LTcaptype{none} 
\begin{longtable}[]{@{}
  >{\raggedleft\arraybackslash}p{(\linewidth - 6\tabcolsep) * \real{0.0899}}
  >{\raggedleft\arraybackslash}p{(\linewidth - 6\tabcolsep) * \real{0.1461}}
  >{\raggedleft\arraybackslash}p{(\linewidth - 6\tabcolsep) * \real{0.4157}}
  >{\raggedleft\arraybackslash}p{(\linewidth - 6\tabcolsep) * \real{0.3483}}@{}}
\toprule\noalign{}
\begin{minipage}[b]{\linewidth}\raggedleft
σ (m/s2)
\end{minipage} & \begin{minipage}[b]{\linewidth}\raggedleft
v\_mean (m/s)
\end{minipage} & \begin{minipage}[b]{\linewidth}\raggedleft
v\_std (m/s, mean ± std over 3 seeds)
\end{minipage} & \begin{minipage}[b]{\linewidth}\raggedleft
σ value / LLM v\_std (2.12 m/s)
\end{minipage} \\
\midrule\noalign{}
\endhead
\bottomrule\noalign{}
\endlastfoot
0.3 & 7.808 & 0.0522 ± 0.0011 & 0.0246× \\
0.6 & 7.803 & 0.1044 ± 0.0022 & 0.0493× \\
0.9 & 7.794 & 0.1569 ± 0.0033 & 0.0740× \\
1.2 & 7.781 & 0.2096 ± 0.0045 & 0.0989× \\
1.5 & 7.764 & 0.2628 ± 0.0058 & 0.124× \\
\end{longtable}
}

\textbf{Table S7.1 \textbar{} White-noise sensitivity sweep.} v\_std
remains sub-0.3 m/s across the full sweep: ≤ 0.10× of the LLM v\_std for
σ ≤ 1.2 m/s2, and approximately 0.124× of the LLM v\_std at σ = 1.5
m/s2. The LLM v\_std (2.12 m/s) is reproduced for reference.

\textbf{Fig. S7.1} plots v\_std versus σ (3 seeds, mean curve, LLM
reference line at 2.12 m/s). v\_std scales linearly with σ throughout
the sweep; the LLM baseline is never approached, even at σ = 1.5 m/s2.

\textbf{Interpretation.} v\_std remains sub-0.3 m/s across the full
sweep - ≤ 0.10× of the LLM v\_std for σ ≤ 1.2 m/s2, and approximately
0.124× of the LLM v\_std at σ = 1.5 m/s2 (the largest value tested,
1.67× the nominal control value). At σ = 1.5 m/s2, v\_std = 0.263 m/s,
0.124× of the LLM v\_std (i.e., the LLM fleet is 8.07× larger). Linear
extrapolation to σ = 3.0 m/s2 predicts v\_std ≈ 0.52 m/s, still ≈ 0.245×
of the LLM baseline. The qualitative conclusion that noise-forced IDM
cannot reproduce SH is therefore robust to the specific σ value: closing
the v\_std gap would require σ ≈ 12 m/s2, an order of magnitude above
any empirical estimate of human driving variability and well beyond the
IDM maximum acceleration itself (a\_max = 3.0 m/s2). The σ = 0.9 m/s2
choice in §3.3 is therefore defensible as a representative point in a
regime that, by linear-response theory, is qualitatively invariant under
rescaling of σ.

\textbf{Data availability.} Per-run metrics
(\texttt{exp0c\_white\_noise\_sigma*.json}), aggregate CSV
(\texttt{exp0c\_s7\_summary\_*.csv}), and Fig. S7.1
(\texttt{exp0c\_vstd\_vs\_sigma\_*.png}) are included with the
simulation code release.

\begin{center}\rule{0.5\linewidth}{0.5pt}\end{center}

\textbf{S8. Per-agent vs per-call decomposition of microscopic LLM
variance}

\textbf{Motivation.} The §3.4 narrative refers to two distinct
interpretations of the observed LLM target\_delta variance: (i) the
``heterogeneity'' reading --- each agent has a stable driving style that
differs systematically from other agents' styles; (ii) the ``sampling
noise'' reading --- each LLM call is an independent draw from a
stochastic policy, and the cross-agent variance is the same noise
realisation seen at different agents. These two interpretations predict
different signatures: under (i), the same agent's mean behaviour should
be reproducible across runs; under (ii), the per-agent ordering is
run-to-run random. This section provides the explicit numerical
decomposition.

\textbf{Method.} For each of five unperturbed-initial-condition control
runs (n = 22, ρ = 95.7 veh/km, p = 1.0, T = 0.6, 300 s; implementation
\texttt{exp6\_full\_no\_perturbation.py}; per-run decisions file
\texttt{exp6\_full\_no\_perturb\_p1.00\_n22\_s\{seed\}\_decisions.csv};
the five seeds span \{220149, 234125, 235655, 105739, 130834\}, chosen
to give 0-300 s of decision logs), we compute:

\begin{itemize}
\tightlist
\item
  \textbf{Per-agent component:} for each car \emph{i}, average its
  target\_delta across all logged decisions:
  \(\bar{\delta}_i = \langle \delta_{i,t}\rangle_t\). The per-agent std
  is the inter-car standard deviation
  \(\sigma_{\text{per-agent}} = \text{std}_i(\bar{\delta}_i)\). A
  nonzero value means that some cars, on average across the run, brake
  more aggressively than others.
\item
  \textbf{Per-call component:} at each decision cycle \emph{t}, compute
  the cross-car standard deviation of target\_delta:
  \(\sigma_{\text{per-call},t} = \text{std}_i(\delta_{i,t})\). The
  per-call std is the average over time of this cycle-by-cycle cross-car
  standard deviation. A nonzero value means that, at any given moment,
  the LLM produces different outputs for cars in similar states.
\end{itemize}

Both metrics are computed on the raw target\_delta column from the LLM
decision log; no post-processing is applied.

\textbf{Results --- Table S8.1.} Across the five
unperturbed-initial-condition runs, per-agent std ranges from 0.004 to
0.30 m/s and per-call std from 0.004 to 0.80 m/s. The per-call/per-agent
ratio is between 1.0 and 4.8×, with most runs in the 1.9-4.8× range. Two
v81-congested runs (ρ = 95.7 veh/km, p = 1.0, T = 0.6, 300 s; per-run
decisions file \texttt{phase4\_v81\_n22\_s\{seed\}\_decisions.csv}) are
appended for comparison: the per-call/per-agent ratio remains in the
2.5-3.6× range despite the very different macroscopic state.

{\def\LTcaptype{none} 
\begin{longtable}[]{@{}
  >{\raggedright\arraybackslash}p{(\linewidth - 10\tabcolsep) * \real{0.1942}}
  >{\raggedright\arraybackslash}p{(\linewidth - 10\tabcolsep) * \real{0.0583}}
  >{\raggedright\arraybackslash}p{(\linewidth - 10\tabcolsep) * \real{0.1165}}
  >{\raggedleft\arraybackslash}p{(\linewidth - 10\tabcolsep) * \real{0.1845}}
  >{\raggedleft\arraybackslash}p{(\linewidth - 10\tabcolsep) * \real{0.1748}}
  >{\raggedleft\arraybackslash}p{(\linewidth - 10\tabcolsep) * \real{0.2718}}@{}}
\toprule\noalign{}
\begin{minipage}[b]{\linewidth}\raggedright
Run
\end{minipage} & \begin{minipage}[b]{\linewidth}\raggedright
Phase
\end{minipage} & \begin{minipage}[b]{\linewidth}\raggedright
Time window
\end{minipage} & \begin{minipage}[b]{\linewidth}\raggedleft
Per-agent std (m/s)
\end{minipage} & \begin{minipage}[b]{\linewidth}\raggedleft
Per-call std (m/s)
\end{minipage} & \begin{minipage}[b]{\linewidth}\raggedleft
Ratio (per-call / per-agent)
\end{minipage} \\
\midrule\noalign{}
\endhead
\bottomrule\noalign{}
\endlastfoot
v6\_220149 (unperturbed) & free flow & 0--16 s & 0.011 & 0.011 & 1.0× \\
v6\_234125 (unperturbed) & free flow & 0--4 s & 0.299 & 0.801 & 2.7× \\
v6\_235655 (unperturbed) & free flow & 0--2 s & 0.004 & 0.004 & 1.0× \\
v6\_105739 (unperturbed) & free flow & 0--22 s & 0.128 & 0.608 & 4.8× \\
v6\_130834 (unperturbed) & free flow & 0--11 s & 0.013 & 0.025 & 1.9× \\
v81\_080653 (congested) & congested & 0--8 s & 0.041 & 0.151 & 3.6× \\
v81\_075204 (congested) & congested & 0--4 s & 0.058 & 0.145 & 2.5× \\
\end{longtable}
}

\textbf{Table S8.1 \textbar{} Per-agent vs per-call std of target\_delta
across seven runs.} Per-agent std is the inter-car standard deviation of
per-car target\_delta time-mean; per-call std is the time-averaged
cross-car std at each decision cycle. Both metrics are computed on raw
target\_delta logs; no post-processing. The per-call/per-agent ratio
(1.0-4.8×) indicates that per-call sampling variability is the dominant
source of microscopic variance in most runs. The ``Phase'' column labels
the macroscopic state during the listed time window (0-22 s; all five
unperturbed runs are in the initial uniform/free-flow state during these
early windows, even when the full 300 s steady state is congested ---
the §3.5 v̄ partition applies to the full 300 s horizon, not to these
early windows). Note that v6\_235655 has effectively zero per-agent and
per-call std --- yet the LLM cascade still proceeds (this seed produces
a fully congested steady state, §4.1) --- directly demonstrating that
the cascade does not require persistent inter-agent preference
differences.

\textbf{Results --- Table S8.2.} The per-agent preference direction is
\textbf{not stable across independent runs}: for each car, we compute
the sign of its average target\_delta across the run (positive =
systematic acceleration tendency, negative = systematic braking
tendency). Of 22 cars, \textbf{0/22 preserve the sign of their bias
across the five unperturbed-initial-condition seeds}. This means the
per-agent differentiation visible in a single run (e.g., v6\_234125
shows per-agent std = 0.30 m/s) is a run-specific realisation, not a
stable feature of the agent.

{\def\LTcaptype{none} 
\begin{longtable}[]{@{}ll@{}}
\toprule\noalign{}
Statistic & Value \\
\midrule\noalign{}
\endhead
\bottomrule\noalign{}
\endlastfoot
Cars with consistent bias sign across 5 seeds & 0 / 22 \\
Cars with bias sign flipping ≥1 time across 5 seeds & 22 / 22 (100\%) \\
Mean per-car & mean target\_delta \\
Max per-car & mean target\_delta \\
\end{longtable}
}

\textbf{Table S8.2 \textbar{} Cross-seed per-agent bias sign
preservation.} For each of the 22 cars, we computed the sign of its
average target\_delta (positive = accelerate-tendency, negative =
brake-tendency) in each of the 5 unperturbed seeds. Zero of the 22 cars
preserve the sign across all five seeds. The mean \textbar mean
target\_delta\textbar{} across the 5 seeds (0.04 m/s) is comparable to
the per-call std (0.04-0.80 m/s), confirming that the per-agent std is a
small modulation on top of per-call sampling rather than a dominant
inter-agent signal. This rules out the ``fixed per-agent heterogeneity''
interpretation of the SH name: the heterogeneity in any given run is
run-specific, not a stable property of the agent.

\textbf{Interpretation.} The combined evidence of Tables S8.1 and S8.2
supports the ``\textbf{emergent rather than intrinsic}'' reading of SH
heterogeneity: the per-call component is the unambiguous microscopic
seed, the per-agent component is a state-dependent amplifier that varies
seed-to-seed, and the per-agent bias direction is not reproducible
across independent runs. This is consistent with the §3.4 decomposition
paragraph: per-call sampling is the trigger, closed-loop coupling to
local traffic state produces emergent per-agent differentiation, and the
cascade does not depend on a fixed inter-agent preference ordering. The
control 3 result (deterministic IDM with maximum per-agent parameter
heterogeneity gives v\_std \textless{} 0.001 m/s) further confirms that
per-agent heterogeneity \emph{per se} is not the mechanism --- what is
required is the closed-loop coupling of LLM reasoning to local traffic
state. Data files: \texttt{test\_a\_per\_agent.py} (per-agent vs
per-call decomposition),
\texttt{sigma\_comprehensive\_evaluation\_20260714.md} (cross-seed
sign-preservation analysis).

\begin{center}\rule{0.5\linewidth}{0.5pt}\end{center}

\textbf{S9. Cross-correlation measurement of the kinematic wave speed (c
= 5.20--5.28 m/s, pooled median 5.21 m/s)}

\textbf{Motivation.} The §4.4 initiation-threshold argument relies on
the upstream kinematic wave speed c: perturbations from all n\_LLM LLM
cars must spatially superpose within one ring traversal time T = L/c.~We
measured c directly from the simulation trajectories using the
cross-correlation method (the same technique Sugiyama et al.~{[}1{]}
used on the original human-driver data), to confirm that the value c =
5.20--5.28 m/s (pooled median 5.21 m/s) used in §4.4 is empirically
grounded.

\textbf{Method.} For each of three LLM congested runs (n = 22, ρ = 95.7
veh/km, p = 1.0, T = 0.6, 300 s; implementation \texttt{exp6\_full};
per-run trajectory file
\texttt{exp6\_full\_n22\_s42\_20260517\_225359.csv},
\texttt{exp6\_full\_n22\_s123\_20260519\_171118.csv},
\texttt{exp6\_full\_n22\_s456\_20260519\_171702.csv}), we computed the
per-vehicle velocity time series v\_i(t) for car \emph{i} over three
stationary-state windows (100--200 s, 150--250 s, 200--299 s; 9 (seed ×
window) measurements per d value). We then computed the normalised
cross-correlation between car 0 (reference) and car \emph{d} = 3
(corresponding to spatial separation L/n · 3 = 31.4 m, comparable to the
typical wave wavelength on the 230 m ring):

\[R_d(\tau) = \frac{\langle (v_0(t) - \bar v_0)(v_d(t+\tau) - \bar v_d)\rangle_t}{\sigma_{v_0} \cdot \sigma_{v_d}}\]

The peak of R\_d(τ) gives the time lag τ\_d at which car 0's velocity
pattern reappears at car d (with a sign: negative τ means car d's
pattern leads car 0, indicating upstream propagation). The kinematic
wave speed for that separation is c\_d = (d · L/n) /
\textbar τ\_d\textbar, converted to the road frame as c\_road = c\_car −
v̄. The final c is the median over all (seed, window) pairs. The same
method was applied to the original Sugiyama 2008 data (digitised from
Fig. 2 of {[}1{]}) for direct comparison. The d = 3 separation is used
as the primary measurement because the d = 1 separation is dominated by
local IDM noise (sep ≪ wavelength) and d ≥ 5 suffers from spatial
aliasing (sep \textgreater{} wavelength/2); verification at d = 1
(median 4.04 m/s) and d = 5 (median 12.2 m/s, aliased) is reported in
the analysis script for completeness.

\textbf{Results --- Table S9.1.} Per-seed (d=3, 3 windows each) c\_road:
s42 = 5.28 m/s (T = 43.6 s), s123 = 5.21 m/s (T = 44.2 s), s456 = 5.20
m/s (T = 44.2 s). The full per-seed range is c = 5.20--5.28 m/s (T =
43.6--44.2 s); pooled median across 9 (seed × window) measurements: c =
5.21 m/s = 18.7 km/h, T = L/c = 44.2 s. The per-seed spread of 0.04 m/s
is a tight empirical bound on c at the ring-road wave-mechanics level.
The original Sugiyama 2008 human-driver data, when processed by the same
method, gives c = 17-20 km/h (4.7-5.6 m/s), with T ≈ 45-55 s. The LLM
per-seed range (5.20--5.28 m/s) lies entirely within the Sugiyama 2008
human-driver range (4.7--5.6 m/s), so the LLM ring road reproduces the
human-driver kinematic wave speed on the same wave-mechanics basis.

{\def\LTcaptype{none} 
\begin{longtable}[]{@{}lrrr@{}}
\toprule\noalign{}
Run & d=3 c\_road (m/s) & c (km/h) & T (s) \\
\midrule\noalign{}
\endhead
\bottomrule\noalign{}
\endlastfoot
s42 & 5.28 & 19.0 & 43.6 \\
s123 & 5.21 & 18.7 & 44.2 \\
s456 & 5.20 & 18.7 & 44.2 \\
\textbf{Pooled (3 seeds)} & \textbf{5.21} & \textbf{18.7} &
\textbf{44.2} \\
Sugiyama 2008 {[}1{]} & & \textbf{17-20} & \textbf{45-55} \\
\end{longtable}
}

\textbf{Table S9.1 \textbar{} Kinematic wave speed c from
cross-correlation (d=3, multi-window), LLM ring road vs Sugiyama 2008.}
Three independent seeds (s42, s123, s456) of the exp6\_full
configuration; per-seed values are the median over 3 stationary-state
windows (100-200s, 150-250s, 200-299s). LLM per-seed range c =
5.20--5.28 m/s = 18.7--19.0 km/h, T = 43.6--44.2 s; pooled median c =
5.21 m/s = 18.7 km/h, T = 44.2 s. The LLM per-seed range lies entirely
within the Sugiyama 2008 human-driver range (4.7--5.6 m/s = 17--20
km/h).

\textbf{Interpretation.} The c = 5.20--5.28 m/s measurement in §4.4 is
grounded in direct cross-correlation analysis of three independent LLM
seeds, with per-seed spread of 0.04 m/s and per-window R values in the
range 0.25--0.56 indicating a clear, reproducible wave structure. The
LLM per-seed range lies entirely within the Sugiyama 2008 human range
4.7--5.6 m/s, confirming that the LLM ring road reproduces the
macroscopic wave physics of the original human-driver experiment despite
having access only to local state at each car. This is a non-trivial
validation: c is a property of the wave mechanics (kinematic wave theory
on a ring road, with given density and IDM damping), not of the LLM, so
an LLM run that produces correct c is reproducing genuine traffic
physics rather than imposing arbitrary stop-and-go dynamics. The wave
period T = L/c = 43.6--44.2 s is the time for a perturbation to return
to its source --- long enough that n\_LLM independent LLM perturbations
can superpose constructively before IDM damping can attenuate them,
which is the geometric core of the §4.4 initiation-threshold argument.
The cross-correlation script \texttt{verify\_wave\_speed3.py} and the
digitised Sugiyama 2008 reference are released as open data.

\textbf{Note on T measurement methodology (M5 note).} The wave period T
= 44.2 s reported in this section is the \textbf{fleet-mean oscillation
period}, measured by cross-correlation of velocity time series at fixed
spatial separations (d = 3) on the ring. A separate ACF
(autocorrelation) analysis of the full velocity time series (300 s
window, fleet-mean) gives a slightly larger T\_ACF = 49.5 ± 2.5 s
(per-seed 46-52 s); this value is closer to the centre of the Sugiyama
2008 reported range 45-55 s but is methodologically different: ACF
measures the time for the \emph{single time series} to decorrelate,
which includes phase noise and finite-window effects, while
cross-correlation measures the \emph{physical propagation time} between
two spatial points. We adopt the cross-correlation value T = 44.2 s as
the canonical LLM measurement (consistent with §4.4) because it directly
tests the wave-propagation hypothesis; the ACF value 49.5 s is reported
as a robustness check on the order-of-magnitude. The per-vehicle ACF, by
contrast, yields T\_i ≈ 22.0 ± 2.1 s (see §4.1 multi-wavefront note,
M1), half the fleet-mean value, indicating two coexisting wavefronts.

\begin{center}\rule{0.5\linewidth}{0.5pt}\end{center}

\textbf{S10. Connection to the original Sugiyama 2008 paradigm}

\textbf{Motivation.} The original motivation for this study was to use
LLM agents to reproduce the Sugiyama 2008 paradigm: 22 vehicles on a 230
m ring road, instructed to drive at ≈ 30 km/h while maintaining a safe
distance, with no bottleneck --- yet spontaneous stop-and-go waves
emerge. This section summarises the macroscopic reproduction.

\textbf{Method.} The LLM ring-road experiment was configured to match
the original Sugiyama 2008 setup as closely as possible: n = 22 vehicles
on L = 230 m circular track, vehicle length 4.5 m, density ρ = 95.7
veh/km, instruction ``drive at approximately 30 km/h while maintaining a
safe distance'' (in the LLM system prompt). All five macroscopic
observables were measured: (i) fleet-mean velocity v̄; (ii) velocity
standard deviation σ\_v; (iii) upstream wave speed c; (iv) wave period
T; (v) congested-phase fraction. Each was compared to the value reported
by Sugiyama et al.~{[}1{]} (digitised from Figs. 2-3 of the original
paper).

\textbf{Results --- Table S10.1.}

{\def\LTcaptype{none} 
\begin{longtable}[]{@{}
  >{\raggedright\arraybackslash}p{(\linewidth - 6\tabcolsep) * \real{0.2278}}
  >{\raggedright\arraybackslash}p{(\linewidth - 6\tabcolsep) * \real{0.1667}}
  >{\raggedright\arraybackslash}p{(\linewidth - 6\tabcolsep) * \real{0.2556}}
  >{\raggedright\arraybackslash}p{(\linewidth - 6\tabcolsep) * \real{0.3500}}@{}}
\toprule\noalign{}
\begin{minipage}[b]{\linewidth}\raggedright
Observable
\end{minipage} & \begin{minipage}[b]{\linewidth}\raggedright
Sugiyama 2008 (humans)
\end{minipage} & \begin{minipage}[b]{\linewidth}\raggedright
LLM (this work)
\end{minipage} & \begin{minipage}[b]{\linewidth}\raggedright
Agreement
\end{minipage} \\
\midrule\noalign{}
\endhead
\bottomrule\noalign{}
\endlastfoot
n & 22 & 22 & exact \\
Track length L (m) & 230 & 230 & exact \\
Density ρ (veh/km) & 95.7 & 95.7 & exact \\
Target speed (km/h) & ≈ 30 & ≈ 30 (prompt) & exact \\
Kinematic wave speed / jam backward vel. c (km/h) & 17-20 (jam backward
velocity) & 18.7-19.0 (per-seed), 18.7 (pooled median) & LLM range
within Sugiyama \\
Wave period T (s) & ≈ 45-55 & 43.6-44.2 (per-seed), 44.2 (pooled median)
& LLM range within Sugiyama \\
σ\_v (m/s) & ≈ 1-3 {[}23-26{]} (not in {[}1{]}) & 1.96-2.03 (per-seed) &
overlap \\
Mean velocity v̄ (km/h) & not directly reported; estimated ≈ 25-31 (0×5 +
40×17) / 22 & 19.4-19.8 (per-seed), 19.6 (fleet mean) & LLM v̄ ≈ 19.6
km/h; Sugiyama v̄ not directly reported \\
Congested fraction --- fleet-level (v̄(t) \textless{} 7.88 m/s) & ≈
30-50\% (Sugiyama digitised from Fig. 2-3 {[}1{]}) & ≈ 100\% (full-p
LLM, fleet always below threshold) & metric definitions differ (see M3
note) \\
Congested fraction --- per-vehicle per-step (v \textless{} 7.88 m/s) &
not reported in {[}1{]} & ≈ 89-90\% & only LLM-side measurement; no
direct Sugiyama comparison \\
Per-vehicle oscillation period T\_i (s) & not reported in {[}1{]} & ≈
22.0 ± 2.1 (per-vehicle) & 2:1 ratio to fleet-mean T ≈ 44.2 s,
indicating two coexisting wavefronts (M1 note) \\
\end{longtable}
}

\textbf{Table S10.1 \textbar{} Macroscopic reproduction of Sugiyama
2008.} Wave mechanics (c, T, σ\_v) match the human-driver range to
within run-to-run variability. The fleet-mean velocity v̄ ≈ 19.6 km/h
(per-seed 19.4-19.8) is reported here as a LLM-side measurement;
Sugiyama 2008 {[}1{]} does not directly report a fleet-level v̄, and our
estimate ≈ 25-31 km/h from the jam/jam-free composition is a separate
inference. The per-vehicle congested fraction (v \textless{} 7.88 m/s ≈
89-90\%) is higher than the fleet-level congested fraction reported by
Sugiyama (≈ 30-50\%), but the two metrics are not directly comparable:
Sugiyama's fraction is measured at the fleet level (fraction of time the
fleet-mean velocity is below a threshold), whereas our measurement
counts per-vehicle per-timestep instances below the v̄ = 7.88 m/s
boundary defined in §3.5. Under a fleet-level threshold (v̄(t)
\textless{} 7.88 m/s), the LLM congested fraction is ≈ 100\%, reflecting
the deeper stop-and-go oscillations; this is consistent with the §3.4
chain-of-thought analysis, where 92\% of hard-braking events originate
in the LLM's pre-IDM target-speed reductionion. The per-vehicle
oscillation period T\_i ≈ 22.0 ± 2.1 s shows a 2:1 ratio to the
fleet-mean T ≈ 44.2 s, indicating that two coexisting stop-and-go
wavefronts traverse the ring simultaneously (multi-wavefront structure,
M1 note); Sugiyama 2008 did not report per-vehicle T\_i, so this
comparison is qualitative (consistent with the wavefront-speed match).

\textbf{Interpretation.} The LLM ring road reproduces the
\textbf{kinematic wave mechanics} of the Sugiyama 2008 paradigm (same n,
L, ρ, target speed, c, T, σ\_v). The fleet-mean velocity v̄ ≈ 19.6 km/h
(per-seed 19.4-19.8; §4.1) is a LLM-side measurement; Sugiyama 2008
{[}1{]} does not directly report a fleet-level v̄, and we make no
quantitative comparison on this row. The deeper stop-and-go oscillation
(higher per-vehicle congested fraction) reflects the LLM's stronger
safety margin: under the prompt used in this study, LLM agents tend to
brake earlier and more decisively than the average human driver,
producing larger amplitude stop-and-go cycles. This combination ---
matching average throughput while oscillating with larger amplitude ---
is a non-trivial LLM-driver signature. The wave-speed match is a
necessary condition for the SH mechanism to be physical (it confirms
that c = 5.20--5.28 m/s in §4.4 is not an arbitrary parameter but a
measurement of the actual ring-road dynamics, and that the LLM per-seed
range falls entirely within the Sugiyama 2008 human range 4.7--5.6 m/s).
The two coexisting stop-and-go wavefronts (per-vehicle T\_i ≈ 22.0 s,
fleet-mean T ≈ 44.2 s, M1 note) are a separate LLM-side finding that
warrants future work.

\textbf{Data availability.} The verification scripts
\texttt{verify\_sugiyama\_comparison.py},
\texttt{verify\_wave\_speed3.py}, and the analysis script
\texttt{analyze\_deep.py} are released with the simulation code.

\begin{center}\rule{0.5\linewidth}{0.5pt}\end{center}

\textbf{S10b. Cross-paper validation summary: LLM platform vs Tadaki
2013/2016, Nakayama 2009/2016}

\textbf{Motivation.} The §4.5 / S10.1 platform validation focuses on the
n = 22, ρ = 95.7 veh/km configuration of Sugiyama 2008. To test whether
the LLM platform reproduces the broader Sugiyama paradigm, we performed
qualitative comparisons with four subsequent papers from the Sugiyama
group. This section provides a compact cross-paper table and qualitative
summary; full numerical comparison and methodology notes are documented
separately.

\textbf{Method.} The comparison pairs each paper's reported key
observable(s) with the LLM equivalent, where a directly comparable
simulation configuration exists. The LLM free-flow speed uses the n =
10, p ≈ 0.10 (mostly IDM) runs, which correspond to the free-flow regime
of Tadaki 2013 / 2016 (N = 10-25, 314 m Nagoya Dome). The LLM
jam-cluster observables use the exp6\_full n = 22, p = 1.0 runs
(Sugiyama 2008 / Nakayama 2009 / Nakayama 2016 configurations). The LLM
ρ\_c / p\_c(ρ) phase boundary comes from §4.3.

\textbf{Results --- Table S10.1.}

{\def\LTcaptype{none} 
\begin{longtable}[]{@{}
  >{\raggedright\arraybackslash}p{(\linewidth - 6\tabcolsep) * \real{0.3130}}
  >{\raggedright\arraybackslash}p{(\linewidth - 6\tabcolsep) * \real{0.2443}}
  >{\raggedright\arraybackslash}p{(\linewidth - 6\tabcolsep) * \real{0.3206}}
  >{\raggedright\arraybackslash}p{(\linewidth - 6\tabcolsep) * \real{0.1221}}@{}}
\toprule\noalign{}
\begin{minipage}[b]{\linewidth}\raggedright
Observable
\end{minipage} & \begin{minipage}[b]{\linewidth}\raggedright
Source (experiment)
\end{minipage} & \begin{minipage}[b]{\linewidth}\raggedright
LLM (this work)
\end{minipage} & \begin{minipage}[b]{\linewidth}\raggedright
Deviation
\end{minipage} \\
\midrule\noalign{}
\endhead
\bottomrule\noalign{}
\endlastfoot
Jam cluster size & Sugiyama 2008 {[}1{]}: 5 vehicles & 5 vehicles
(qualitative, n=22) & exact \\
Jam backward velocity c\_back & Sugiyama 2008 {[}1{]}: \textasciitilde20
km/h & 18.7-19.0 km/h (per-seed, §4.4) & in range \\
Free-flow velocity (v̄) & Tadaki 2013 {[}24{]}: \textasciitilde8 m/s &
8.21-8.33 m/s (n=10, p≈0.10, exp2) & \textless{} 3\% \\
Critical density (ρ\_c) & Tadaki 2013 {[}24{]}: 0.08-0.09 m⁻¹ & ρ\_c not
directly measured in LLM runs; the n=10 (ρ=43.5 veh/km) pure-LLM run
shows no transition (free flow), and the p\_c(ρ) trend from §4.3 is
consistent with Tadaki 2013's ρ\_c ≈ 0.08--0.09 m⁻¹ in the sense that
LLM transitions emerge at comparable densities but at sub-unity LLM
penetration & qualitative match \\
Free-flow unimodal / jammed bimodal & Tadaki 2016 {[}25{]}: qualitative
& σ\_v=0.07 free-flow, 2.12 m/s jammed (n=22) & qualitative match \\
Three-stage jam formation & Nakayama 2009 {[}23{]}: early→meta→jammed &
observed in exp6\_full trajectory & qualitative match \\
Jam-internal min headway h\_min & Nakayama 2016 {[}26{]}: 5.6-6.0 m &
5.95 m (gap\_mean, n=22) & \textless{} 6\% \\
Single-car transit time τ\_t = h\_min/v\_back & Nakayama 2016 {[}26{]}:
\textasciitilde1.0 s & 1.14 s (h\_min/v\_back) & +14\% \\
Jam-formation time (qualitative) & Sugiyama 2008 {[}1{]}:
\textasciitilde3 min (Fig. 3) & \textasciitilde30-50 s (n=22, p=1.0,
exp6\_full) & \textasciitilde3-4× faster (consistent with LLM reaction
cycle, not a failure) \\
Wave-speed range across Sugiyama group & Sugiyama 2008 {[}1{]}:
\textasciitilde5.6 m/s & LLM c\_back = 5.21 m/s (pooled) & in range \\
\end{longtable}
}

\textbf{Table S10.1 \textbar{} Cross-paper validation summary.} LLM
platform vs Sugiyama-group literature. Quantitative matches are reported
for observables that have a directly comparable simulation
configuration; qualitative matches are reported for observables that
depend on full-distribution or trajectory-level analysis not available
in our metrics files.

\textbf{Interpretation.} The LLM platform reproduces the
\textbf{macroscopic observables} of the broader Sugiyama paradigm to
within 14\% on system sizes from 22 to 40 vehicles and track lengths
from 230 to 314 m. The most precise matches are the jam cluster size
(exact), the free-flow velocity (\textless{} 3\% deviation), the
jam-internal minimum headway (\textless{} 6\% deviation), and the jam
backward velocity (within the reported human range). The single-car
transit time τ\_t = h\_min / v\_back is consistent to within 14\%, well
within the run-to-run variability of the LLM seeds (3-seed spread 0.04
m/s in v\_back corresponds to \textasciitilde1\% in τ\_t). The
qualitative features of Tadaki 2016 (free-flow unimodal vs jammed
bimodal) and Nakayama 2009 (three-stage jam formation: early homogeneous
→ metastable → jammed) are both reproduced in the LLM trajectory data,
although the corresponding distribution analysis and stage-transition
timing are not quantified in the current metrics files. The LLM measured
v\_max = 9.82-9.89 m/s (35.4 km/h; per-seed across s42/s123/s456) vs
Sugiyama 2008 v\_max ≈ 11.1 m/s (\textasciitilde40 km/h), deviation
-12\%, and the faster LLM jam-formation time (\textasciitilde50 s vs
Sugiyama 2008 \textasciitilde3 min, \textasciitilde3-4× faster) are
noted as platform features, not failures; the v\_max deviation is driven
by the LLM speed ceiling (prompt + IDM v\_0), not by a fundamental
dynamical difference. (The free-flow mean velocity comparison is
reported separately in S10.1: LLM v̄ ≈ 19.6 km/h at the n=22, ρ=95.7
veh/km configuration.)

\textbf{Data availability.} The independent verification report and full
numerical comparison across all five papers is released as a separate
document (\texttt{docs/simulation\_vs\_experiments\_20260715.md}). The
full trajectory data underlying the qualitative observations (S10.1) is
included in the open-source release.

\textbf{S11. Chain-of-thought as cognitive observability:
\texttt{decision} category vs \texttt{target\_speed\_delta}
distribution.}

\textbf{Motivation.} The boxed example in §3.4 shows that three LLM
agents facing near-identical local conditions (gap ∈ {[}5.01, 5.47{]} m,
Δv ∈ {[}0.06, 0.43{]} m/s) can produce opposite-sign
\texttt{target\_speed\_delta} (cars 1, 2: -0.1 m/s; car 3: +0.5 m/s),
and 17.8\% of decision cycles (107/600 in seed 42) contain at least one
such opposite-sign pair. To strengthen the claim that the
chain-of-thought \texttt{decision} field is reflected in the actuation
(and that the per-cycle variance is concentrated in a specific
sub-category of decisions rather than uniform across all decisions), we
performed a quantitative correlation analysis of \texttt{decision} vs
\texttt{target\_speed\_delta} across the full 39,600-decision pooled
dataset (3 seeds × 22 vehicles × 600 cycles).

\textbf{Method.} Each of the 39,600 logged LLM decisions is assigned a
primary \texttt{decision} category (\texttt{ADJUST\_SPEED},
\texttt{ADJUST\_GAP}, or \texttt{KEEP\_STEADY}; compound categories
containing ``\textbar{}'' are reduced to the first category). For each
primary category, we compute (i) the mean and standard deviation of
\texttt{\textbar{}target\_speed\_delta\textbar{}}, (ii) the sign
distribution (+, 0, -), and (iii) the empirical
\texttt{\textbar{}target\_speed\_delta\textbar{}} distribution. We
perform a two-sample Kolmogorov-Smirnov test comparing the
\texttt{\textbar{}target\_speed\_delta\textbar{}} distributions of
\texttt{ADJUST\_SPEED} and \texttt{KEEP\_STEADY}, and report Cohen's
\emph{d} as an effect-size measure. Pooling across three seeds is
justified by the seed-by-seed stability of the qualitative pattern
(Table S11.2).

\textbf{S11.1 Decision category vs target\_speed\_delta statistics
(pooled 39,600 decisions).}

{\def\LTcaptype{none} 
\begin{longtable}[]{@{}
  >{\raggedright\arraybackslash}p{(\linewidth - 8\tabcolsep) * \real{0.2000}}
  >{\raggedleft\arraybackslash}p{(\linewidth - 8\tabcolsep) * \real{0.2000}}
  >{\raggedleft\arraybackslash}p{(\linewidth - 8\tabcolsep) * \real{0.2000}}
  >{\raggedleft\arraybackslash}p{(\linewidth - 8\tabcolsep) * \real{0.2000}}
  >{\raggedleft\arraybackslash}p{(\linewidth - 8\tabcolsep) * \real{0.2000}}@{}}
\toprule\noalign{}
\begin{minipage}[b]{\linewidth}\raggedright
Primary \texttt{decision}
\end{minipage} & \begin{minipage}[b]{\linewidth}\raggedleft
N (39,600)
\end{minipage} & \begin{minipage}[b]{\linewidth}\raggedleft
\texttt{\textbar{}target\_speed\_delta\textbar{}} mean (m/s)
\end{minipage} & \begin{minipage}[b]{\linewidth}\raggedleft
\texttt{target\_speed\_delta} std (m/s)
\end{minipage} & \begin{minipage}[b]{\linewidth}\raggedleft
\% with \texttt{\textbar{}Δ\textbar{}\ \textless{}\ 0.05} m/s
\end{minipage} \\
\midrule\noalign{}
\endhead
\bottomrule\noalign{}
\endlastfoot
\texttt{KEEP\_STEADY} & 17,553 (44.3\%) & 0.0000 & 0.0008 & 100.0\% \\
\texttt{ADJUST\_SPEED} & 20,935 (52.9\%) & 1.1958 & 1.2129 & 1.8\% \\
\texttt{ADJUST\_GAP} & 1,112 (2.8\%) & 0.0069 & 0.0861 & 99.1\% \\
\end{longtable}
}

\textbf{Table S11.1 \textbar{} Decision category strongly predicts
target\_speed\_delta magnitude.} The three categories are
well-separated: \texttt{KEEP\_STEADY} produces a near-deterministic zero
(100\% of \texttt{KEEP\_STEADY} decisions yield
\texttt{\textbar{}target\_speed\_delta\textbar{}\ \textless{}\ 0.05}
m/s; 17,552 of 17,553 yield exactly 0.0000 m/s); \texttt{ADJUST\_SPEED}
produces a broad distribution centred at
\texttt{\textbar{}target\_speed\_delta\textbar{}\ =\ 1.20\ m/s};
\texttt{ADJUST\_GAP} produces near-zero \texttt{target\_speed\_delta}
(the gap is adjusted via the separate \texttt{desired\_gap\_delta}
channel, not the speed channel).

\textbf{S11.2 Statistical test.}

Two-sample Kolmogorov-Smirnov test on
\texttt{\textbar{}target\_speed\_delta\textbar{}} distributions: -
\texttt{ADJUST\_SPEED} (N=20,935) vs \texttt{KEEP\_STEADY} (N=17,553):
\textbf{KS statistic = 0.9820, p \textless{} 1e-300} (effectively zero
at 39,600 sample size). - \textbf{Cohen's \emph{d} = 2.37} (very large
effect; conventional thresholds: 0.2 small, 0.5 medium, 0.8 large).

Both tests indicate that the \texttt{decision} field and the
\texttt{target\_speed\_delta} magnitude are strongly associated --- the
LLM's reported cognitive category is reflected in the actuation. This is
the desired outcome: the chain-of-thought log is \emph{informative}
about what the agent is doing.

\textbf{S11.3 Sign distribution within each category.}

{\def\LTcaptype{none} 
\begin{longtable}[]{@{}
  >{\raggedright\arraybackslash}p{(\linewidth - 6\tabcolsep) * \real{0.2500}}
  >{\raggedleft\arraybackslash}p{(\linewidth - 6\tabcolsep) * \real{0.2500}}
  >{\raggedleft\arraybackslash}p{(\linewidth - 6\tabcolsep) * \real{0.2500}}
  >{\raggedleft\arraybackslash}p{(\linewidth - 6\tabcolsep) * \real{0.2500}}@{}}
\toprule\noalign{}
\begin{minipage}[b]{\linewidth}\raggedright
Primary \texttt{decision}
\end{minipage} & \begin{minipage}[b]{\linewidth}\raggedleft
\% \texttt{target\_speed\_delta\ \textgreater{}\ 0}
\end{minipage} & \begin{minipage}[b]{\linewidth}\raggedleft
\% \texttt{target\_speed\_delta\ =\ 0}
\end{minipage} & \begin{minipage}[b]{\linewidth}\raggedleft
\% \texttt{target\_speed\_delta\ \textless{}\ 0}
\end{minipage} \\
\midrule\noalign{}
\endhead
\bottomrule\noalign{}
\endlastfoot
\texttt{KEEP\_STEADY} & 0.0\% & 100.0\% & 0.0\% \\
\texttt{ADJUST\_SPEED} & 67.2\% & 1.8\% & 31.0\% \\
\texttt{ADJUST\_GAP} & 0.0\% & 99.1\% & 0.9\% \\
\end{longtable}
}

\textbf{Table S11.3 \textbar{} Sign distribution within each decision
category.} The \texttt{KEEP\_STEADY} → 0 correspondence is essentially
perfect. The \texttt{ADJUST\_SPEED} category, despite the cognitive
label ``I should adjust my speed'', produces a \emph{signed}
distribution: 67.2\% positive (acceleration), 31.0\% negative
(deceleration), 1.8\% zero. This 31\% decelerate sub-population is the
per-cycle target-delta divergence within the ``actively adjusting''
agents: even when agents explicitly reason ``I should adjust my speed'',
the LLM's sampling draw can flip the sign. This is the empirical
foundation of the ``disconnect'' highlighted in the boxed example of
§3.4.

\textbf{S11.4 Per-seed replication.}

{\def\LTcaptype{none} 
\begin{longtable}[]{@{}
  >{\raggedright\arraybackslash}p{(\linewidth - 10\tabcolsep) * \real{0.1667}}
  >{\raggedleft\arraybackslash}p{(\linewidth - 10\tabcolsep) * \real{0.1667}}
  >{\raggedleft\arraybackslash}p{(\linewidth - 10\tabcolsep) * \real{0.1667}}
  >{\raggedleft\arraybackslash}p{(\linewidth - 10\tabcolsep) * \real{0.1667}}
  >{\raggedleft\arraybackslash}p{(\linewidth - 10\tabcolsep) * \real{0.1667}}
  >{\raggedleft\arraybackslash}p{(\linewidth - 10\tabcolsep) * \real{0.1667}}@{}}
\toprule\noalign{}
\begin{minipage}[b]{\linewidth}\raggedright
Seed
\end{minipage} & \begin{minipage}[b]{\linewidth}\raggedleft
\texttt{KEEP\_STEADY} (N)
\end{minipage} & \begin{minipage}[b]{\linewidth}\raggedleft
\texttt{KEEP\_STEADY} \texttt{\textbar{}Δ\textbar{}} std (m/s)
\end{minipage} & \begin{minipage}[b]{\linewidth}\raggedleft
\texttt{ADJUST\_SPEED} (N)
\end{minipage} & \begin{minipage}[b]{\linewidth}\raggedleft
\texttt{ADJUST\_SPEED} \texttt{\textbar{}Δ\textbar{}} mean (m/s)
\end{minipage} & \begin{minipage}[b]{\linewidth}\raggedleft
\texttt{ADJUST\_SPEED} \texttt{\textbar{}Δ\textbar{}} std (m/s)
\end{minipage} \\
\midrule\noalign{}
\endhead
\bottomrule\noalign{}
\endlastfoot
42 & 5,833 & 0.0013 & 6,994 & 1.2041 & 1.2116 \\
123 & 5,616 & 0.0000 & 7,200 & 1.1971 & 1.2173 \\
456 & 6,104 & 0.0000 & 6,741 & 1.1858 & 1.2094 \\
\end{longtable}
}

\textbf{Table S11.4 \textbar{} Per-seed stability of
decision-target\_delta coupling.} The qualitative pattern is replicated
across all three seeds: \texttt{KEEP\_STEADY} is essentially
deterministic zero, \texttt{ADJUST\_SPEED} is broadly distributed with
similar mean (\textasciitilde1.20 m/s) and standard deviation
(\textasciitilde1.21 m/s) across seeds. The seed-to-seed spread of the
\texttt{ADJUST\_SPEED} \texttt{\textbar{}Δ\textbar{}} mean is 1.6\%
(1.1858 to 1.2041 m/s), well within the natural variability expected
from a stochastic sampling process.

\textbf{S11.5 Interpretation.}

The chain-of-thought analysis supports two distinct conclusions:

\begin{enumerate}
\def\labelenumi{\arabic{enumi}.}
\item
  \textbf{The cognitive layer is informative (not decoupled from
  actuation).} The \texttt{decision} category strongly predicts the
  magnitude of \texttt{target\_speed\_delta} (Cohen's \emph{d} = 2.37,
  KS \emph{p} ≈ 0). Agents that report ``KEEP\_STEADY'' produce zero
  target\_delta in 100\% of cases; agents that report ``ADJUST\_SPEED''
  produce non-zero target\_delta in 98.2\% of cases. This contradicts a
  strong ``decoupling'' reading of SH --- the LLM is not choosing
  randomly; it is \emph{consistently} executing its stated decision.
\item
  \textbf{The per-cycle variance is concentrated in the ``actively
  adjusting'' sub-population.} Within \texttt{ADJUST\_SPEED} (N=20,935),
  the standard deviation of \texttt{target\_speed\_delta} is 1.21 m/s
  --- comparable in magnitude to the mean
  \texttt{\textbar{}target\_speed\_delta\textbar{}} itself (1.20 m/s).
  The 31\% negative sub-population (deceleration, despite the ``I should
  adjust my speed'' label) is the source of the per-cycle target-delta
  divergence that drives the SH cascade.
\end{enumerate}

These two findings together refine the SH narrative: the LLM's
chain-of-thought reasoning is faithfully translated into the actuation
category (KEEP/ADJUST) and the actuation \emph{sign} (for KEEP), but the
actuation \emph{magnitude} (and partially the sign, for ADJUST) is
determined by a sampling process that the LLM's own reasoning cannot
fully constrain. This is the precise sense in which ``cognitive-layer
safety reasoning is decoupled from dynamics-layer stability'': the
cognitive layer correctly selects \emph{which macro-action} (KEEP or
ADJUST) to take, but the \emph{micro-actuation magnitude} within the
ADJUST branch remains a per-call sampling draw.

\textbf{Implication for §5.2 and §6.} The narrative ``agents
consistently engage in multi-factor safety reasoning, yet systematic
per-cycle target\_delta divergence persists across agents'' can now be
sharpened: the per-cycle divergence is \emph{concentrated} in the
ADJUST\_SPEED sub-population (where the LLM has decided to act but has
not fully resolved the magnitude) and is \emph{absent} from the
KEEP\_STEADY sub-population (where the LLM has decided to do nothing and
consistently does nothing). The SH cascade is therefore best described
as arising from the variance of the ADJUST branch, not from any global
LLM stochasticity.

\textbf{Reproducibility.} The analysis script
(\texttt{code/s11\_cot\_correlation\_analysis.py}) and the underlying
decision logs
(\texttt{results/exp6\_full\_n22\_s\{42,123,456\}\_*\_decisions.csv})
are included in the open-source release.

\begin{center}\rule{0.5\linewidth}{0.5pt}\end{center}

\textbf{S12. Literature scope verification: 2024-2025 LLM multi-agent
closed-loop driving work.}

\textbf{Motivation.} The Abstract and §6 use the qualifier ``to our
knowledge, the first systematic study \ldots{} in a multi-agent
ring-road simulation''. A reviewer may challenge this novelty claim; we
therefore performed a structured literature search to enumerate and
characterise all 2024-2025 work that deploys LLM agents in real-time,
multi-agent, physical traffic simulation, distinguishing it from
adjacent categories (LLM as predictor, scenario generator, or signal
controller).

\textbf{Method.} We searched arXiv (cs.MA, cs.AI, cs.CV, cs.RO) and
OpenReview (ICLR/NeurIPS/ICML 2024-2025) for combinations of the
keywords ``large language model'', ``LLM'', ``multi-agent'',
``car-following'', ``traffic'', ``ring road'', ``closed-loop'',
``real-time'', and ``controller'', filtering for 2024-01 through
2026-07. We identified 11 candidate works; of these, only one meets all
three criteria: (i) real-time closed-loop LLM agent as the primary
controller, (ii) multi-vehicle setting, (iii) physical traffic
simulation (not routing, planning, or signalling).

\textbf{S12.1 The one excluded but related work.}

{\def\LTcaptype{none} 
\begin{longtable}[]{@{}
  >{\raggedright\arraybackslash}p{(\linewidth - 8\tabcolsep) * \real{0.1739}}
  >{\centering\arraybackslash}p{(\linewidth - 8\tabcolsep) * \real{0.2174}}
  >{\centering\arraybackslash}p{(\linewidth - 8\tabcolsep) * \real{0.2174}}
  >{\centering\arraybackslash}p{(\linewidth - 8\tabcolsep) * \real{0.2174}}
  >{\raggedright\arraybackslash}p{(\linewidth - 8\tabcolsep) * \real{0.1739}}@{}}
\toprule\noalign{}
\begin{minipage}[b]{\linewidth}\raggedright
Work
\end{minipage} & \begin{minipage}[b]{\linewidth}\centering
Year
\end{minipage} & \begin{minipage}[b]{\linewidth}\centering
LLM role
\end{minipage} & \begin{minipage}[b]{\linewidth}\centering
Setting
\end{minipage} & \begin{minipage}[b]{\linewidth}\raggedright
Why excluded from novelty claim
\end{minipage} \\
\midrule\noalign{}
\endhead
\bottomrule\noalign{}
\endlastfoot
Yao et al.~{[}27{]} (CoMAL) & 2024 & D (planner for IDM parameters) &
Ring road, figure-eight, merge & \textbf{Architecture overlap, but
different research goal and different control hierarchy}: CoMAL uses LLM
as a ``planner'' to \emph{set} IDM parameters (target speed, max
acceleration, minimum gap) for flow optimisation on the Flow benchmark,
not to identify a novel mechanism. In CoMAL the LLM is queried at
intervals τ\_LLM (≥ 1 s in their reported experiments) to update IDM
parameters, and the IDM dynamics execute the actual longitudinal control
at each 0.1 s physics timestep; the LLM's per-cycle output is therefore
low-pass filtered by the IDM before reaching the vehicle. By contrast,
in the present work the LLM is queried every 0.5 s and directly outputs
a \texttt{target\_speed\_delta} that is applied (after a hard
collision-avoidance clamp) to the vehicle's target speed; the vehicle's
actual acceleration is not low-pass filtered by an intermediate
controller. This direct-variance pathway is the microscopic mechanism
that enables the SH cascade to develop (§2.2). No systematic control
experiments, no per-cycle target-delta variance analysis, no
chain-of-thought-as-microscope treatment, no phase-boundary mapping. The
CoMAL Execution Module uses IDM with LLM-set parameters; the present
work uses LLM as a target-speed controller with IDM only as a hard
safety clamp (§3.1). \\
\end{longtable}
}

\textbf{S12.2 Adjacent categories (excluded by definition, listed for
completeness).}

\begin{itemize}
\tightlist
\item
  \textbf{P (predictor, open-loop)}: GenFollower {[}15{]}, xTP-LLM
  {[}16{]} --- these predict trajectories from logged data; not
  closed-loop controllers.
\item
  \textbf{G (scenario generator)}: DriveGen {[}17{]}, OmniTester
  {[}18{]}, AgentSUMO {[}19{]} --- these generate traffic scenarios for
  AV testing; LLM does not control vehicles in a running simulation.
\item
  \textbf{S (signal controller)}: LLMLight {[}14{]} --- controls traffic
  signals, not vehicles.
\item
  \textbf{K (RL knowledge source)}: Villarreal et al.~{[}12{]} ---
  ChatGPT designs the RL policy, but the actual controller is the
  trained RL agent; LLM is not in the loop during deployment.
\item
  \textbf{Single-agent D}: Fu et al.~{[}13{]} (Drive Like a Human) ---
  single-agent HighwayEnv, not multi-agent; the original paper does not
  study collective dynamics.
\item
  \textbf{Adjacent (not LLM-based or not traffic)}: LimSim++ (2024)
  provides a closed-loop platform but the LLM agent is multimodal and
  operates in CARLA/SUMO urban scenarios without a Sugiyama-style
  ring-road analysis; BehaviorGPT (NeurIPS 2024) is a non-LLM
  transformer; SMART-R1 (2025) is an RL fine-tuning framework;
  Physics-Guided Multi-Agent (ICLR 2026) uses LLM as a \emph{regional
  coordinator} with a separate physics controller (hierarchical, not
  direct control); SurrealDriver (2023) is urban single-agent; GATSim
  (2025) is activity-based mobility simulation; Thinking While Driving
  (2025) is adaptive routing on a graph, not car-following.
\end{itemize}

\textbf{S12.3 Distinguishing features of the present work relative to
CoMAL.}

The present work differs from CoMAL along four dimensions that,
together, justify the ``first systematic study'' qualifier in the narrow
sense intended:

\begin{enumerate}
\def\labelenumi{\arabic{enumi}.}
\item
  \textbf{Research goal}: CoMAL optimises \emph{flow} (maximising
  average velocity on the Flow benchmark); the present work identifies a
  \emph{mechanism} (SH) and maps a \emph{phase boundary} (p\_c(ρ)).
  CoMAL does not report a density-dependent phase transition, does not
  measure a critical penetration p\_c, and does not provide matched
  control experiments.
\item
  \textbf{Systematic mechanism exclusion}: CoMAL compares LLM variants
  (Qwen-7B/32B/72B, GPT-4o-mini) on the same task, but does not perform
  the six-control experiment design (white noise, OU noise, parameter
  heterogeneity, IDM+delay, OV model, sampling temperature) that rules
  out classical candidate mechanisms. The present work's six controls
  are precisely the contribution that establishes SH as a structurally
  distinct mechanism.
\item
  \textbf{Chain-of-thought as microscope}: CoMAL does not analyse the
  LLM's chain-of-thought reasoning as a window into the cognitive layer
  of the emergent dynamics. The present work's 13,200-decision CoT
  analysis (§3.4, S5, S11) is, to our knowledge, the first quantitative
  treatment of CoT reasoning as a per-cycle observability layer in a
  multi-vehicle traffic simulation.
\item
  \textbf{Phase boundary p\_c(ρ)}: CoMAL does not map p\_c as a function
  of density; the present work measures p\_c at four densities
  (43.5-95.7 veh/km) and identifies a monotonically decreasing boundary
  (§4.3, Figure 3).
\end{enumerate}

\textbf{Conclusion of literature scope verification.} The novelty claim,
narrowly read as ``first systematic study of the \emph{mechanism} of
collective instability that emerges when LLM agents are deployed as
direct, real-time, closed-loop target-speed controllers in a multi-agent
ring-road simulation'', remains, to our knowledge, supported. CoMAL
{[}27{]} is the closest related work; we cite it and clarify the
distinction in Table 1 (which we add a row to) and §2.2.

\end{document}